%% file: authordraft2.tex
\documentclass[acmlarge,nonacm]{acmart}

\setcopyright{none}
\AtBeginDocument{%
  \providecommand\BibTeX{{%
    \normalfont B\kern-0.5em{\scshape i\kern-0.25em b}\kern-0.8em\TeX}}}

\acmSubmissionID{}

\usepackage{pdfpages}
\usepackage{dcolumn}
\usepackage{multirow}
\definecolor{darkgreen}{rgb}{0.0, 0.5, 0.0}
\definecolor{darkred}{rgb}{0.82, 0.1, 0.26}

\begin{document}

\title[Shape of You: Implications of Social Context and Avatar Design in a VR Workout]{Shape of You: Implications of Social Context and Avatar Body Shape on Relatedness, Emotions, and Performance in a Virtual Reality Workout}
\author{Jana Franceska Funke}
\email{jana.funke@uni-ulm.de}
\orcid{0000-0002-0635-5078}
\affiliation{%
  \institution{Institute of Media Informatics}
  \streetaddress{James-Franck Ring}
  \city{Ulm}
  \state{}
  \country{Germany}
  \postcode{89081}
}

\author{Ria Matapurkar}
\email{ria.matapurkar@uni-ulm.de}
\orcid{0000-0002-3126-7096}
\affiliation{%
  \institution{Institute of Media Informatics, Ulm University}
  \city{Ulm}
  \country{Germany}
}

\author{Enrico Rukzio}
\email{enrico.rukzio@uni-ulm.de}
\orcid{0000-0002-4213-2226}
\affiliation{%
  \institution{Institute of Media Informatics, Ulm University}
  \city{Ulm}
  \country{Germany}
}

\author{Teresa Hirzle}
\email{tehi@di.ku.dk}
\orcid{0000-0002-7909-7639}
\affiliation{%
  \institution{Department of Computer Science, University of Copenhagen}
  \city{Copenhagen}
  \country{Denmark}
}

\renewcommand{\shortauthors}{Funke, et al.}

\begin{abstract} 
Virtual reality (VR) enables individuals to customize avatars to their liking, for example, by giving them a muscular or slim body shape. Such body differences may influence interactions with others, which is particularly relevant in exercise contexts, where physique may represent athletic abilities. Yet, while \textit{embodying} avatars is well-studied, the influence of working out with \textit{another} avatar remains little explored. To better understand such social VR exercise scenarios, we conducted a user study (N=48), investigating how body type (muscular, slim, obese), sex body features (female, male), and social context (teammate, opponent) influence a person's feelings of relatedness, emotions, and performance while working out in VR. Our results indicate that body type strongly influences relatedness, emotions, and performance, but female/male features have little relevance. Furthermore, teammate conditions positively influence these variables. The results suggest implications for designing immersive fitness experiences that cater to diverse user preferences and needs.


\end{abstract}



\begin{teaserfigure}
  \includegraphics[width=\textwidth]{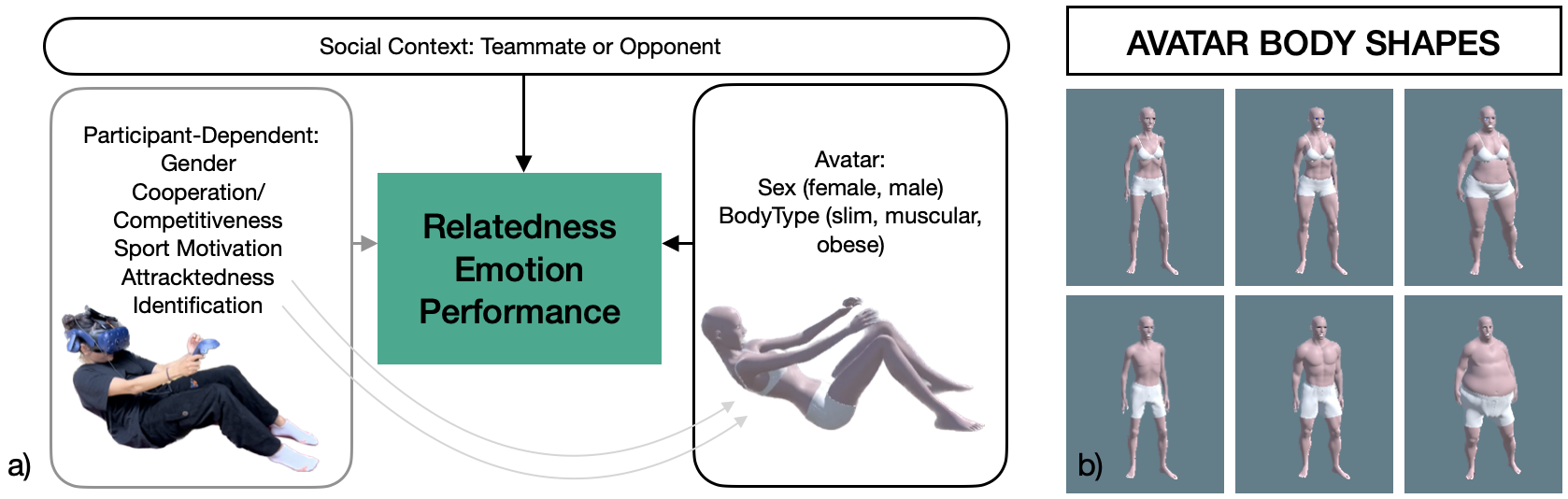}
  \caption{On the left side, (a) the research objective is graphically presented. A participant and a virtual avatar are doing sit-ups next to each other. The participant has characteristics like chosen gender, cooperation and competitiveness scores, general sport motivation, attractedness towards the virtual avatar and identification with the virtual avatar. The virtual avatar has a visual representation that changes throughout the study according to the six avatars presented on the right (b). The setup happens in two different social contexts where the participant and avatar are either teammates or opponents. All these factors from the avatar, the scenario and the participant influence relatedness, emotion, and performance.}
  \Description{In picture (a) on the left, there is a graphic with a long bar above three boxes that says social context: Teammate or Opponent. The left box shows a participant in VR doing sit-ups with a headset on. Above, there is a list saying gender, cooperation/competitiveness, sport motivation, attractedness, and identification. In the right box there is a virtual avatar doing sit-ups as well. Above the avatar, it says sex (female, male) and body type (slim, muscular, obese). From the top left and right box, there is an arrow towards the box in the middle where it says: Relatedness, Emotion, Performance. The picture (b) shows six avatar body shapes. In the top row, they have feminine visual manifestations in slim, muscular, and obese variants. Below, there are masculine visual manifestations in slim, muscular, and obese variations.}
  \label{fig:teaser}
\end{teaserfigure}

\received{}
\maketitle

\section{Introduction}


Virtual reality (VR) workouts offer a cost-effective, time-saving, travel-free, thus an uncomplicated way to exercise.
Furthermore, VR can provide a safe space for individuals who may feel uncomfortable with their body or with going to public fitness centers.
Nevertheless, users may prefer social interaction over isolation during exercise (e.g., to escape loneliness~\cite{savikkoPsychosocialGroupRehabilitation2010,vallerandIntegrativeAnalysisIntrinsic1999}, experience motivational support~\cite{weinsteinYouHaveHear2018,mcwhorterObeseChildMotivation2003,vallerandIntegrativeAnalysisIntrinsic1999,RelativeEffectsPositive}, or engage in competition ~\cite{pengPlayingParallelEffects2013,frankenWhyPeopleCompetition1995,dimenichiPowerCompetitionEffects2015,brondiEvaluatingEffectsCompetition2015}). Consequently, we see a demand for controlled virtual environments that facilitate social exercise experiences in VR - as an area of study that is currently under-explored.

In VR, users act through an avatar whose body can be created and customized to their liking. For example, they can make their avatar more or less muscular than they actually are or change features often attributed to biological sex.
In social sport and exercise scenarios, such body shapes and features are particularly relevant, as they may imply certain abilities. For instance, muscular body shapes may be perceived as more athletic than slim body shapes~\cite{kocurFlexingMusclesVirtual2020b}. Indeed, initial studies indicate that the visual appearance of another person in the room can affect performance~\cite{fittonDancingAvatarsMinimal2023}. For example, \citet{fittonDancingAvatarsMinimal2023} found that performance is improved when interacting with an exercise instructor of the same hair and skin color. However, a systematic analysis of how body shape features influence a person's exercise experience is currently missing.

While it has been investigated how these body features may influence an exercise experience when \textit{embodied}~\cite{kocurFlexingMusclesVirtual2020b}, it is little known about how body types and body features 
of \emph{another} avatar influences a person's exercise experience in VR.
For example, a muscular self-representation may enhance perceived muscle strength~\cite{kocurFlexingMusclesVirtual2020b} and exertion~\cite{kocurPhysiologicalPerceptualResponses2021a}. Still, we do not know how these properties are affected by working out with someone else in the room being represented as a muscular avatar. Yet, understanding such dynamics is essential for designing social VR exercise scenarios.
Therefore, in this paper, we investigate how body shape and body features, like the visual gender of another avatar in the room, influence a person's exercise experience.

By presenting another avatar in the room, the exercise experience becomes a social scenario. Besides the appearance of the avatar, literature suggests that the social context is another factor that influences people's workout experience: 
Working out together as teammates and working out in a competitive scenario influences performance but also other factors like motivation and enjoyment \cite{frankenWhyPeopleCompetition1995,dimenichiPowerCompetitionEffects2015,shawCompetitionCooperationVirtual2016,shahSocialVRbasedCollaborative2022,hoegBuddyBikingUser2023}. 
Often, studies are restricted to one social context, where people either work out in a competitive or collaborative scenario.
A study by~\citet{penaAmWhatSee2016}, for example, indicates that when people competed against an obese avatar while having a normal-weight avatar themselves on a Nintendo Wii, physical activity and performance increased compared to the opposite scenario.
Therefore, since adding another avatar to the workout experience inherently creates a social context and is thus relevant for designing social VR exercise scenarios, we include social context in our study design.


For a better understanding of complex interrelationships in social VR exercise experiences, we present a within-participant user study (N=48). We examine how \emph{social context} (i.e., working out in a collaborative scenario with a teammate or a competitive scenario with an opponent), \emph{body type} (slim, muscular, or obese), and \emph{features traditionally attributed to biological sex} (cis-female, cis-male features)
\footnote{The World Health Organization's definition of sex indicates that \textit{``sex refers to the biological characteristics that define humans as female or male. While these sets of biological characteristics are not mutually exclusive, as there are individuals who possess both, they tend to differentiate humans as males and females.}''~\cite{who_sex}. 
Therefore, we refer to the body features of our avatars as ``traditionally attributed to female and male body features''. Furthermore, we acknowledge that the bodies of the avatars used in the study are cis-male and cis-female and we are aware that this does not sufficiently cover the diverse multitude of gender identities. As this is an initial investigation into the influences of avatars' body shapes and features on participants, we settled to include only cis-female and cis-male bodies to reduce complexity in the study design. We do, however, point out that to gain a comprehensive understanding of the influences, future studies need to consider more body shapes and features.}
influence a person's VR workout experience.

Hereby, we focus on the \textit{social} aspects of a workout experience and operationalize workout experience by measuring a person's feeling of \emph{relatedness} to the teammate/opponent, their \emph{emotional experience} during the workout in addition to \emph{performance}.
To study this, we designed and implemented a VR workout scenario where a person performed sit-ups with a virtual avatar (\autoref{fig:teaser} (a) and (b)).

Our results indicate that teammate social contexts are highly relevant for social VR exercise scenarios. Besides supporting relatedness towards the avatar of the teammate, working out with a teammate can have a relaxing effect on heart rate and reduce negative emotions like jealousy and embarrassment.
Furthermore, while working out with slim or obese avatars increased performance, working out with an obese avatar reduced negative emotions like jealousy, intimidation, and shame. We also found that working out with a muscular avatar triggered even more negative emotions while reducing the feeling of responsibility towards the avatar. 
We found that the features attributed to biological sex had rather little influence on a person's exercise experience overall. However, working out with a cis-male avatar reduced psychological distress.
Interestingly, being attracted towards the avatar had more significantly positive influences on a person than identifying oneself with the avatar. This is an interesting avenue for future work.


In summary, our work makes the following contributions:
\begin{enumerate} 
 \item We investigate the effects of body type (slim, muscular, obese) and features attributed to biological sex (cis-male, cis-female) of a teammate or opponent avatar in a VR workout on feelings of relatedness, emotional experience, and performance. (empirical contribution)
 \item We demonstrate how these effects are influenced by participants' self-described gender and their perceived attraction, identification towards the avatar, and cooperative, competitive, and sport motivation scores. (empirical contribution)
\item We summarize our results in an implication table and discuss key findings for participant-independent (e.g., avatar body type) and participant-dependent (e.g., participant gender) variables, including a prediction model. (theoretical contribution)
\item Finally, we discuss the practical and ethical implications of our work, such as how social context can influence someone's experience positively or negatively in VR workouts. We also present a recommendation list to inform the design of future VR workout experiences and to promote the understanding of the complex relations of social context and avatar body features for the HCI Community. (theoretical contribution)
 \end{enumerate}






\section{Related Work}
Although exercising has benefits for mental~\cite{taylorRelationPhysicalActivity1985} and physical health~\cite{xiaoExerciseCardiovascularProtection2021}, sedentary behaviors \cite{hamiltonTooLittleExercise2008} and obesity \cite{mansonEscalatingPandemicsObesity2004} are concerning health threads in our digital societies. 
To help counter this, developments of using VR as an at-home exercise device \cite{mouattUseVirtualReality2020} are on the rise \cite{hamalainenMartialArtsArtificial2005,FitXRBoxenHIITa}.
We add to these works by investigating how different aspects of working out with another avatar in VR influence a person's workout experience in a social VR workout. In doing so, our research is mainly related to four fields of research, covering \emph{Motivational Factors for Exercise}, \emph{Competition and Collaboration During Exercising},\emph{Effects of Virtual Avatars and Embodiment}, and \emph{Influence Visual Sex and Gender preferences}.

\subsection{Motivational Factors for Exercise}
Exercising can be intrinsically or extrinsically motivated, or both. \citet{vallerandIntegrativeAnalysisIntrinsic1999} mention the three psychological needs, ``autonomy, competence, and relatedness'' that impact a person's motivation to exercise. In detail, they mention social factors like competitiveness and cooperation, coaches' behaviors, and success or failure. 

\citet{louwExerciseMotivationBarriers2012} investigated general motives for exercise: general health, maintaining fitness, feeling good, controlling weight, strength, endurance, and feeling energized. Barriers included lack of time, focus on other priorities, lack of energy, health issues, or missing an exercise partner (younger groups).
Certain personal characteristics influence the motivation to exercise as well. For example, the gender people identify themselves with or their age: While boys between 12-17 were more motivated to physically exercise than girls of the same age in a study by~\citet{portela-pinoGenderDifferencesMotivation2020}, the motivation to exercise of women grew when they grew older through affiliations, competition, and health motivators~\cite{kowalczykAgerelatedDifferencesMotives2017}. Furthermore, men seem more strongly motivated through competitiveness than women~\cite{gillGenderDifferencesCompetitive1988}.

\subsubsection{Social Factors and Relatedness}
Previous research underscores the significance of relatedness in motivating exercise outside virtual reality (VR) contexts~\cite{allenSocialMotivationYouth2003}. Within VR, social effects have been explored in studies on treating social anxiety~\cite{dechantHowAvatarCustomization2021}. Additionally, avatar design's impact on social factors has been examined~\cite{rahillEffectsAvatarPlayersimilarity2021, praetoriusUserAvatarRelationshipsVarious2021,dechantHowAvatarCustomization2021}. Notably, \citet{murrayEffectsPresenceOthers2016} found that working out with a virtual companion in VR led to higher performance and heart rate than exercising alone, highlighting the positive influence of social presence on exercise outcomes.

Studies in augmented reality (AR) environments, such as \citet{millerSocialInteractionAugmented2019}, have observed the influence of virtual agents on task performance and social connectedness. However, AR scenarios demonstrated less social connectedness compared to real-world interactions.



Our work builds on these studies, focusing on a person's social exercise experience in the VR environment and exploring the dynamics of working out with or against another virtual participant.

\subsubsection{Well-being and Emotions}
The role of emotions has been shown to have an important influence on performance in non-sport contexts, such as at work~\cite{popaEMOTIONSROLEMOTIVATION2013} or social interaction in VR~\cite{deighanSocialVirtualReality2023}. It has also been shown that user expectations regarding emotional states in VR affect VR properties, such as presence. \citet{jicolPredictiveModelUnderstanding2023}, for example, propose a refined model considering emotional arousal levels and personality traits in predicting presence in VR. However, while emotions have been studied in VR and for performance-related tasks, it is currently unclear how social VR exercise experiences influence a person's emotional state.

\subsubsection{Performance}
While the previously discussed emotional and social factors lack an exploration in VR, performance influences are better understood. For example, success or failure during a workout does influence future performance. \citet{edwardsImpactActivePassive2018} found that encouragement improved performance and motivation in a sprint and endurance activity. 
A study by~\citet{hudsonAvatarTypesMatter2016} investigated avatars as learning agents in VR. They showed that in different contexts, avatar appearance influenced attitude and motivation by affecting the behavior or the psychology of a user and thus subsequently raising or degrading performance.
\citet{bornMotivatingPlayersPerform2021} used virtual performance augmentation (VPA)to engage players in a strenuous activity. VPA significantly increased the time players interacted with the activity; however, enjoyment and perceived exertion were not significantly influenced.

\citet{michaelRaceYourselvesLongitudinal2020} introduce a VR cycling exergame allowing non-athletes to compete against "ghost" avatars that include past and projected future performance. This method enhanced physical performance and motivation. \citet{mcneillSelfmodelledSkilledpeerModelled2021,mcneillSelfmodelledSkilledpeerModelled2020} suggest self-modeled action observation benefits skilled golfers in refining motor control. \cite{edwardsImpactActivePassive2018b} note verbal encouragement positively impacts exercise power and motivation, potentially benefiting participants' health and exercise motivation.

\subsection{Competition and Collaboration During Exercising}
Choosing between competitive and collaborative contexts significantly impacts workout and exercise experiences \citep{frankenWhyPeopleCompetition1995}. 

In a study by \citet{pengPlayingParallelEffects2013a}, parallel competition in separated physical spaces resulted in high enjoyment, future play motivation, and physical intensity compared to single-player or cooperative conditions in the same physical space. Conversely, a Nintendo Wii study by \citet{staianoAdolescentExergamePlay2013a} demonstrated more weight loss and self-efficacy in a cooperative exergame, with no significant effects in the competitive condition. Both cooperative and competitive exergames increased peer support more than the control group.

Studies such as \citet{dimenichiPowerCompetitionEffects2015} and \citet{penaIncreasingExergamePhysical2014a} highlight the positive impact of competition on performance in physical tasks, but competition was found to be detrimental to memory tasks and decreased physical activity when the opponent avatar was obese. In contrast, recent studies emphasize the positive effects of collaboration in VR exergames, including enhanced valance and enjoyment \cite{hoegBuddyBikingUser2023} and increased intrinsic motivation \cite{shahSocialVRbasedCollaborative2022}.

Exergames like Running Wheel \cite{nunesMotivatingPeoplePerform2014} leverage competition through real-time metrics, demonstrating strong participant motivation. Collaborative games, as found by \citet{brondiEvaluatingEffectsCompetition2015}, foster high social presence and emotional connection, while competitive games enhance user performance. \citet{satoCollaborativeDigitalSports2014a} proposed collaborative digital fitness solutions, emphasizing motivation and a shared sense of caring. Additionally, \citet{partonEffectsCompetitivenessChallenge2019} emphasized the importance of challenge levels in virtual reality-based rowing exercises.

In a tennis Wii exergame by \citet{penaAmWhatSee2016}, avatar body size influenced physical activity levels, suggesting that using a normal-weight avatar enhances exercise performance. \citet{pataneExploringEffectCooperation2020} explored the impact of cooperative and non-cooperative activities in VR on racial bias reduction, finding lower implicit bias scores in the cooperative group. Moreover, \citet{sekhavatCollaborationBattleMinds2020} showed that a competitive reinforcement approach effectively increased attention levels in a multiplayer attention training game compared to a collaborative approach.

\subsection{Effects of Visual Body Shape of Virtual Avatars}

A systematic exploration by \citet{rheuEnhancingHealthyBehaviors2020} on health interventions utilizing avatars identified avatar body size, self-domain (e.g., ideal-self), customizability, body transformation, and avatar behaviors as influential characteristics. Additionally, \citet{rodriguesPersonalizationImprovesGamification2021} found that students using personalized avatars exhibited increased motivation driven by intrinsic motivation and identified regulation.

The Proteus Effect, emphasizing the impact of digital representation on behavior, was explored in studies such as \citet{sadekSuperheroPoseEnhancing2022}, where embodying a professional soccer player avatar resulted in enhanced performance. Avatars embodying physics-awareness \citet{taoEmbodyingPhysicsAwareAvatars2023} and iconic figures like Einstein \citet{banakouVirtuallyBeingEinstein2018} demonstrated positive effects on task performance, or cognitive workload in the work of \citet{kocurEffectsSelfExternal2020}. \citet{kocurEffectsSelfExternal2020} emphasizes the importance of considering both user and external avatar appearance in VR experience when designing for cognitively demanding tasks.

Studies delving into the impact of avatar body shape revealed intriguing findings. \citet{piryankovaOwningOverweightUnderweight2014} demonstrated that owning an overweight or underweight avatar led to significantly overestimating affordance and body size estimation. In contrast, \citet{kocurFlexingMusclesVirtual2020b} observed increased strength performance and decreased effort perception when embodying a muscular avatar, particularly among male participants. \citet{kocurPhysiologicalPerceptualResponses2021a} extended this by showing that an athletic body shape in a cycling scenario positively influenced heart rate and perceived exertion.

Regarding embodiment effects, \citet{dollingerAreEmbodiedAvatars2023} discovered that performing a meditation task with body movement in VR negatively affected body awareness. Avatars in uncomfortable postures led to lower subjective body ownership but increased physiological responses like heart rate \citep{bergstromFirstPersonPerspectiveVirtual2016a}.

Studies examining the impact of other people's visual representations on physical activities in VR are limited. However, environmental visual cues, such as a fire or ice world, were found to influence thermal perception and skin temperature \citep{kocurEffectsAvatarEnvironment2023}.
In a recent study, \cite{clarkeFakeForwardUsingDeepfake2023} could show that an avatar with Deep Fake (a virtual trainer having the participant's face) improves performance in physical exercise when guiding the sports exercise. Another study by \citet{fittonDancingAvatarsMinimal2023} showed that the customization of a dancing instructor with hair and skin color improved visual imagery of the dance moves.




While the existing literature provides valuable insights into the impact of avatar body shapes and self-representation in VR, there is a noticeable gap in understanding how different body types and features influence another person's exercise experience in VR.

\subsection{Influence of Visual Sex and Gender in VR}
Effects of visual sex and gender on VR exercise experiences have been explored only very broadly. \citet{planteDoesVirtualReality2003} conducted a study involving 121 college students exposed to different exercise conditions in VR, finding that VR exercise may enhance post-exercise energy and tiredness levels in females. However, overall, VR exercise resulted in reduced tiredness. Additionally, \citet{tammylinExercisingEmbodiedYoung2021} focused on female elderly individuals, revealing that those embodying young avatars, particularly if they did not engage in vigorous exercise, reported greater self-efficacy for future exercise and higher physical activity during the exercise phase compared to those embodying older avatars.

The influence of virtual embodiment on biased groups was investigated by \citet{maisterChangingBodiesChanges2015}, who found a significant reduction in biases against outgroups when participants were embodied in VR characters of different genders, ages, or races. 
\citet{panVirtualCharacterPersonality2015} studied human interaction with virtual female characters exhibiting different personalities, discovering that participants responded differently based on virtual character personality, indicating the impact of personality on subjective and behavioral responses in human-virtual interactions.

\section{Implementation}
In this section, we specify the implementation and hardware setup. 

\subsection{Hardware and Software Setup}

\begin{figure}[!h]
    \centering
    \includegraphics[width=0.6\textwidth]{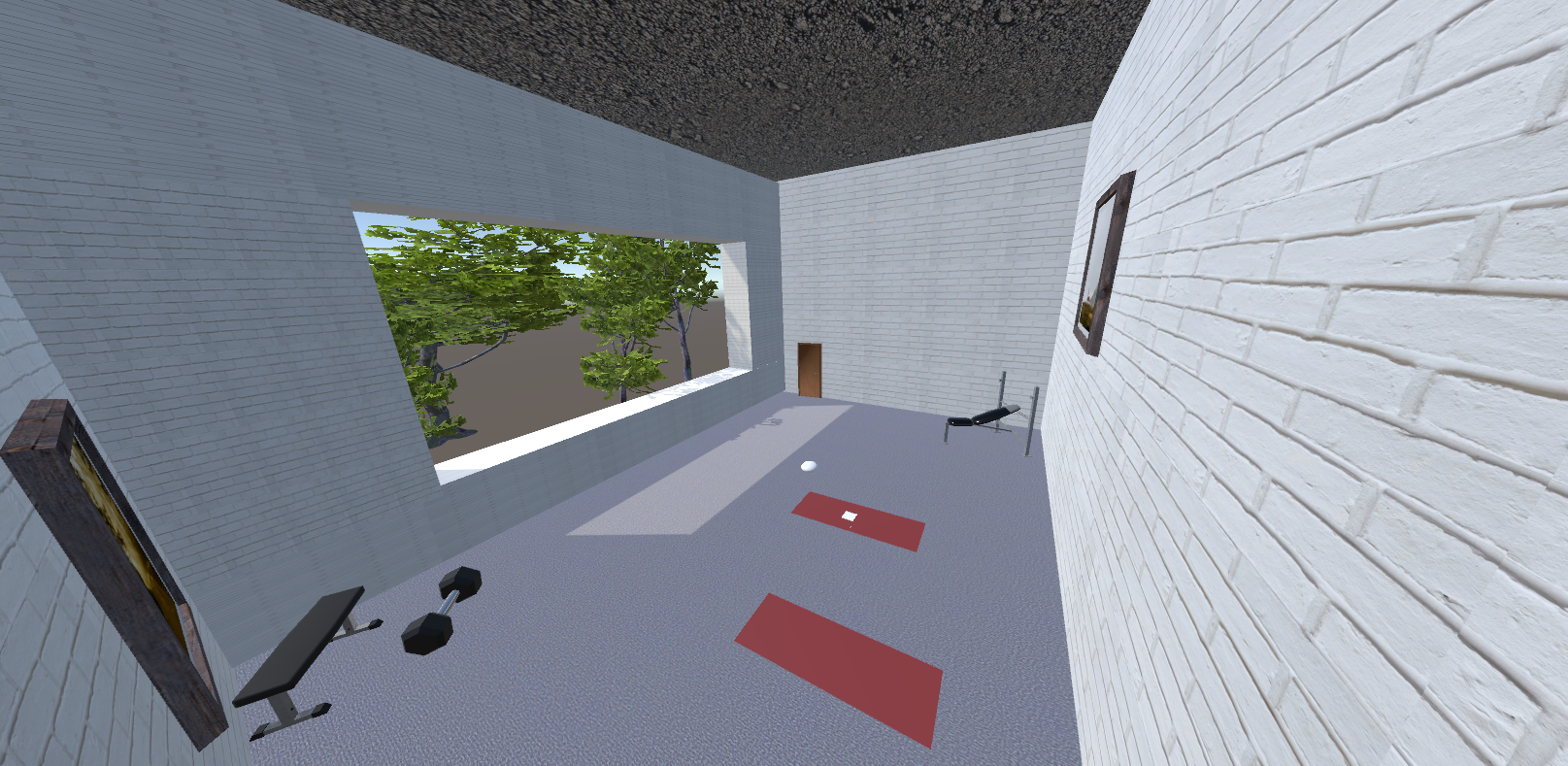}
    \caption{The Figure shows the VR scene of the study. The virtual room was designed to be calm and not distracting but at the same time, not boringly empty.}
    \label{fig:VRscene}
    \Description{The Figure shows the virtual reality scene where the study took place. The virtual room was designed to be calm and not distracting, having a window with trees behind it, a few fitness tools in the corners, and two exercise mats in the middle.}
\end{figure}

We used a wireless HTC Vive Pro VR headset with two handheld controllers to represent the user's hands in VR. 
Additionally, a yoga mat was set up in the middle of the VR lab, corresponding to a virtual mat in the scene (figure \autoref{fig:VRscene}). An X on the real-world mat marked the sitting position for the user, which corresponded to a white square on the virtual yoga mat. 

The scene was designed to be calm and not distracting. In the beginning, a voice introduced the social context, and the word ``teammate'' or ``opponent'' was written on the wall accordingly.
A sphere above the user's yoga mat was used to count the number of sit-ups. Participants had to touch the sphere with the controller for each sit-up; it then turned from green to red for 100 ms, emitting a beep sound to signal the registration of the sit-up.  
The button was active for 15 seconds; then, a bell automatically rang to signal that the time was up. 
Background music was not included as we did not want music preferences to influence our results.
The other avatar was positioned on the second yoga mat, next to the user's yoga mat.

\subsection{Agent Body Design}
\begin{figure}[!h]
    \centering
    \includegraphics[width=\textwidth]{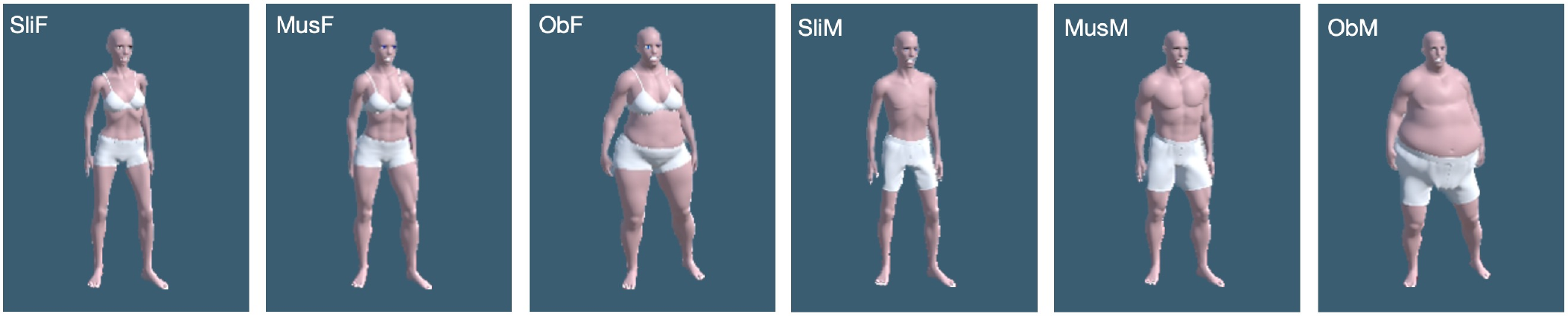}
    \caption{The six agents participants had to compete against or work out with. The upper row shows the female-shaped agents, and the bottom row shows the male-shaped agent. From left to right, the agents represent a slim, muscular, and obese body type. The scanty clothing was chosen to emphasize the different body shapes.}
    \label{fig:SixAvatars}
    \Description{The figure shows the six avatars participants competed against or worked out during the study. The upper row shows the avatars with female features, and the bottom row shows the avatars with male features. From left to right, the avatars' body types are slim, muscular, and obese. The avatars with male features wear shorts, and the avatars with female features wear shorts and bras to emphasize their body types.}
\end{figure}

We designed six avatars \autoref{fig:SixAvatars} based on the Basemeshes big pack 3D asset by turbosquid\footnote{https://www.turbosquid.com/3d-models/basemeshes-big-pack-3d-1760254}. We chose the clothing to emphasize the different body types.

Once appearing in the room, the avatars were standing in a quiet position for five seconds. Then, they performed a stand-to-sit movement and waited to sit on the map until the workout started. A countdown of three was given before the exercise started, and the avatar performed the sit-ups next to the user.
We put the avatars next to the participants since the focus should not be forced unnaturally onto them. 

\section{Evaluation}

Based on the presented setup, we conducted a within-participant study with three independent variables: the \textit{social context} (two levels: teammate, opponent), the \textit{body type} (three levels: slim, muscular, obese), and the \textit{features attributed to biological sex}, represented through a cis-female and a cis-male avatar. In the following, we will refer to the latter as ``female'' or ``male'' body features.

\subsection{Research Questions}
We formulated two research questions guiding our work:
\begin{itemize}
    \item [\textbf{RQ1}] \emph{How does \textbf{a)} social context (teammate, opponent), \textbf{b)} avatar body type (slim, muscular, obese) and \textbf{c)} visual sex (cis-female, cis-male) of another avatar influence performance, emotions, and relatedness in a social VR exercise?}
    \item [\textbf{RQ2}] \emph{How do participant-dependent variables (i.e., gender, cooperation/competitiveness tendencies, fitness level, their level of attraction and identification towards another avatar's body) additionally to social context, body type and visual sex, influence performance, emotions, and relatedness in a social VR exercise?}
\end{itemize}

\subsection{Conditions}

Our setup resulted in the following twelve conditions, which we split up into two times six according to the social context. 

\begin{table}[h!]
\begin{flushleft}
\begin{tabular}{llllllll}
\textbf{Teammate Conditions} & \textbf{Opponent Conditions} \\
 \emph{(1) Teammate Slim Female (TSliF)}: \autoref{fig:SixAvatars} a & \emph{(7) Opponent Slim Female (OSliF)}: \autoref{fig:SixAvatars}a \\
\emph{(2) Teammate Muscular Female (TMusF)}: \autoref{fig:SixAvatars} b & \emph{(8) Opponent Muscular Female (OMusF)}: \autoref{fig:SixAvatars}b \\
\emph{(3) Teammate Obese Female (TObF)}: \autoref{fig:SixAvatars} c & \emph{(9) Opponent Obese Female (OObF)}: \autoref{fig:SixAvatars}c \\
\emph{(4) Teammate Slim Male (TSliM)}: \autoref{fig:SixAvatars} d & \emph{(10) Opponent Slim Male (OSliM)}: \autoref{fig:SixAvatars}d \\
\emph{(5) Teammate Muscular Male (TMusM)}: \autoref{fig:SixAvatars}e & \emph{(11) Opponent Muscular Male (OMusM)}: \autoref{fig:SixAvatars}e\\
\emph{(6) Teammate Obese Male (TObM)}: \autoref{fig:SixAvatars}f & \emph{(12) Opponent Obese Male (OObM)}: \autoref{fig:SixAvatars}f \\
\end{tabular}
\end{flushleft}
\end{table}

\begin{figure}[!h]
    \centering
    \includegraphics[width=\textwidth]{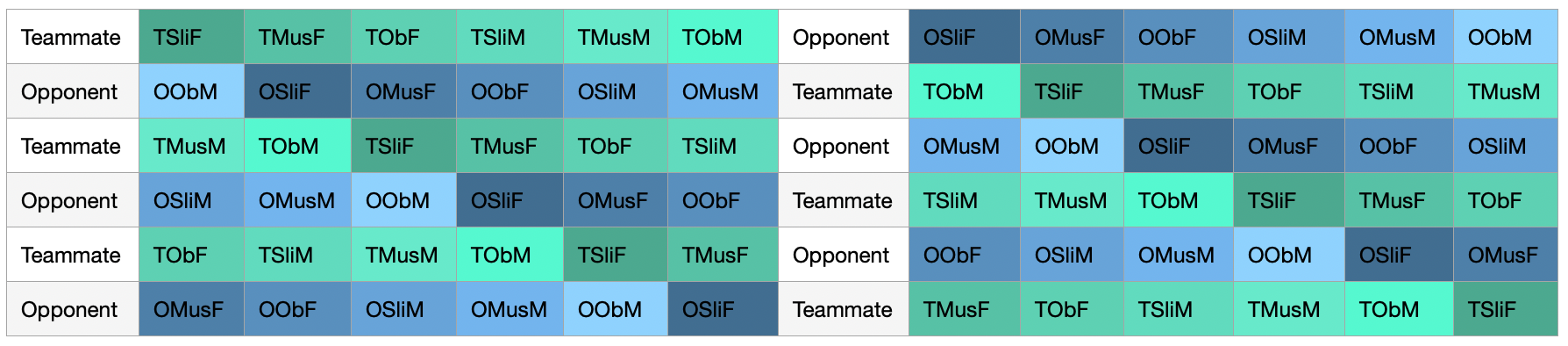}
    \caption{The figure shows how we counterbalanced our study conditions to avoid learning and tiredness effects. While the teammate and opponent social contexts were swapped in blocks for each participant, the six different avatars were rotated according to a Latin square within these blocks. Each row represents a participant's run. After the sixth row, it was repeated, starting from the first row with the next participant. The green color refers to the Teammate scenarios, while the blue refers to the Opponent scenarios. After the twelve rows, the next participant started with the first row again.}
    \label{fig:latinSquare}
    \Description{The figure shows how we counterbalanced our study conditions to avoid learning and tire effects. While Teammate and Opponent were swapped in blocks for each participant, the six different avatars were rotated as a Latin square within these blocks. Each row represents a participant's run. After the sixth row, it was repeated, starting from the first row with the next participant. The green color refers to the Teammate scenarios, while the blue one refers to the Opponent scenarios.}
\end{figure}

We separated the social context (teammate, opponent) into two blocks to avoid participants having to switch between empathizing with a different social context too often. 
This would likely have led to mixing up the social contexts, as there was no visual difference between the avatars in each context, except the text on the wall. The social context participants started with was counterbalanced, and inside a social context block, the six avatars were rotated according to the Latin-square in~\autoref{fig:latinSquare}. 

The teammate's social context was introduced with the following statement: \emph{``You will be presented with a teammate now. In this condition, the other person will be your teammate, and together, you must try to do as many sit-ups as possible.''}
The opponent's social context was introduced with this statement: \emph{``You will be presented with an opponent now. In this condition, the other person will be your opponent, and you are competing through the number of sit-ups against this person.''}
Furthermore, the social context was written on the wall in each scenario, and at the beginning of a scene, a voice introduced the social context verbally.
We carefully ensured that the teammate/opponent avatar was always spoken of as \textit{another person}. We wanted to give the participants the impression that they were working out with another person, while in reality, the avatar was implemented as a non-player character (NPC).

\subsection{Measures}
To measure our dependent variables \textit{relatedness}, \textit{emotions}, and \textit{performance}. We used objective measures (in-game log files and a smartwatch) and subjective measures in the form of questionnaires that were polled after each round of sit-ups, as well as after each social context.

\subsubsection{Relatedness} \label{sct:relatedness}
Relatedness was measured as post-condition questions and questionnaires that were asked after each round of sit-ups:

\paragraph{Inclusion of Other in the Self}
The Inclusion of Other in the Self (IOS) \cite{woosnam2010inclusion} assesses the connectedness of oneself to another in the style of a Venn diagram with one circle representing the person and one circle representing the ``other''. The level of overlap indicates the level of connectedness.

\paragraph{Relatedness (IMI)}
The relatedness sub-scale of the Intrinsic Motivation Inventory (IMI) \cite{ostrow2018testing} assesses responses to eight questions about trust, the feeling of being distant, interaction, and friendship with another individual (7-point scale ranging from 1 = Not at all true to 7 = very true). 

\paragraph{Custom Likert questions}
Additionally we used custom 7-point Likert (\cite{battertonLikertScaleWhat2017}) questions (1 = Strongly Disagree to 7 = Strongly Agree ):
For both social contexts, we asked if participants were attracted to the other person, felt responsible for the other person, were repulsed, were focused on the other person, were focused on themselves, and if they were satisfied with their performance.
Specifically for the teammate social context, we asked ``I thought of the other person as my teammate.'', and ``I sometimes forgot I was working in a team.''.
Specifically for the opponent social context, we asked ``I felt motivated to win against the person'', ``I thought of the other person as an opponent'', and ``I felt like I competed more with myself rather than my opponent''.

\subsubsection{Emotional Experience} 
We measured the emotional experience in post-condition questions and questionnaires that were asked after each round of sit-ups:

\paragraph{Affective State}
The Self-Assessment Manikin (SAM)~\cite{bradley1994measuring} is a 5-point image scale comprising of three subscales: valence (1 = happy to 5 = unhappy), arousal (1 = excited to 5 = calm), and dominance (1 = controlled to 5 = in-control).

\paragraph{Exercise Experience}
The three sub-scales of the Subjective Exercise Experiences Scale (SEES)~\cite{meauley1994subjective} evaluate positive well-being (strong, great, positive, terrific), psychological distress (crummy, awful, miserable, discouraged), and fatigue (exhausted, fatigued, tired, drained).
They comprise a 7-point scale (1 = Not At All to 7 = Very Much So) to assess a participant's emotional state after the exercise. 

\paragraph{Social Emotions}
To further evaluate a state of social well-being, we assessed social emotions \citet{lewisSelfconsciousEmotionsEmbarrassment2008, burnettDevelopmentAdolescenceNeural2009, hareliWhatSocialSocial2008}. With ten custom questions on a 7-point Likert scale (1 = Strongly Disagree to 7 = Strongly Agree ), we asked if participants felt intimidated, confident, jealous, empathy, envious, embarrassed, guilty, pride, shame, and admiration.

\subsubsection{Performance}
The in-game measures were extracted through log files, while the heart rate was measured with external sensors:

We measured the sit-up completion time after each sit-up and counted the overall amount of sit-ups (log entry when the controller triggered contact with the white sphere).
We measured participants' heart rate using a Samsung heart rate sensor for the wrist as a second, continuous performance measure. 
Additionally we used custom 7-point Likert questions (1 = Strongly Disagree to 7 = Strongly Agree ):
For both social contexts, we asked if participants were satisfied with their performance and how much effort they put in.

\subsubsection{Final Questions and Demographics}
A final questionnaire was presented at the very end after all twelve rounds.

\paragraph{Presence, Sport Motivation, Cooperative, and Competitiveness}
Presence was measured using the Slater-Usoh-Seed Questionnaire (SUS) (7-point Scale ranging from 1 = very present to 7 = very much present)~\cite{slater2000virtual}. 

We asked participants to fill out all sub-scales of the Sports Motivation Scale (SMS -6) \cite{mallett2007sport}: Intrinsic Motivation, Integrated Regulation, Identified Regulation, External Regulation, Amotivation, and Introjected Regulation. They are rated on a 6-point scale (1 = Does Not Correspond At All to 6 = Corresponds exactly).

The Cooperative and Competitive Personality Scale CCPS \cite{lu2013cooperativeness} assesses a person's views and beliefs on cooperation and competition in three parts: cognition, behavior, and affect (7-point scale ranging from 1 = Do Not Agree At All to 7 = Totally Agree). 

\paragraph{Demographics}
We recorded participants' age, gender, sexual gender interests, and occupation. Furthermore, we asked participants how frequently they do sports, how much they like sports, and what sport they do, as well as their experience with VR. 
We also asked participants to rate which of the six avatars they identified most with and which one they felt the most attracted to.

\subsection{Procedure}
Our national ethics, data protection, and health and safety regulations require the involvement of the Ethics Commission only in special medical or ethical studies. Studies like the one conducted in the paper are not subject to consultation. Even without the explicit involvement of the Ethics Commission, we are obliged to adhere to the national  Guidelines for Safeguarding Good Research Practice, the ACM Code of Ethics and Professional Conduct, and our University Statutes for Safeguarding Good Scientific Practice. 
Participants were recruited from the local university through email lists and posters. They were informed on the posters and emails about tasks, risks, and confidential data that will be assessed anonymously.
After signing an informed consent of participation, the participant was equipped with a VR headset and a smartwatch to measure heart rate. 
For the baseline measurement of how many sit-ups participants can do while wearing the headset, the participant performed as many sit-ups as possible in the home screen environment for 15 seconds. 
The measured number was used to set the number of sit-ups the avatar was performing to have an opponent or teammate that is almost equal in performance.
After pre-measurements, participants conducted the twelve main conditions, separated into the two social context blocks. 

\begin{figure}[!h]
    \centering
    \includegraphics[width=\textwidth]{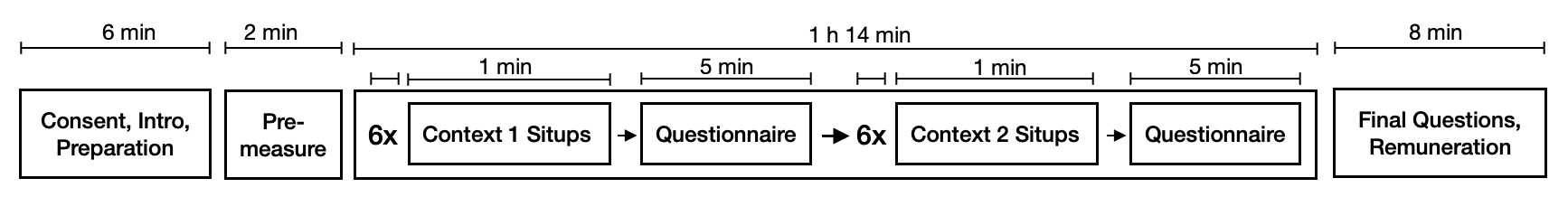}
    \caption{The figure shows a visualized version of the procedure as a flow chart.}
    \label{fig:procedure}
    \Description{The figure shows a visualized version of the procedure as a flow chart.}
\end{figure}

In each round, participants saw the avatar for 5 seconds, sat on a mat next to them, started the heart rate measure, and did 15 seconds of sit-ups. Afterward, they put down the headset and answered the post-condition questionnaire. After all twelve conditions were finished, the final demographics questionnaire was presented.  
A remuneration of 15 Euro was given to the participants in the end. 
In total, the procedure took about 1h and 30 minutes for each participant (see \autoref{fig:procedure}).

\subsection{Participants}

An a priori power analysis was conducted using G*Power version 3.1.9.6 (Faul et al., 2007) for sample size estimation. The chosen effect size was 0.15, which is considered small according to Cohen's (1988) criteria. With a significance criterion of $\alpha$ = .05, power = .95, and number of measurements = 12, the minimum sample size needed with this effect size is N = 48 for within repeated measures ANOVA. 
For within-between interaction (RQ2), however, the effect size 0.25 was used, resulting in  N = 24 instead of N = 63 participants (we address this in limitations).

We recruited 48 participants (21 identified as female, 26 as male, and one person indicated their gender as ``other'' in the demographics questionnaire, M$_{age}$ = 23.67, SD = 2.29, Min-Max= 19--30). 
87.50\% of the participants were university students, and 16.67\% were employed. The remaining ones are currently unemployed.
Over 75\% were exercising regularly, while about 20\% never or rarely exercise. 
About 75\% liked exercising (spread across the levels of ``somewhat agree'' and ``strongly agree'').
77\% of the participants had never used VR headsets before, but about 75\% liked using headsets (ratings between ``somewhat agree'' and ``strongly agree'').
We asked the participants to indicate with which avatar body they most identified with.
12 participants identified themselves mostly with the SliF avatar (\autoref{fig:SixAvatars}a), 8 with the MusF avatar (\autoref{fig:SixAvatars}b), 2 with ObF (\autoref{fig:SixAvatars}c), 19 with SliM (\autoref{fig:SixAvatars}d) and 7 with MusM (\autoref{fig:SixAvatars}e).
When we asked which gender(s) they felt attracted to, seven indicated they were attracted to the same gender as themselves, while 38 indicated they were attracted to the opposite gender (female, male), and three were attracted to both female and male gender.  

When asked which body type they found most attractive, eleven of the participants who indicated they were attracted to male body types selected the SliM (\autoref{fig:SixAvatars} d) and twelve the MusM (\autoref{fig:SixAvatars} e) body type. 

Of the 26 participants who indicated being primarily attracted to female body types, 11 indicated they were most attracted to the SliF (\autoref{fig:SixAvatars} a), and 14 to the MusF (\autoref{fig:SixAvatars} b).

\section{Analysis \& Results}

We structure the reporting of our results into two parts. We start by reporting participant-independent results and answering RQ1. 
Then, we report participant-dependent results, including participant-dependent demographics and information like gender, the body type they most identified with, the body type they were most attracted to, sports motivation, and their cooperation and competitive score. By doing so, we answer RQ2. 
A correlation Matrix of all dependent and participant-related data can be found in the appendix. 


\subsection{Participant-Independent Effects on Relatedness, Emotion, and Performance}

We analyze and report our study results within the research question RQ1 according to relatedness, emotions, and performance.
We used linear mixed models (LMM) to analyze our data.  
Since the variable \textit{body type} has three levels and linear mixed models work with baseline comparisons (i.e., comparing one condition against one baseline), we applied contrast coding to identify which of the levels we should use as a baseline. The contrast coding identified \textit{obese} as the body type baseline (obese - muscular, obese - slim), \textit{female} as the baseline for sex, and \textit{opponent} as the baseline for social context. 

To avoid over-fitting, we compared different models (Appendix \autoref{ModelComparisionPI}) and used the best-fitting model (i.e., the model with the smallest Akaike information criterion (AIC)) for each dependent variable. 
All models and their AIC values are shown in Appendix \autoref{tab:models3L}.


\subsubsection{Relatedness}
For relatedness, we report the results for \emph{IOS} (ordinal), \emph{IMI Relatedness} (interval), as well as six custom Likert scale questions. We refer to the Likert questions as indicated in \autoref{sct:relatedness}: \emph{Level of Responsibility} (ordinal), \emph{Level of Attraction} (ordinal), \emph{Level of Focus on Other Person} (ordinal), \emph{Level of Focus on Self} (ordinal), and \emph{Satisfaction with Performance} (ordinal).
The results of the linear mixed models indicate the following statistically significant differences for relatedness towards the intercept, reported in \autoref{tab:lmm_relatedness}. 

The best-fitting models included social context for relatedness variables (except for the \emph{attracted} question). 
For \emph{attractedness}, only body type and sex were relevant in the model. 
The avatar's sex was only important for \emph{IMI} and \emph{focused}, while body type was included in all models.

\paragraph{Main effects} 
We found significant main effects for \emph{IOS}, indicating that participants felt closer to the other avatar in the teammate conditions (M:2.61, SD:1.67)\footnote{We use M for the mean and SD for the standard deviation} compared to the opponent conditions (M:2.11, SD:1.58). 
Further, they felt closer to the other avatar for slim (M: 2.48, SD: 1.68) and muscular (M: 2.48, SD: 1.69) body types, compared to the obese body type (M: 2.12 SD: 1.54). 

We also found a significant main effect of body type on the \emph{IMI Relatedness subscale}, indicating that participants felt significantly less related to the obese body type (M: 3.63, SD: 1.26) compared to the slim (M: 4.06, SD: 1.34) or muscular body types (M: 3.97, SD: 1.35).

\begin{figure}[!h]
    \centering
    \includegraphics[width=\textwidth]{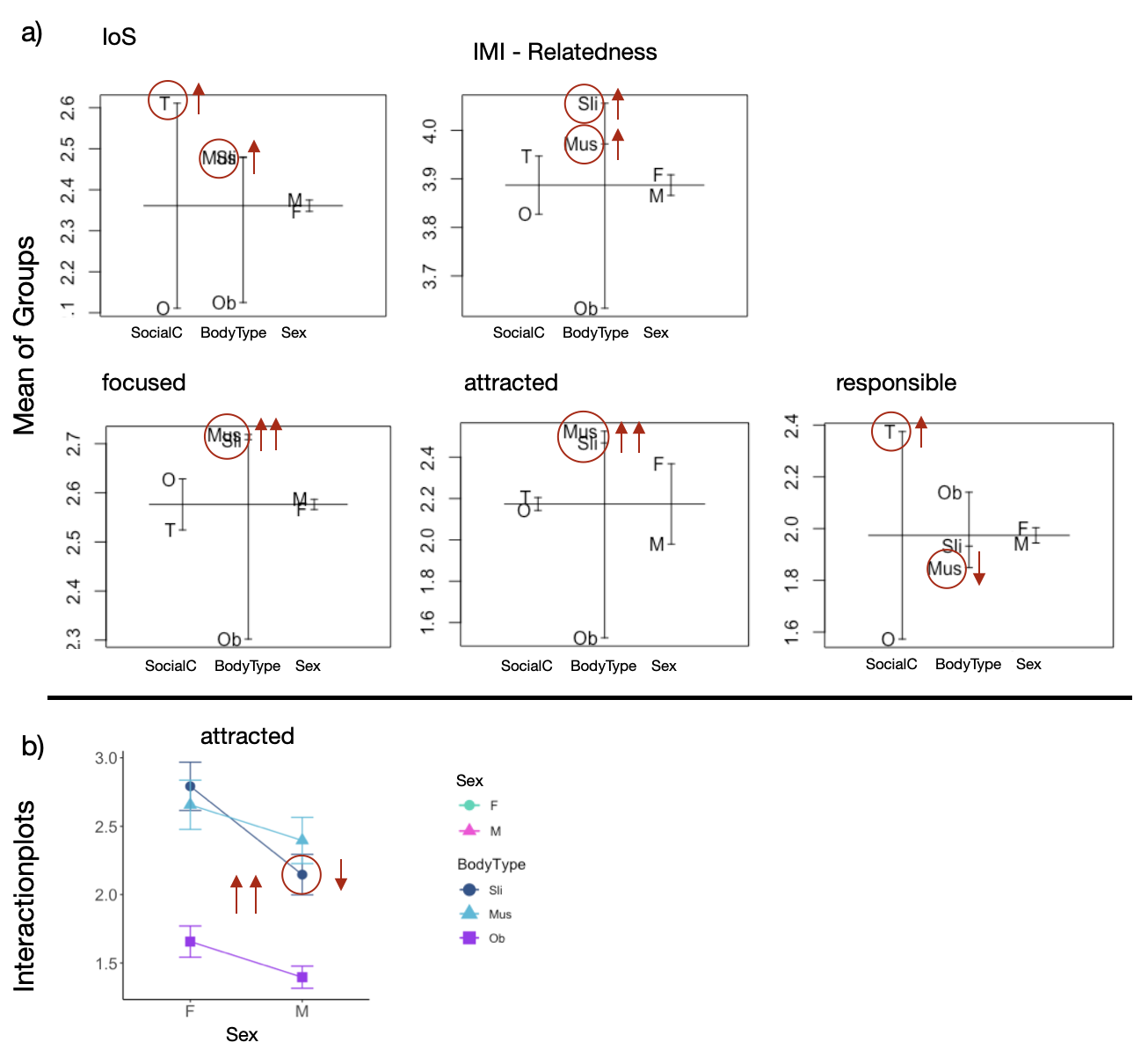}
    \caption{a) The plots show different mean values for each group in relatedness variables. The long line in the middle is the overall mean, while the group means of the two levels from social context, the three levels of body type and the two levels of sex are shown as differences towards the overall mean. The red circle marks where main effects occurred and if they were positive or negative. b) The plots show the interaction between two of the three levels. The red circle marks where interaction effects occurred and if they were positive or negative.}
    \label{fig:R_Plots}
    \Description{The plot is separated into two parts. The upper part is part a, and the part below the line is part b. In a) the first graph shows the means of IoS. While the teammate mean is high above the mean line and circled red with an up arrow, the opponent line is low beneath the mean. Obese is low below the mean line and the group means of muscular and slim are above the line, circled red with arrows upwards. In the next graph, IMI-Relatedness, the muscular and slime body type means are above the general mean line far from the obese mean, circled red with arrows pointing upwards. The next graph is about the focus on the other avatar. Here as well slim and muscular body type is are high, circled red with an arrow up. For graph attracted toward the other avatar, again slim and muscular body types are high, circled red with an arrow up. The last graph in a, is feeling of responsibility towards the other avatar. Here the teammate in social context is very high, circled red with an arrow upwards as well as muscular body type as an low mean circled red with an arrow down. In b the interaction plot is about the attraction towards the other avatar. Its shows the interaction between mean of body type and sex. Here the slim male point is circled red with an arrow down.}
\end{figure}

There were also significant main effects for the \emph{Level of Responsibility}; participants felt more responsible for the other avatar in the teammate conditions (M: 2.38, SD: 1.68) than in the opponent conditions (M: 1.57, SD: 0.91). Also, participants felt less responsible for the muscular avatar (M: 1.85, SD: 1.28) compared to the obese ones (M: 2.14, SD: 1.56).
Furthermore, we found a significant main effect of \emph{Level of Attraction}. More precisely, the participants felt less attracted to the obese body shapes (M: 1.53, SD: 0.98) compared to the slim (M: 2.47, SD: 1.62) and muscular ones (M: 2.53, SD: 1.71). 
Finally, for \emph{Level of Focus on Other Person}, we found that participants focused less on avatars with body type obese (M: 2.30, SD: 1.65) than on avatars with body type slim (M: 2.71, SD:1.80) or on avatars with body type muscular (M: 2.72, SD: 1.81).
Main effects are marked with red circles in \autoref{fig:R_Plots} a).

\paragraph{Interaction effects}
Taking a look at interaction effects, we found that participants were less attracted towards male and slim body-shaped avatars. 
Interaction effects are marked with red circles in \autoref{fig:R_Plots} b).


\begin{table}[]
  \caption{The table shows all significant results of main and interaction effects of relatedness variables. The Intercept (Intcp) shows which variables and what baseline were included in the best-fitting model (O = Opponent: Teammate; Ob = Obese: BodyType; F = Female: Sex).}
\label{tab:lmm_relatedness}
\resizebox{\textwidth}{!}{%
\begin{tabular}{lllllllll}
\textbf{Relatedness} & \textbf{Effect} & \textbf{Beta} & \textbf{95\% CI} & \textbf{t()} & \textbf{p-value} & \textbf{Std. Beta} & \textbf{Std. 95\% CI} \\
\hline
IOS & Intcp (O, Ob) & 1.88 & (1.49, 2.26) & t(562) = 9.51 & \textless .001 & - & - \\
       & SocialC (T) & 0.50 & (0.18, 0.82) & t(562) = 3.10 & 0.002 & 0.30 & (0.11, 0.50) \\
       & BT (Sli) &  0.35 & (0.12, 0.59) & t(562) = 2.97 & 0.003 & 0.22 & (0.07, 0.36) \\
       & BT (Mus) &  0.35 & (0.09, 0.61) & t(562) = 2.67 & 0.008 & 0.22 & (0.06, 0.37) \\
 \hline
IMI & Intcp (O, Ob, F) & 3.59 & (3.29, 3.90) & t(558) = 22.83 & \textless .001 & - & - \\
 & BT (Sli) & 0.42 & (0.17, 0.67) & t(558) = 3.31 & \textless .001 & 0.32 & (0.13, 0.51) \\
  & BT (Mus) & 0.34 & (0.08, 0.59) & t(558) = 2.62 & 0.009 & 0.25 & (0.06, 0.45) \\
 \hline
responsible &  Intcp (O, Ob) & 1.74 & (1.50, 1.98) & t(562) = 14.13 & \textless .001 & - & - \\
 & SocialC (T) & 0.80 & (0.47, 1.13) & t(562) = 4.74 & \textless .001 & 0.57 & (0.33, 0.80) \\
 & BT (Mus) & -0.29 & (-0.53, -0.05) & t(562) = -2.38 & 0.018 & -0.21 & (-0.38, -0.04) \\
 \hline
attracted & Intcp (Ob, F) & 1.66 & (1.33, 1.98) & t(560) = 10.11 & \textless .001 & - & - \\
 & BT (Sli) & 1.14 & (0.82, 1.45) & t(560) = 6.99 & \textless .001 & 0.74 & (0.53, 0.94) \\
 & BT (Mus) & 1.00 & (0.67, 1.33) & t(560) = 5.96 & \textless .001 & 0.65 & (0.44, 0.86) \\
  & BT (Sli) × Sex (M) &  -0.39 & (-0.75, -0.02) & t(560) = -2.05 & 0.041 & -0.25 & (-0.49,-0.01) \\
 \hline 
focused & Intcp (O, Ob, F) & 2.34 & (1.94, 2.75) & t(558) = 11.44 & \textless .001 & - & - \\
 & BT (Sli) & 0.41 & (0.19, 0.62) & t(558) = 3.68 & \textless .001 & 0.23 & (0.11, 0.35) \\
 & BT (Mus) & 0.42 & (0.20, 0.63) & t(558) = 3.77 & \textless .001 & 0.24 & (0.11, 0.36) \\
\end{tabular}}
\end{table}

\subsubsection{Emotional Experience}
The dependent variables considered for emotional experience are: \emph{Intimidated} (ordinal), \emph{Jealousy} (ordinal), \emph{Empathy} (ordinal), \emph{Envy} (ordinal), \emph{Embarrassed} (ordinal), \emph{Guilty} (ordinal), \emph{Pride} (ordinal), \emph{Shame} (ordinal), \emph{Admire} (ordinal), \emph{SEES PositiveWellBeing} (interval), \emph{SEES PsychologicalDistress} (interval), and \emph{Enjoyed Activity} (ordinal). 

The results of the linear mixed models indicate the following statistically significant differences for emotion towards the intercept, reported in \autoref{tab:lmm_emotions1} and \autoref{tab:lmm_emotions2}.
For social emotion, the best-fitting models never included the factor sex. 
For \emph{positive well-being}, \emph{psychological distress}, and \emph{enjoyment}, sex was relevant again. The affective states (\emph{arousal, dominance, and valance}) were only influenced by social context.

\paragraph{Main effects} 

We found many positive main effects in the variables for emotional experience. 
Participants felt less \emph{intimidated} and \emph{jealous} when working out with a muscular avatar (Intimidated -> M: 2.41, SD: 1.70; Jealous -> M: 2.11, SD: 1.59) and a slim avatar (Intimidated -> M: 2.03, SD: 1.44; Jealous -> M: 1.88, SD: 1.37) compared to an obese avatar (Intimidated -> M: 1.68 SD: 1.14; Jealous -> M: 1.45, SD: 0.91)
They also felt less \emph{embarrassed}, and less \emph{shame} with an muscular avatar(Embarrassed -> M: 1.89, SD: 1.28; Shame -> M: 1.70, SD: 1.13) compared to an obese avatar (Embarrassed -> M: 1.47, SD: 0.90; Shame -> M: 1.51, SD: 0.93)
On the other side, participants \emph{envy} and \emph{admire} the slim (Envy -> M: 1.97, SD: 1.34; Admire -> M: 2.82, SD: 1.73) and muscular (Envy -> M: 2.16, SD: 1.47; Admire -> M: 3.04, SD: 1.72) body types significantly more than the obese body type (Envy -> M: 1.58, SD: 1.04; Admire -> M: 2.37, SD: 1.74)

\begin{figure}[t]
    \centering
    \includegraphics[width=0.75\textwidth]{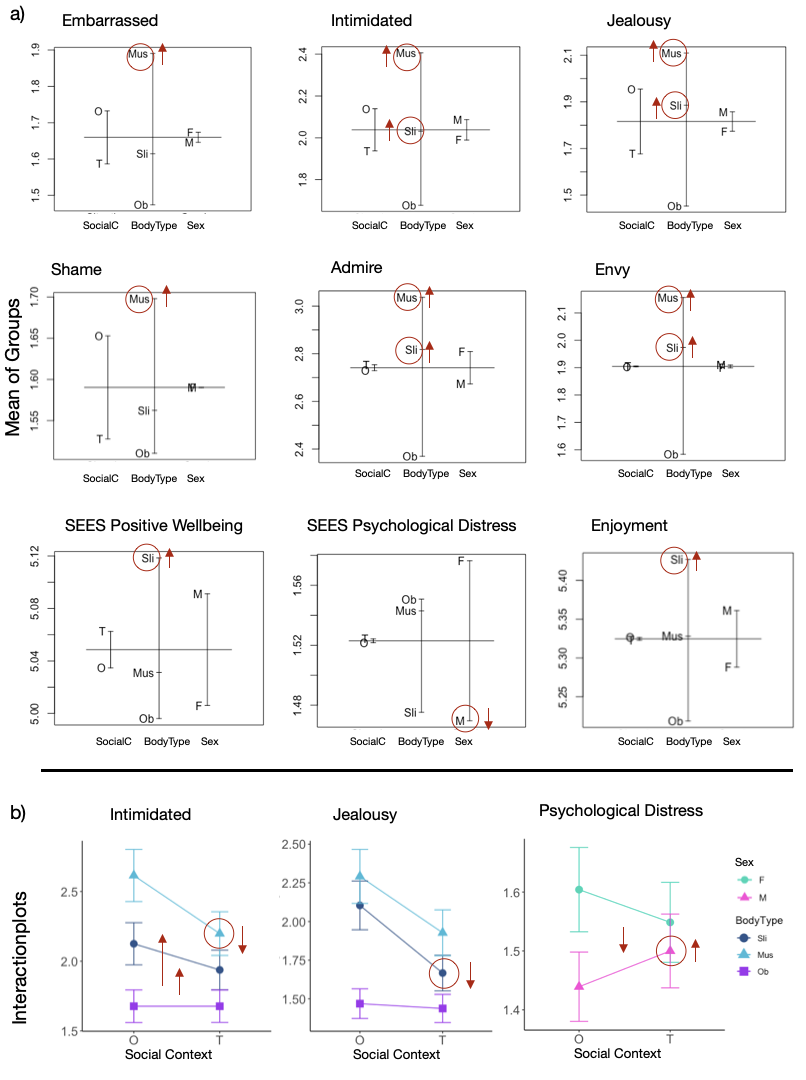}
    \caption{a) The plots show different mean values for each group in emotion variables. The long vertical line in the middle is the overall mean, while the group means of the two levels from social context, the three levels of body type, and the two levels of sex are shown as differences towards the overall mean. The red circle marks where the main effects occurred and if they were positive or negative. b) The plots show the interaction between two of the three levels. The red circle marks where interaction effects occurred and if they were positive or negative.}
    \label{fig:EM_Plots}
    \Description{a) The first figure is for embarrassment, where muscular is above the overall mean line, circled in red with an arrow pointing upwards. For Intimidated slim is almost on top of the line and musular above, circled red with an arrow upwards, almost similar for Jelousy, Admire and Envy. 
    For Shame, there is only muscular high above the mean line, circled red with an arrow facing upwards.
    In positive well-being and enjoyment there is slim high above the mean line and circled red with an arrow facing upwards.
    In psychological distress, male is below the mean and circled red with an arrow facing downwards. 
    b) The first interaction plot is between social context and body type. The teammate muscular is circled red with an arrow facing downwards, and the obese points are lower than all the muscular and slim interaction points. 
    For Jealousy, the slim teammate interaction point is circled red. It can be seen that the opponent slim point is much higher.
    The last interaction plot is between sex and social context, where teammate male is circled red facing upwards. Both teammate and opponent male points are below the female interaction points.}
\end{figure}

Participants \emph{enjoyed} the activity significantly more when the other avatar had a slim-shaped body (M: 5.43, SD: 1.40) compared to an obese-shaped body (M: 5.22, SD: 1.54). 
Furthermore, participants' \emph{Positive Well-being} (SEES) was significantly higher when working out with a slim avatar (M: 5.12, SD: 0.95) compared to an obese avatar (M: 5.00, SD: 0.96).
Finally, participants were significantly more \emph{distressed} when working out with a cis-male avatar (M: 1.47, SD: 0.73) than a cis-female avatar (M: 1.58, SD: 0.84).
Main effects are marked with red circles in \autoref{fig:EM_Plots} a).

\paragraph{Interaction effects} 

In teammate conditions, participants felt significantly more \emph{intimidated} when working out with an avatar of obese body type than with a muscular body type.
Participants were significantly less \emph{jealous} when working out with teammate avatars of obese body types, than with slim body types. 
Finally, participants experienced lower \emph{psychological distress} when working out with an avatar of male body features in teammate condition. 
Main effects are marked with red circles in \autoref{fig:EM_Plots} b).





\begin{table}[]
  \caption{The table shows all significant results of main and interaction effects of social emotion variables.The Intercept (Intcp) shows which variables and what baseline were included in the best-fitting model (O = Opponent: Teammate; Ob = Obese: BodyType; F = Female: Sex).}
\label{tab:lmm_emotions1}
\resizebox{\textwidth}{!}{%
\begin{tabular}{lllllllll}
\textbf{Social Emotions} & \textbf{Effect} & \textbf{Beta} & \textbf{95\% CI} & \textbf{t()} & \textbf{p-value} & \textbf{Std. Beta} & \textbf{Std. 95\% CI} \\
 \hline
Intimidated & Intcp (O, Ob) & 1.68 & (1.37, 1.98) & t(560) = 10.84 & \textless .001 & - & - \\
 & BT (Sli) & 0.45 & (0.20, 0.70) & t(560) = 3.51 & \textless .001 & 0.30 & (0.13, 0.48) \\
 & BT (Mus) & 0.94 & (0.58, 1.30) & t(560) = 5.09 & \textless .001 & 0.64 & (0.39, 0.88) \\
 & SocialC(T) × BT (Mus)  & -0.42 & (-0.75, -0.08) & t(560) = -2.46 & 0.014 & -0.28 & (-0.51, -0.06) \\
  \hline
Jealousy & Intcp (O, Ob) &  1.47 & (1.19, 1.75) & t(560) = 10.34 & \textless .001 & - & - \\
 & BT (Sli) & 0.64 & (0.36, 0.91) & t(560) = 4.50 & \textless .001 & 0.47 & (0.27, 0.68) \\
 & BT (Mus) & 0.82 & (0.47, 1.17) & t(560) = 4.63 & \textless .001 & 0.61 & (0.35, 0.87) \\
 & SocialC(T) × BT (Sli) & -0.41 & (-0.74, -0.07) & t(560) = -2.37 & 0.018 & -0.30 & (-0.55, -0.05) \\
  \hline
Empathy & Intcp (O, Ob) & 3.11 & (2.61, 3.60) & t(562) = 12.28 & \textless .001 & - & - \\
  \hline  
Envy & Intcp (O, Ob) & 1.58 & (1.34,1.82) & t(562) = 12.99, & \textless .001 & - & - \\
 & BT (Sli) & 0.39 & (0.16, 0.62) & t(562) = 3.31 & 0.001 & 0.30 & (0.12, 0.47) \\
 & BT (Mus) & 0.57 & (0.30, 0.85) & t(562) = 4.07 & \textless .001  & 0.44 & (0.23, 0.65) \\
  \hline
Embarrassed & Intcp (O, Ob) & 1.55 & (1.31, 1.79) &  t(562) = 12.60 & \textless .001 & - & - \\
 & BT (Mus) & 0.42 & (0.21, 0.63) & t(562) = 3.88 & \textless .001 & 0.38 & (0.19, 0.57) \\
  \hline
Guilty & Intcp (O, Ob) & 1.54 & (1.30, 1.77) & t(562) = 12.82 & \textless .001 & - & - \\
  \hline
Pride & Intcp (O) & 2.95 & (2.51, 3.40) & t(570) = 13.07 & \textless .001 & - & - \\
  \hline
Shame & Intcp (O, Ob) & 1.57 & (1.33,1.81) & t(562) = 12.81 & \textless .001 & - & - \\
 & BT (Mus) & 0.19 & (0.01, 0.36) & t(562) = 2.13 & 0.033 & 0.18 & (0.01, 0.35) \\
  \hline
Admire & Intcp (Ob) & 2.37 & (1.95, 2.79) & t(566) = 11.15 & \textless .001 & - & - \\
 & BT (Sli) & 0.45 & (0.18, 0.72) & t(566) = 3.26 & 0.001 & 0.26 & (0.10, 0.41) \\
 & BT (Mus) & 0.67 & (0.29, 1.04) & t(566) = 3.49 & \textless .001 & 0.38 & (0.17, 0.60) \\
  \end{tabular}}
\end{table}

\begin{table}[]
  \caption{The table shows all significant results of the main and interaction effects of further emotion variables. The Intercept (Intcp) shows which variables and what baseline were included in the best-fitting model (O = Opponent: Teammate; Ob = Obese: BodyType; F = Female: Sex).}
\label{tab:lmm_emotions2}
\resizebox{\textwidth}{!}{%
\begin{tabular}{lllllllll}
\textbf{Emotions} & \textbf{Effect} & \textbf{Beta} & \textbf{95\% CI} & \textbf{t()} & \textbf{p-value} & \textbf{Std. Beta} & \textbf{Std. 95\% CI} \\
 \hline  
PosWellbeing & Intcp (O, Ob, F) & 4.94 & (4.68, 5.20) & t(558) = 36.79, & \textless .001 & - & - \\
 & BT (Sli) & 0.12 & (0.02, 0.22) & t(558) = 2.41 & 0.016 & 0.12 & (0.02, 0.23) \\
 \hline
PsyDistress & Intcp (O, F) & 1.60 & (1.37, 1.84) & t(565) = 13.38 & \textless .001 & - & - \\
 & Sex (M) &  -0.16 & (-0.25, -0.08) & t(565) = -4.00 & \textless .001 & -0.21 & (-0.31, -0.11) \\
 & SocialC(T) × Sex (M) & 0.12 & (0.02, 0.22) & t(565) = 2.27 & 0.023 & 0.15 & (0.02,0.28) \\
  \hline
Enjoyment & Intcp (O, Ob, F) & 5.18 & (4.83, 5.54) & t(558) = 28.84 & \textless .001 & - \\
 & BT (Sli) & 0.21 & (0.02, 0.40) & t(558) = 2.15 & 0.032 & 0.14 & (0.01, 0.27) \\
  \hline
Arousal & Intcp (O) & 1.77 & (1.58, 1.95) & t(570) = 19.00 & \textless .001 & - \\
    \hline
Dominance & Intcp (O) & 3.43 & (3.12, 3.74) & t(570) = 21.64 & \textless .001 & - \\
   \hline
Valance & Intcp (O) & 1.77 & (1.58, 1.95) & t(570) = 19.00 & \textless .001 & - & - \\
  \hline
\end{tabular}}
\end{table}

\subsubsection{Performance}
The dependent variables we measured for performance are: \emph{NumberOfSitUps (NoS)} (ordinal), \emph{Time} (interval), \emph{HeartRate (HR)} (interval), \emph{Effort} (ordinal), and \emph{SEES Fatigue} (interval) 
The results of the linear mixed models indicate the following statistically significant differences for performance towards the intercept, reported in \autoref{tab:lmm_performance}.

The best-fitting models included Body Type for all performance-related dependent variables.
Social context was relevant for \emph{HR}, \emph{Effort}, and \emph{SEES Fatigue}, but sex was only relevant for \emph{NumberOfSitUps} and \emph{SEES Fatigue}.
The \emph{HR} model has to be considered with caution since the AIC value is very high.   

\paragraph{Main effects} 
The \emph{number of sit-ups} does not vary much; participants were only positively affected in the number of sit-ups by the body type muscular (M: 7.38, SD: 2.01) and slim (M: 7.39, SD: 1.89) in comparison to the obese body type (M: 7.16, SD: 1.87).
The \emph{time for one sit-up} and \emph{heart rate} of participants were not mainly affected by 
factors of sex, body shape, or social context.
Participants felt that they put significantly more \emph{effort} into their workout when an avatar with slim body shape (M: 5.14, SD:1.70) and muscular body shape (M: 5.21, SD: 1.73) 

\begin{figure}[!h]
    \centering
    \includegraphics[width=0.8\textwidth]{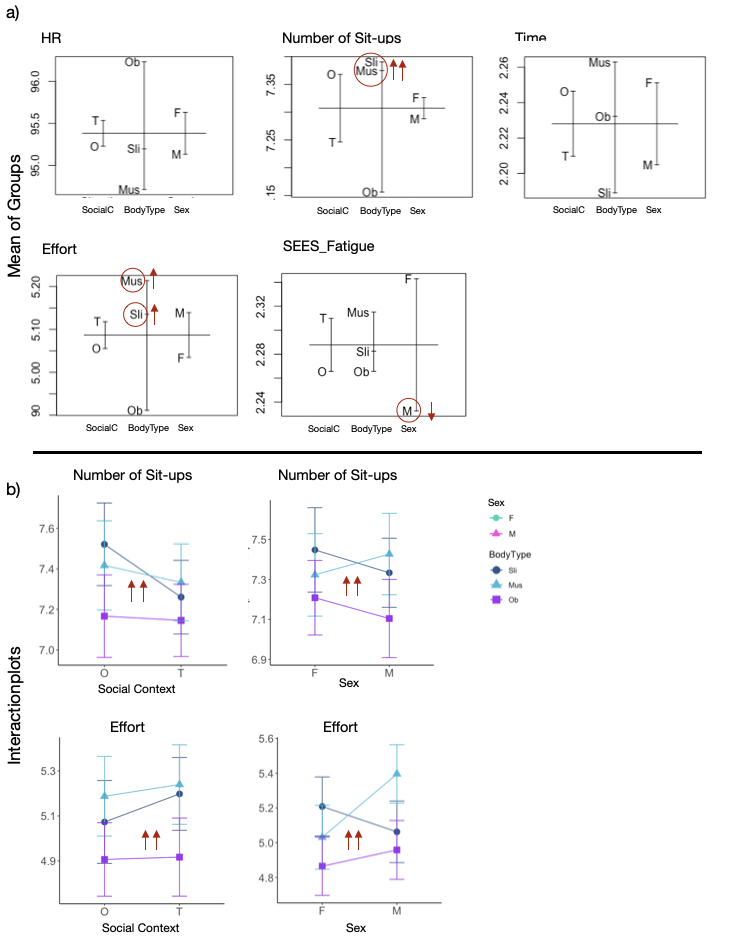}
    \caption{a) The plots show different mean values for each group in performance variables. The long line in the middle is the overall mean, while the group means of the two levels from social context, the three levels of body type and the two levels of sex are shown as difference towards the overall mean. The red circle marks where main effects occurred and if they were positive or negative. b) The plots show the interaction between two of the three levels. The red arrows indicate the main effects that can also be seen here.}
    \label{fig:PPlots}
    \Description{The upper plots (a) are mean plots of groups. For number of sit-ups, muscular and slim are above the overall mean, circled red with arrows facing upwards. Obese is below the line.  For effort muscular is high and slim is a bit are above the mean circled red with a upwards arrow. For Fatigue, while female sex is high above the overall mean, the male mean is below, circled red with a downwards arrow. 
    The plots for (b) are interaction plots. The first two for Number of Sit-ups, the second two for effort. All four show the line of obese avatars below the other two lines connecting the values between Social Context (the two left plots) and Sex (the two right plots)
    
    }
\end{figure}

With an avatar with male features (M:2.23, SD:1.11) in the room, participants also felt lower \emph{fatigue} compared to an avatar with female features (M:2.34, SD:1.26). 
Main effects are marked with red circles in \autoref{fig:PPlots} a).

No interaction effects were found for performance results.




\begin{table}[htbp]
  \caption{The table shows all significant results of main and interaction effects of performance variables. The Intercept (Intcp) shows which variables and what baseline were included in the best-fitting model (O = Opponent: Teammate; Ob = Obese: BodyType; F = Female: Sex).}
\label{tab:lmm_performance}
\resizebox{\textwidth}{!}{%
\begin{tabular}{lllllllll}
\textbf{Performance} & \textbf{Effect} & \textbf{Beta} & \textbf{95\% CI} & \textbf{t()} & \textbf{p-value} & \textbf{Std. Beta} & \textbf{Std. 95\% CI} \\
 \hline
NumSitups & Intcp (Ob, F) & 7.18 & (6.66, 7.69) & t(562) = 27.40 & \textless .001 & - & - \\
 & BT (Sli) & 0.23 & (0.07, 0.40) & t(562) = 2.80 & 0.005 & 0.12 & (0.04, 0.21) \\
 & BT (Mus) & 0.22 & (0.05, 0.38) & t(562) = 2.59 & 0.010 & 0.11 & (0.03, 0.20) \\
  \hline
Time & Intcp (Ob) & 2.23 & (2.05, 2.41) & t(566) = 24.69 & \textless .001 & - & - \\
 \hline 
HR & Intcp (O, Ob) & 96.08  & (93.90, 98.26) & t(562) = 86.46 & \textless .001 & - & -\\
\hline
Effort & Intcp (O, Ob) & 4.88 & (4.45, 5.31) & t(562) = 22.09 & \textless .001 & - & - \\
 & BT (Sli) & 0.22 & (0.03, 0.42) & t(562) = 2.27 & 0.024 & 0.13 & (0.02, 0.25) \\
 & BT (Mus) & 0.30 & (0.10, 0.50) & t(562) = 2.96 & 0.003 & 0.18 & (0.06, 0.30) \\
 \hline
Fatigue & Intcp (O, F) & 2.32 & (1.99,2.65) & t(566) = 13.93 & \textless .001 & - & - \\
 & Sex (M) & -0.11 & (-0.22, -5.04e-03) & t(566) = -2.06 & 0.040 & -0.09 & (-0.18, -4.24e-03) \\
\end{tabular}}
\end{table}

\subsection{Participant-Dependent Effects on Relatedness, Emotion, and Performance}

In this section, we report our study results within the research question (RQ2) according to relatedness, emotional experience, and performance.
For the analysis of the data we used linear mixed models. For this model we not only included social context, body type and female/male body features, but further included participants' \emph{gender}, \emph{identification} (\emph{true} for the avatar participants identified most with, \emph{false} for the avatars participants did not identify with) and attractedness (true or false from the chosen avatar they selected as being attracted most to). 
As mentioned before, we used these variables to test different models (Appendix \autoref{ModelComparisionPD}) and used the models that fitted best (AIC small) (see Appendix \autoref{tab:resultsAll9}).
We removed one participant for the analysis since this participant did not identify as female or male and we did not have enough data to form a complete group (F=21, M=26). 
We acknowledge that the results of this model will be used and discussed with care, since the model is very complex. However, we believe it is necessary to show that the influences of participants' gender, attractedness, and identification has a large significant relevance, and there are interesting influences by these factors.




\subsubsection{Relatedness}
Taking a closer look into the models that fit best, we can see that for \emph{IMI} and \emph{IoS}, all nine variables seem to have a strong influence. \emph{Attracted} was influenced by the overall most attractive avatar for the participant. And the gender of the participant influenced the \emph{attracted} results.

For relatedness, we report the results for \emph{IOS} (ordinal), \emph{IMI Relatedness} (interval), as well as six custom Likert scale questions. We refer to the Likert questions as indicated in \autoref{sct:relatedness}: \emph{Level of Responsibility} (ordinal), \emph{Level of Attraction} (ordinal), \emph{Level of Focus on Other Person} (ordinal), \emph{Level of Focus on Self} (ordinal), and \emph{Satisfaction with Performance} (ordinal).
The results of the linear mixed models indicate the following statistically significant differences for relatedness towards the intercept, reported in \autoref{tab:lmm_relatedness_6}.

\begin{table}[]
  \caption{The table shows all significant results of main and interaction effects of relatedness variables. The Intercept (Intcp) shows which variables and what baseline were included in the best-fitting model (O = Opponent: Teammate; Ob = Obese: BodyType; F = Female: Sex; Attr = Being attracted towards the avatar; Ident = Identification with the avatar).}
\label{tab:lmm_relatedness_6}
\resizebox{\textwidth}{!}{%
\begin{tabular}{lllllllll}
\textbf{Relatedness} & \textbf{Effect} & \textbf{Beta} & \textbf{95\% CI} & \textbf{t()} & \textbf{p-value} & \textbf{Std. Beta} & \textbf{Std. 95\% CI} \\
\hline
IOS & SocialC (T) × BT (Mus)                                                    & 55.84  & [8.92, 102.77]    & 2.35   & 0.020  & 1.27  & [-0.58, 3.13] \\
&BT (Sli) × Gender (M)                                                           & -38.26 & [-71.74, -4.78]   & -2.25  & 0.025  & 0.46  & [-0.58, 1.51] \\
&BT (Mus) × Gender (M)                                                           & 67.61  & [7.70, 127.52]    & 2.22   & 0.027  & -0.09 & [-0.23, 0.04] \\
&Sex (M) × Ident (True)                                                     & 112.69 & [40.00, 185.38]   & 3.06   & 0.003  & -1.16 & [-1.77, -0.55] \\
&Gender (M) × Ident (True)                                                  & 1.90   & [0.98, 2.82]      & 4.07   & < .001 & 1.16  & [0.60, 1.72] \\
&Attr (True) × Ident (True)                                       & -11.95 & [-19.74, -4.16]   & -3.02  & 0.003  & -0.33 & [-0.61, -0.04] \\
& ... & ... & ... & ... & ... & ... & ... \\
 \hline
IMI & BT (Mus)× Sex (M) & -107.73 & [-190.29, -25.17] & -2.57 & 0.011 & -0.85 & [-1.49, -0.20] \\
&SocialC (T)× Gender (M) & 63.61 & [4.01, 123.22] & 2.10 & 0.037 & -0.13 & [-0.48, 0.21] \\
&SocialC (T)× Attr (True) & 189.09 & [24.63, 353.56] & 2.27 & 0.024 & 0.45 & [-0.96, 1.85] \\
&Sex (M) × Ident (True) & 505.79 & [34.77, 976.80] & 2.12 & 0.035 & -4.79 & [-13.67, 4.08] \\
&Gender (M) × Ident (True) & 6.01 & [2.12, 9.90] & 3.05 & 0.003 & 4.53 & [1.60, 7.46] \\
& ... & ... & ... & ... & ... & ... & ... \\
\hline
responsible &  Intcp (O,Ob,F) & 1.77 & [1.51, 2.02] & 13.55 & < .001 & -0.15 & [-0.33, 0.03] \\
&SocialC (T)& 0.83 & [0.49, 1.16] & 4.87 & < .001 & 0.58 & [0.35, 0.82] \\
&BT (Mus)& -0.31 & [-0.55, -0.07] & -2.49 & 0.013 & -0.22 & [-0.39, -0.05] \\
\hline
attracted & Intcp (O,Ob,F) & 1.62 & [1.24, 2.01] & 8.29 & < .001 & -0.35 & [-0.61, -0.10] \\
&BT (Sli)& 0.70 & [0.44, 0.95] & 5.36 & < .001 & 0.46 & [0.29, 0.63] \\
&BT (Mus)& 0.71 & [0.43, 0.99] & 5.02 & < .001 & 0.46 & [0.28, 0.65] \\
&Attr (True) & 0.93 & [0.70, 1.16] & 8.08 & < .001 & 0.61 & [0.46, 0.76] \\
\hline
focused & Intcp (O,Ob,F) & 2.37 & [1.96, 2.78] & 11.45 & < .001 & -0.13 & [-0.36, 0.10] \\
&BT (Sli)& 0.41 & [0.19, 0.63] & 3.64 & < .001 & 0.23 & [0.11, 0.36] \\
&BT (Mus)& 0.40 & [0.18, 0.63] & 3.60 & < .001 & 0.23 & [0.10, 0.35] \\
\hline
\end{tabular}}
\end{table}

\paragraph{Main effects}
For IOS and IMI, we did not find main effects. However, the teammate condition positively influenced responsibility, while a muscular body type influenced responsibility negatively.
Slim and Muscular Body Type influence attractedness significantly positive and both also influence the focus on the other person significantly positive. 

\paragraph{Interaction effects} 
Positive influences for IOS Relatedness are teammate muscular, muscular male, male identification and male avatar identification. 
Negative influence for IOS are slim male as well as attraction and identification towards the other avatar. 
IMI Relatedness is negatively influenced for muscular males avatar , but positively influenced by male participants in the teammate condition, being attracted in the teammate condition, male identification and male avatar identification. 

\subsubsection{Emotional Experience}

While \emph{Empathy, Envy and Shame} are influenced by all nine variables, many others were only influenced by the other avatar and scenario, like in \emph{Intimidation, Jealousy, Embarrassment, Pride, Admire, Positive Well-being, Enjoyment, Dominance and Valance}. 
\emph{Guilty} was additionally influenced by the sport motivation score, while \emph{Arousal} was additionally influenced by gender, and \emph{Psychological Distress} was influenced by gender and attractive avatar.
The results of the linear mixed models indicate the following statistically significant differences for emotions are reported in \autoref{tab:lmm_emotions1_6} and \autoref{tab:lmm_emotions2_6}.

\begin{table}[]
  \caption{The table shows all significant results of main and interaction effects of social emotion variables.The Intercept (Intcp) shows which variables and what baseline were included in the best-fitting model (O = Opponent: Teammate; Ob = Obese: BodyType; F = Female: Sex; Attr = Being attracted towards the avatar; Ident = Identification with the avatar).}
\label{tab:lmm_emotions1_6}
\resizebox{\textwidth}{!}{%
\begin{tabular}{lllllllll}
\textbf{Social Emotions} & \textbf{Effect} & \textbf{Beta} & \textbf{95\% CI} & \textbf{t()} & \textbf{p-value} & \textbf{Std. Beta} & \textbf{Std. 95\% CI} \\
 \hline
Intimidated & Intcp (O,Ob,F) & 2.04 & [1.73, 2.36] & 12.79 & < .001 & -4.00e-15 & [-0.21, 0.21] \\
&BT (Sli)   & -0.36 & [-0.52, -0.21] & -4.55 & < .001 & -0.24 & [-0.35, -0.14] \\
\hline
Jealousy & Intcp (O,Ob,F)    & 1.82 & [1.56, 2.09] & 13.65 & < .001 & 1.41e-15 & [-0.19, 0.19] \\
&SocialC (T) & 0.14 & [0.05, 0.23] & 3.21 & 0.001  & 0.10 & [0.04, 0.17] \\
&BT (Sli)      & -0.37 & [-0.53, -0.21] & -4.47 & < .001 & -0.27 & [-0.39, -0.15] \\
\hline
Empathy & SocialC (T) & 249.50 & [139.03, 359.97] & 4.45 & < .001 & 5.10 & [2.98, 7.22] \\
&Sex (M) & -246.46 & [-380.40, -112.53] & -3.63 & < .001 & -1.81 & [-3.59, -0.02] \\
&SocialC (T) × BT (Sli) & 532.50 & [281.47, 783.52] & 4.18 & < .001 & 6.96 & [3.89, 10.02] \\
&SocialC (T) × Gender (M) & 39.99 & [9.97, 70.02] & 2.62 & 0.009 & 0.16 & [-0.67, 1.00] \\
&BT (Sli) × Gender (M) & -69.22 & [-123.04, -15.41] & -2.53 & 0.012 & -1.74 & [-3.22, -0.26] \\
&SocialC (T) × Attr (True) & -172.09 & [-267.39, -76.78] & -3.56 & < .001 & -0.56 & [-0.94, -0.18] \\
&BT (Sli) × Attr (True) & 298.97 & [94.94, 503.00] & 2.89 & 0.004 & 0.18 & [-0.29, 0.66] \\
&Sex (M) × Attr (True) & 218.39 & [93.20, 343.57] & 3.44 & < .001 & 1.91 & [0.56, 3.25] \\
&Gender (M) × Attr (True) & 10.48 & [2.12, 18.84] & 2.47 & 0.014 & 1.79 & [0.36, 3.21] \\
&SocialC (T) × Ident (True) & -65.47 & [-121.32, -9.61] & -2.31 & 0.022 & -4.85 & [-6.96, -2.73] \\
&SocialC (T) × Cooperation & -45.06 & [-65.15, -24.97] & -4.42 & < .001 & -8.42 & [-11.99, -4.86] \\
&Sex (M) × Cooperation & 44.77 & [20.21, 69.32] & 3.59 & < .001 & 0.84 & [0.03, 1.64] \\
&SocialC (T) × Competition & -42.94 & [-68.00, -17.88] & -3.38 & < .001 & 0.49 & [0.15, 0.82] \\
&Sex (M) × Competition & 60.37 & [29.56, 91.18] & 3.86 & < .001 & -0.24 & [-0.67, 0.19] \\
&SocialC (T) × SMS & -43.57 & [-79.91, -7.23] & -2.36 & 0.019 & 1.19 & [0.73, 1.65] \\
&Sex (M) × SMS & 79.57 & [34.01, 125.14] & 3.44 & < .001 & -0.79 & [-1.35, -0.23] \\
& ... & ... & ... & ... & ... & ... & ... \\
\hline
Envy & SocialC (T) & 133.33 & [31.09, 235.56] & 2.57 & 0.011 & 4.65 & [2.00, 7.30] \\
&BT (Mus) & -104.24 & [-183.16, -25.31] & -2.60 & 0.010 & -6.01 & [-10.85, -1.17] \\
&Sex (M) & -209.09 & [-322.49, -95.68] & -3.63 & < .001 & 1.78 & [-0.20, 3.77] \\
&Ident (True) & -110.86 & [-187.40, -34.33] & -2.85 & 0.005 & -5.31 & [-10.10, -0.51] \\
&SocialC (T) × BT (Sli) & 238.86 & [11.14, 466.57] & 2.07 & 0.040 & 7.31 & [3.47, 11.15] \\
&BT (Mus) × Sex (M) & 83.30 & [27.37, 139.23] & 2.94 & 0.004 & 0.18 & [0.03, 0.32] \\
&BT (Sli) × Attr (True) & 282.05 & [111.73, 452.37] & 3.26 & 0.001 & -0.48 & [-1.04, 0.08] \\
&Sex (M) × Attr (True) & 138.22 & [31.52, 244.92] & 2.55 & 0.011 & -0.05 & [-1.51, 1.40] \\
&BT (Sli) × Ident (True) & -465.92 & [-843.29, -88.56] & -2.43 & 0.016 & -8.48 & [-15.36, -1.59] \\
&BT (Mus) × Ident (True) & 108.59 & [19.56, 197.63] & 2.40 & 0.017 & 6.04 & [1.19, 10.89] \\
&Sex (M) × Ident (True) & 99.62 & [20.04, 179.21] & 2.47 & 0.014 & -1.64 & [-2.56, -0.72] \\
&Gender (M) × Ident (True) & 2.56 & [1.42, 3.70] & 4.41 & < .001 & 1.94 & [1.07, 2.80] \\
&Attr (True) × Ident (True) & -10.71 & [-20.26, -1.16] & -2.21 & 0.028 & -0.61 & [-1.07, -0.16] \\
&SocialC (T) × Cooperation & -23.23 & [-41.82, -4.64] & -2.46 & 0.015 & -7.76 & [-12.15, -3.38] \\
&BT (Mus) × Cooperation & 15.56 & [2.14, 28.98] & 2.28 & 0.023 & 9.78 & [1.51, 18.05] \\
&Sex (M) × Cooperation & 37.95 & [17.10, 58.80] & 3.59 & < .001 & 0.86 & [-0.02, 1.74] \\
&Ident (True) × Cooperation & 17.11 & [3.83, 30.39] & 2.54 & 0.012 & 8.35 & [-0.23, 16.93] \\
&Sex (M) × Competition & 48.48 & [22.35, 74.61] & 3.66 & < .001 & -0.44 & [-0.93, 0.05] \\
&Sex (M) × SMS & 75.42 & [37.08, 113.76] & 3.88 & < .001 & -0.65 & [-1.27, -0.03] \\
& ... & ... & ... & ... & ... & ... & ... \\
  \hline
Embarrassed & Intcp (O,Ob,F)           & 1.66                 & [1.43, 1.90]           & 14.20           & < .001       & -6.12e-17           & [-0.21, 0.21]          \\
&SocialC (T)        & 0.08                 & [0.00, 0.15]           & 1.99            & 0.047        & 0.07                & [0.00, 0.14]           \\
&BT (Sli)             & -0.19                & [-0.30, -0.07]         & -3.27           & 0.001        & -0.17               & [-0.27, -0.07]         \\
\hline
Guilty & Intcp (O,Ob,F)           & 0.94                 & [0.12, 1.77]           & 2.24            & 0.025        & 1.09e-15            & [-0.20, 0.20]          \\
\hline
Pride & Intcp (O,Ob,F)           & 5.06                 & [4.82, 5.31]           & 41.12           & < .001       & 3.42e-15            & [-0.25, 0.25]          \\
& BT (Mus)             & 0.07                 & [0.02, 0.12]           & 2.82            & 0.005        & 0.07                & [0.02, 0.12]           \\
\hline
Shame & Intcp (O,Ob,F)                                       & -250.46              & [-340.88, -160.04]            & -5.46           & < .001       & 0.26                 & [-3.82, 4.35]             \\
&BT (Sli)                                         & -285.94              & [-559.83, -12.05]             & -2.06           & 0.041        & 2.10                 & [-3.81, 8.01]             \\
&Sex (M)                                              & -195.80              & [-268.72, -122.88]            & -5.29           & < .001       & 0.60                 & [-1.03, 2.23]             \\
&Gender (M)                                           & 47.91                & [12.01, 83.82]                & 2.63            & 0.009        & -1.44                & [-2.83, -0.04]            \\
&Attr (True)                                & 235.37               & [168.61, 302.13]              & 6.95            & < .001       & 1.38                 & [0.78, 1.99]              \\
&Cooperation                                       & 46.49                & [30.18, 62.80]                & 5.62            & < .001       & -1.22                & [-8.23, 5.80]             \\
&Competition                                       & 62.21                & [42.28, 82.15]                & 6.15            & < .001       & -0.19                & [-0.71, 0.32]             \\
&SMS                                               & 87.79                & [58.61, 116.96]               & 5.93            & < .001       & -1.07                & [-1.74, -0.41]            \\
&BT (Sli) × Sex (M)                                  & -271.40              & [-402.15, -140.64]            & -4.09           & < .001       & -0.19                & [-2.81, 2.42]             \\
&BT (Mus) × Sex (M)                                  & 48.50                & [13.16, 83.84]                & 2.70            & 0.007        & -0.18                & [-0.30, -0.05]            \\
&BT (Mus) × Gender (M)                               & 39.49                & [2.78, 76.20]                 & 2.12            & 0.035        & -0.02                & [-0.16, 0.13]             \\
&BT (Sli) × Attr (True)                    & 288.27               & [166.33, 410.21]              & 4.66            & < .001       & 0.59                 & [0.08, 1.09]              \\
&Sex (M) × Attr (True)                         & 153.13               & [86.01, 220.24]               & 4.50            & < .001       & 0.10                 & [-1.14, 1.33]             \\
&Sex (M) × Ident (True)                         & 60.93                & [14.53, 107.33]               & 2.59            & 0.010        & -0.69                & [-1.52, 0.15]             \\
&BT (Sli) × Cooperation                           & 54.71                & [17.10, 92.33]                & 2.87            & 0.005        & -1.72                & [-15.48, 12.05]           \\
&Sex (M) × Cooperation                                & 36.94                & [23.52, 50.37]                & 5.42            & < .001       & -1.68                & [-2.50, -0.86]            \\
&Gender (M) × Cooperation                             & -8.67                & [-15.36, -1.98]               & -2.55           & 0.011        & -0.85                & [-1.40, -0.31]            \\
&Attr (True) × Cooperation                  & -43.02               & [-55.26, -30.78]              & -6.93           & < .001       & 0.55                 & [-0.44, 1.55]             \\
&BT (Sli) × Competition                           & 69.60                & [36.88, 102.31]               & 4.19            & < .001       & -0.14                & [-1.30, 1.03]             \\
&BT (Mus) × Competition                           & -10.11               & [-18.46, -1.77]               & -2.39           & 0.018        & -0.17                & [-0.33, -0.01]            \\
&Sex (M) × Competition                                & 29.19                & [11.86, 46.52]                & 3.32            & 0.001        & 0.39                 & [-0.04, 0.81]             \\
&Attr (True) × Competition                  & -57.27               & [-73.49, -41.05]              & -6.96           & < .001       & 0.46                 & [0.06, 0.85]              \\
&Ident (True) × Competition                  & -7.45                & [-14.52, -0.39]               & -2.08           & 0.039        & -0.33                & [-0.49, -0.18]            \\
&Cooperation × Competition                         & -11.25               & [-14.87, -7.64]               & -6.14           & < .001       & 0.70                 & [0.11, 1.30]              \\
&BT (Sli) × SMS                                   & 109.83               & [53.25, 166.40]               & 3.83            & < .001       & -1.42                & [-2.75, -0.09]            \\
&BT (Mus) × SMS                                   & -18.31               & [-31.05, -5.57]               & -2.83           & 0.005        & -0.10                & [-0.39, 0.19]             \\
&Sex (M) × SMS                                        & 63.20                & [38.32, 88.09]                & 5.00            & < .001       & 0.66                 & [0.11, 1.22]              \\
&Gender (M) × SMS                                     & -14.30               & [-26.68, -1.91]               & -2.28           & 0.024        & -0.19                & [-0.52, 0.13]             \\
&Attr (True) × SMS                          & -81.81               & [-104.65, -58.97]             & -7.06           & < .001       & 1.16                 & [0.64, 1.67]              \\
&Ident (True) × SMS                          & -13.19               & [-24.31, -2.06]               & -2.34           & 0.020        & -0.10                & [-0.46, 0.25]             \\
&Cooperation × SMS                                 & -16.12               & [-21.42, -10.83]              & -6.01           & < .001       & -1.03                & [-1.82, -0.24]            \\
&Competition × SMS                                 & -21.02               & [-28.01, -14.03]              & -5.92           & < .001       & -0.23                & [-0.83, 0.36]             \\
& ... & ... & ... & ... & ... & ... & ... \\
\hline
Admire & Intcp (O,Ob,F)                               & 2.74                 & [2.37, 3.12]               & 14.29           & < .001        & 2.92e-15              & [-0.22, 0.22]              \\
&BT (Sli)                                 & -0.35                & [-0.55, -0.16]             & -3.55           & < .001        & -0.20                 & [-0.31, -0.09]             \\
&SocialC (T) × BT (Sli)                & -0.13                & [-0.24, -0.01]             & -2.15           & 0.032         & -0.07                 & [-0.14, -0.01]             \\
\hline
\end{tabular}}
\end{table}

\begin{table}[]
  \caption{The table shows all significant results of main and interaction effects of further emotion variables. The Intercept (Intcp) shows which variables and what baseline were included in the best-fitting model (O = Opponent: Teammate; Ob = Obese: BodyType; F = Female: Sex; Attr = Being attracted towards the avatar; Ident = Identification with the avatar).}
\label{tab:lmm_emotions2_6}
\resizebox{\textwidth}{!}{%
\begin{tabular}{lllllllll}
\textbf{Emotions} & \textbf{Effect} & \textbf{Beta} & \textbf{95\% CI} & \textbf{t()} & \textbf{p-value} & \textbf{Std. Beta} & \textbf{Std. 95\% CI} \\
 \hline  
PosWellbeing & Intcp (O,Ob,F)         & 5.06                  & [4.82, 5.31]               & 41.12           & < .001        & 3.42e-15              & [-0.25, 0.25]             \\
&BT (Mus)           & 0.07                  & [0.02, 0.12]               & 2.82            & 0.005         & 0.07                  & [0.02, 0.12]              \\
\hline
& BT (Sli) & 0.07 & [0.02, 0.12] & t(546) = 2.81 & 0.005 & 0.07 & [0.02, 0.12] \\
\hline
PsyDistress & Intcp (O,Ob,F)                 & 1.46                 & [1.25, 1.68]               & 13.25           & < .001        & -0.06                 & [-0.34, 0.22]             \\
&BT (Mus)                   & -0.04                & [-0.07, 0.00]              & -2.05           & 0.041         & -0.05                 & [-0.09, 0.00]             \\
&Sex (M)                        & 0.05                 & [0.02, 0.08]               & 3.24            & 0.001         & 0.07                  & [0.03, 0.11]              \\
&Gender (M)                     & -0.18                & [-0.35, -0.01]             & -2.09           & 0.037         & -0.23                 & [-0.45, -0.01]            \\
&Attr (True)          & 0.04                 & [0.00, 0.08]               & 2.19            & 0.029         & 0.05                  & [0.01, 0.10]              \\
\hline
Enjoyment & Intcp (O,Ob,F)         & 5.33                  & [4.97, 5.69]               & 28.78           & < .001        & -1.65e-14             & [-0.24, 0.24]             \\
\hline
Arousal & Intcp (O,Ob,F)         & 2.96                  & [2.71, 3.21]               & 23.54           & < .001        & 0.04                  & [-0.19, 0.27]             \\
&BT (Sli)           & 0.08                  & [0.02, 0.14]               & 2.81            & 0.005         & 0.08                  & [0.02, 0.13]              \\
&BT (Mus)           & -0.07                 & [-0.13, -0.02]             & -2.50           & 0.013         & -0.07                 & [-0.12, -0.01]            \\
&Gender (M)             & 0.39                  & [0.14, 0.63]               & 3.14            & 0.002         & 0.36                  & [0.13, 0.58]              \\
\hline
Dominance & Intcp (O,Ob,F)         & 3.45                  & [3.14, 3.75]               & 22.42           & < .001        & -1.04e-14             & [-0.25, 0.25]             \\
&BT (Sli)           & 0.07                  & [0.01, 0.14]               & 2.15            & 0.032         & 0.06                  & [0.01, 0.11]              \\
\hline
Valance & Intcp (O,Ob,F)                                & 1.74                  & [1.56, 1.92]               & 18.92           & < .001        & -5.11e-15             & [-0.23, 0.23]             \\
&SocialC (T) × Sex (M)                      & 0.04                  & [0.01, 0.08]               & 2.32            & 0.021         & 0.06                  & [0.01, 0.11]              \\
&SocialC (T) × BT (Sli) × Sex (M)          & -0.05                 & [-0.11, 0.00]              & -2.03           & 0.042         & -0.07                 & [-0.14, 0.00]             \\
&SocialC (T) × BT (Mus) × Sex (M)          & 0.06                  & [0.00, 0.11]               & 2.10            & 0.036         & 0.07                  & [0.00, 0.14]              \\
\hline
\end{tabular}}
\end{table}

\paragraph{Main effects}
Intimidation is negatively influenced by a slim body type as well as jealousy. Jealousy on the other hand, is positively influenced by the teammate condition.
Empathy is positively influenced by the teammate condition but negatively influenced by a male avatar. Envy is positively influenced by teammate condition but negatively influenced by a muscular avatar or a male avatar or identification with the other avatar.
A teammate condition significantly heightens embarrassment but lowers embarrassment significantly with a slim avatar.
Pride is significantly positive influenced by working out with a muscular avatar.
Shame is negatively influenced (lower) by working out with a slim body type or a male avatar, but higher when they identify as male, are attracted towards the other avatar or have high cooperation, competition, or sports motivation values.

Positive Well-being is positively influenced by a muscular or a slim body type of the avatar. 
A muscular body type avatar also lowers psychological distress, together with identifying as a male.
However, psychological distress is significantly higher when working out with a male avatar or being attracted towards the avatar.
A muscular body type lowers arousal, while a slim avatar body type heightens arousal together with identifying as male.
Dominance is positively influenced by working out with a slim body type.

\paragraph{Interaction effects}
Combinations of teammate with slim avatar or male gender influence empathy positively, like combinations with being attracted towards slim avatar or male avatars.
Male participants working out with slim body type avatars or teammate conditions where participants are attracted toward the avatar, lower empathy. 
Male avatars with participants with high levels of cooperation, high level of competition, or high level of sports motivation heighten empathy.
Meanwhile, teammate conditions with participants that identify with the avatar or have high levels of Cooperation, competition, or sports motivation lower empathy, except of when they are attracted towards a slim body type avatar.

A teammate condition with a slim body type avatar and a male muscular avatar heighten envy.
Envy is positively influenced (higher) by being attracted towards the male avatar or the slim body shape. 
Identification with the avatar is hightening envy in conditions where additionally the body type is muscular, the sex of the avatar male or the gender of the participant male. 
Identification with the avatar lowers envy if the body type is slim or participants are attracted to the avatar. 
High cooperation characteristics in the teammate condition lower envy, while high cooperation characteristics heighten envy when the avatar is muscular or male or if they identify with the avatar. 
Working out with male avatar and being competitive or sport motivated hightens envy as well.

Shame is lower for male slime avatars but higher for male muscular avatars. 
Furthermore, working out with male muscular avatars, being attracted to slim avatars, being attracted to male avatars, and identifying with the male avatar heightens shame.
Shame is further heightened by being cooperative and working out with a slim body shape or a male avatar. 

Valence is heightened for teammate conditions where the avatar is male or additionally slim or muscular. 

\subsubsection{Performance}

In Performance, the \emph{NumberOfSitups} was additionally influenced by the sport motivation score of the participant but when having more granular data through \emph{Time} for each sit-up, all nine variables seem to influence the results.
The \emph{HR} model has to be considered with caution since AIC value is very high.
The results of the linear mixed models indicate the following statistically significant differences for performance are reported in \autoref{tab:lmm_performance_6}.

\begin{table}[htbp]
  \caption{The table shows all significant results of main and interaction effects of performance variables. The Intercept (Intcp) shows which variables and what baseline were included in the best-fitting model (O = Opponent: Teammate; Ob = Obese: BodyType; F = Female: Sex; Attr = Being attracted towards the avatar; Ident = Identification with the avatar).}
\label{tab:lmm_performance_6}
\resizebox{\textwidth}{!}{%
\begin{tabular}{lllllllll}
\textbf{Performance} & \textbf{Effect} & \textbf{Beta} & \textbf{95\% CI} & \textbf{t()} & \textbf{p-value} & \textbf{Std. Beta} & \textbf{Std. 95\% CI} \\
 \hline
NumSitups & Intcp (O,Ob,F)             & 4.72                 & [2.92, 6.51]              & 5.17            & < .001        & -0.03                & [-0.30, 0.23]             \\
& BT (Sli)         & 0.22                 & [0.09, 0.36]              & 3.19            & 0.001         & 0.12                 & [0.04, 0.19]              \\
& BT (Mus)         & 0.21                 & [0.07, 0.35]              & 2.95            & 0.003         & 0.11                 & [0.04, 0.18]              \\
\hline
Time & Attr (True)                            & -754.90              & [-815.14, -694.65]             & -24.69          & < .001        & -7.21                & [-8.01, -6.41]             \\
&SocialC (T)× Attr (True)        & 813.96               & [733.80, 894.11]               & 20.01           & < .001        & 7.81                 & [6.75, 8.88]               \\
&BT (Sli)× Attr (True)           & 763.87               & [606.44, 921.30]               & 9.56            & < .001        & 6.71                 & [3.19, 10.22]              \\
&Sex (M) × Attr (True)                  & 905.27               & [750.19, 1060.35]              & 11.50           & < .001        & 7.91                 & [6.94, 8.88]               \\
&Gender (M) × Attr (True)               & 14.45                & [6.50, 22.41]                  & 3.58            & < .001        & 5.04                 & [4.33, 5.75]               \\
&Gender (M) × Ident (True)               & -6.22                & [-7.94, -4.49]                 & -7.10           & < .001        & -7.32                & [-9.35, -5.29]             \\
&Attr (True) × Cooperation              & 150.10               & [138.68, 161.52]               & 25.90           & < .001        & -2.37                & [-3.63, -1.11]             \\
&Attr (True) × Competition              & 202.87               & [189.75, 215.99]               & 30.47           & < .001        & -0.37                & [-0.67, -0.06]             \\
&Attr (True) × SMS                      & 213.12               & [192.91, 233.33]               & 20.78           & < .001        & -3.47                & [-3.88, -3.06]             \\
& ... & ... & ... & ... & ... & ... & ... \\
\hline
HR & Intcp (O,Ob,F)                                             & 94.15                & [91.09, 97.20]            & 60.55           & < .001        & -0.10                & [-0.35, 0.15]             \\
&BT (Sli)                                         & 4.68                 & [0.77, 8.59]              & 2.35            & 0.019         & 0.38                 & [0.06, 0.71]              \\
&SocialC (T)× BT (Sli)                     & -8.49                & [-13.98, -3.00]           & -3.04           & 0.002         & -0.70                & [-1.15, -0.25]            \\
&BT (Sli)× Sex (M)                               & -8.96                & [-14.45, -3.47]           & -3.21           & 0.001         & -0.74                & [-1.19, -0.28]            \\
&(SocialC (T)× BT (Sli)) × Sex (M)         & 11.81                & [4.05, 19.57]             & 2.99            & 0.003         & 0.97                 & [0.33, 1.61]              \\
\hline
Effort & Intcp (O,Ob,F)             & 4.87                 & [4.40, 5.35]              & 20.19           & < .001        & -0.12                & [-0.40, 0.16]             \\
\hline
Fatigue & Intcp (O,Ob,F)             & 2.37                 & [1.96, 2.78]              & 11.45           & < .001        & -0.13                & [-0.36, 0.10]             \\
&BT (Sli)         & 0.41                 & [0.19, 0.63]              & 3.64            & < .001        & 0.23                 & [0.11, 0.36]              \\
&BT (Mus)         & 0.21                 & [0.07, 0.35]              & 2.95            & 0.003         & 0.11                 & [0.04, 0.18]              \\
\hline
\end{tabular}}
\end{table}

\paragraph{Main effects}
Number of Situps is positively influenced by avatar body types slim and muscular, while the time per situp was significantly negative influenced (faster) by being attracted towards the avatar. 
Heart rate was positively influenced (higher) by slim body type avatars. 
A feeling of fatigue was significantly higher for slim and muscular avatar body types. 

\paragraph{Interaction effects} 
Time had a lot of interaction effects, but we only report the ones with two levels. 
Male avatar identification negative influenced (faster) time, while positively (slower) was influenced by attracted teammate, slim attracted, being attracted to the male avatar,  
being male and being attracted towards the avatar, and being attracted together with cooperation, competition or sports motivation. However, results with so many significant effects have to be considered carefully because of the data and the model.
HR is negatively influenced (lower) with teammate slim avatars and male slim avatars. However, slim teammate avatars that are male influence HR positively (higher).

\section{Discussion} 

To get a better overview, we summed up all the results of RQ1 in~\autoref{tab:results3F} and the ones for RQ2 in Appendix~\autoref{tab:DiscEffectPartDependent}. Next, we discuss details separately for both research questions. We conclude by proposing practical implications for researchers and practitioners and by pointing out the ethical implications of our work.

\begin{table}[!ht]
\caption{The table shows an overview of all significant results for the linear mixed model with social context (SocialC), body type (BT), and sex of another avatar. A plus in the front indicates an overall positive effect, and a plus in front of the columns of relatedness, emotional experience, and performance indicates an overall positive effect for this independent variable. A minus refers to a negative effect. Bold effects match with results from models including participant independent variables.}
\label{tab:results3F}
    \centering
    \resizebox{\textwidth}{!}{%
    \begin{tabular}{l|l|l|l|p{2.3cm}|l|p{2cm}|l|p{2.3cm}}
     SocialC & BT & Sex & & Relatedness & & Performance & & Emotion \\
\hline
Team & & & + & \textcolor{darkgreen}{IOS $\uparrow$, \textbf{responsible $\uparrow$}} &  & &  & \\
 & SLI & & + & \textcolor{darkgreen}{IOS $\uparrow$, IMI $\uparrow$, \textbf{attracted $\uparrow$, focused $\uparrow$}} & + & \textcolor{darkgreen}{\textbf{NoS $\uparrow$}, Effort $\uparrow$} & ? & \textcolor{darkred}{\textbf{Intimidated $\uparrow$, Jealousy $\uparrow$, Envy $\uparrow$}}, \textcolor{darkgreen}{ \textbf{Admire $\uparrow$, Pos Wellbeing $\uparrow$}, Enjoyment $\uparrow$} \\
 & Mus &  & ? & \textcolor{darkgreen}{IOS $\uparrow$, IMI $\uparrow$, \textbf{attracted $\uparrow$, focused $\uparrow$,}} \textcolor{darkred}{\textbf{responsible $\downarrow$}} & + & \textcolor{darkgreen}{NoS $\uparrow$, Effort $\uparrow$} & - & \textcolor{darkred}{\textbf{Intimidated $\uparrow$, Jealousy $\uparrow$}, Envy $\uparrow$ , Embarrassed $\uparrow$, Shame $\uparrow$},\textcolor{darkgreen}{ \textbf{Admire $\uparrow$}} \\
 & & Male & + & \textcolor{darkgreen}{attracted $\uparrow$} & + & \textcolor{darkgreen}{Fatigue $\downarrow$} & + & \textcolor{darkgreen}{\textbf{PsyDistress $\downarrow$}} \\
\hline
Team & & Male & & & & & - & \textcolor{darkred}{PsyDistress $\uparrow$} \\
Team & Mus & & & & & & + & \textcolor{darkgreen}{Intimidated $\downarrow$} \\
Team & Sli & & & & & & + & \textcolor{darkgreen}{Jealousy $\downarrow$} \\
 & Sli & Male & - & \textcolor{darkred}{attracted $\downarrow$} & & & & \\
\hline
    \end{tabular}}
\end{table}


\subsection{Participant-Independent Effects}
We discuss our results on participant-independent effects, as designers may not always have access to the user data that would allow them to understand effects more specifically.
We discuss our findings based on the selected models (Appendix \autoref{ModelComparisionPI}) and on the overview of significant results in \autoref{tab:results3F}.

\subsubsection{Influence of Social Context}

First of all, we want to point out that social context was relevant for almost all models, indicating an influence on almost all dependent variables. 

For Relatedness, we found significant positive effects for inclusion of the other in the self (IoS) and in feeling responsible in the teammate condition.
We expected these results, as teammates need a positive mindset towards the other person to work together, resulting in a feeling of closeness and responsibility for them \cite{satoCollaborativeDigitalSports2014a,brondiEvaluatingEffectsCompetition2015}. 
It is interesting that only the mindset really invoked these significant effects since participants only had to immerse into a teammate or opponent mindset instead of really being in different scenarios.


\textit{F1a: Teammate conditions are helpful to create feelings of relatedness through inclusion as well as a feeling of responsibility.}

 The social context (teammate/opponent) did not significantly influence participant's emotional experience. However, working out with a teammate reduced feelings of intimidation and jealousy. When working out with a muscular avatar, this was most notable in the feeling of intimidation, while when working out with a slim one, this was most notable for the feeling of jealousy. 

\textit{G1b: Teammate conditions can have positive effects on a person's emotional experience (e.g., reducing feelings of intimidation and jealousy) when working out with a slim or muscular avatar.}

While male avatars overall lowered psychological distress, in teammate scenarios, males create significantly higher psychological distress compared to males in opponent scenarios. 
Reasons why male teammates create more psychological distress than male opponents are unclear, and more future work is needed.
However, carefully assumed, a traditional male picture could invoke a feeling of physical strength for the team while, at the same time, participants feel uncomfortable about relying on someone. 

\textit{F1c: Male teammates increase psychological distress significantly more than male opponents.}

Working out with a teammate vs working out with an opponent did not have a significant influence on performance when not including participant-dependent values. 
However, since all models include social context, it can be assumed that the effect is there but too small to be significant here.
This should be investigated further in future work.

\subsubsection{Influence of Body Shape}
Our results show that body shape had a strong influence on feelings of relatedness, emotional experience, and performance.

Slim and muscular avatars have a positive effect on different attributes of relatedness, like inclusion (IOS) and relatedness (IMI), but also attractiveness and more focus on the other person. 
However, responsibility significantly lowers with muscular avatars. We assume a muscular avatar looks as being able to handle him or herself.


\textit{F1d: Relatedness can be heightened through slim and muscular body-shaped avatars}

The emotional experience for slim avatars is very twisted.
While a slim avatar positively influences well-being and enjoyment, it also heightens intimidation, jealousy, and envy. 
A muscular avatar heightens intimidation, jealousy, envy, embarrassment, and shame.
Slim and muscular body types heighten admiration. Which we decided to see as a positive emotion, turning envy and jealousy into a positive feeling towards the other person. 

However, the effect of muscular body type being intimidated switches to lower intimidation in teammate condition. For the slim body type jealousy is lower in teammate condition. 

From another perspective, we could say that the obese body type could counteract negative emotions that could also decrease motivation to exercise \cite{vallerandIntegrativeAnalysisIntrinsic1999}.

\textit{F1e: Emotional experiences are more negative for muscular and slim body shapes, having additional positive emotional experiences with slim body shapes. Slim and muscular body types get less negative emotional experiences in teammate condition.}


The number of sit-ups and the perceived effort are heightened through muscular and slim body-shaped avatars. 
It can be assumed that it was motivating to see a visual representation of what could be if the workout were continued regularly as a positive rather than a negative incentive.
Conversely, it means that an obese avatar lowers performance and effort.
Our results align with the work of \cite{penaAmWhatSee2016}, where an obese opponent lowered performance compared to a 'normal' opponent.

\textit{F1f: Performance can be heightened through slim and muscular body-shaped avatars}

\subsubsection{Influence of Visual Sex}
Visual sex is very seldom included in the best-fitting model. Indicating that visual sex is not an important factor for relatedness, performance and emotion.

A male avatar positively influences attractiveness. However, this main effect only indicates that we had a sample size where more people were attracted to male body shapes. 
As mentioned before, 
A slim male, on the other side, participants were less attracted to. 
\textit{F1g: In our sample, participants were more attracted to male avatars.}

Since the AIC model fit value is very low for psychological distress, we we can give more weight to these results.
A male avatar lowers psychological distress compared to a female avatar; however, in the teammate condition, the effect is more negative than in the opponent condition. 

\textit{F1h: A male avatar lowers psychological distress, though the effect is lower in the teammate condition than in the opponent condition.}

A male avatar made participants feel less fatigued while working out, which could be helpful if participants get demotivated from fatigue. 
We could assume that a male avatar is potentially seen as strong in general, and while in a teammate situation, the male avatar could carry the team, in a competitive scenario, it would not be too disgraceful to lose against him.
Here, we assume that perceptions about the sex/gender of a person play a role in our results.

\textit{F1i: A male avatar can lower fatigue.}

\subsection{Participant-Dependent}
In the following subsection, we discuss participant-dependent variables.
If there is access to participant-related data, the following results allow us to understand the effects more specifically.
Whether this should be the case at all, we discuss in ethical implications.
We discuss our findings based on the selected models (Appendix \autoref{ModelComparisionPI}) and on the overview of significant results in Appendix \autoref{tab:DiscEffectPartDependent}. The table is reduced to only presenting the main effects and interaction effects of two levels, including social context, body type, and sex, since they are the variables we were interested in.

\subsubsection{Influence of Social Context}
Including participant-dependent variables has increased the significance of social context. Similar to the earlier models, the teammate condition enhances the sense of responsibility while lowering heart rate and negative emotions like jealousy and embarrassment, improving overall valence. The results suggest that participants are more relaxed and emotionally balanced when in a team scenario.

Additionally, we found interesting interaction effects, that could enhance or lower effects. For example, being attracted to a teammate heightens relatedness and performance time and lowers envy and shame, but being a female participant lowers relatedness. 
A male teammate lowers shame, and a slim teammate lowers heart rate, envy, and shame while improving valance. 

Attraction to a teammate in the team condition heightens relatedness, improves performance time, and reduces envy and shame. A male teammate also lowers shame, and a slim teammate lowers heart rate, envy, and shame while improving valence. These results reflect the complex nature of social interactions in exercise settings, with team dynamics both enhancing emotional experiences and improving performance. The competitive scenario, however, does not show a clear performance improvement, which contrasts with findings from other studies that show competition typically boosts performance in VR exercises \cite{dimenichiPowerCompetitionEffects2015,mouattUseVirtualReality2020,shawCompetitionCooperationVirtual2016}. Collaboration appears to be an equally strong motivator in team-based settings, and further investigation is needed to understand if exercise type (endurance vs. strength training) or the use of virtual trainers instead of teammates influences these outcomes.

\textit{F2a: Teammate conditions improve responsibility, performance (especially lowering heart rate), induce positive emotions like valence, and reduce jealousy and embarrassment.}

\subsubsection{Influence of Body Type}
Consistent with previous findings, slim and muscular body types improve attractiveness and focus, especially in team settings. The muscular body type also lowers responsibility, perhaps due to participants' assumption that muscular avatars are capable of completing tasks like sit-ups without needing support. Both slim and muscular avatars have a strong positive effect on performance, further emphasizing their impact.

Main effects: Slim and muscular avatars continue to heighten negative emotions like intimidation and jealousy, but at the same time, they increase positive well-being. Slim avatars, in particular, improve positive emotions while maintaining responsibility. This dual effect of body types is likely influenced by societal perceptions of slim and muscular physiques, which are often promoted as ideal body types\cite{RODGERS2022284}.

Interaction effects: Interaction effects show that muscular avatars, especially male ones, heighten empathy, while identification with muscular male avatars also increases envy and shame. Slim avatars, on the other hand, reduce these negative emotions when combined with strong cooperation, competition, or sport motivation scores. However, being attracted to slim avatars tends to heighten envy and shame. These nuanced effects highlight the complexity of avatar-body-type dynamics and suggest that the muscular and slim avatars evoke strong emotional reactions, both positive and negative, based on participant interactions.

Muscular body types, however, heighten empathy in many interaction effects, but identifying with a muscular male heightens envy, and a muscular male heightens shame. These interaction effects are likely to be related to our sample and have to be taken with caution. 

Interaction effects between slim males or slim body types with strong cooperation, competition, and sport motivation scores lower envy and shame, except of being attracted to a slim avatar seems to heighten envy and shame.

\textit{F2b: Slim and muscular body shapes boost performance but often heighten negative emotions. Slim body types perform slightly better, also increasing positive emotions while not lowering responsibility.}

\subsubsection{Influence of Virtual Sex}
Male avatars or identifying as male continues to reduce psychological distress, which aligns with the previous results. They can further heighten or lower emotions such as empathy, shame, and envy. 

A muscular male teammate may lead to reduced effort on the part of participants, possibly due to their reliance on the perceived strength and capability of the muscular avatar. This aligns with our previously mentioned findings, suggesting that participants in team-based scenarios feel more relaxed and delegate responsibility to their (muscular) teammate.
While male avatars can heighten empathy when participants are attracted to or identify with them, they also increase envy in such cases. This is particularly true for muscular male avatars, which seem to evoke strong emotional reactions related to participants’ attraction or identification with the avatar. 


\textit{F2c: Male avatars reduce psychological distress. In combination with body types or social context, they can heighten or lower emotions such as empathy, shame, and envy}

\subsubsection{Influence of Being Attracted towards the Avatar}
Attraction towards avatars continues to significantly influence emotional and performance outcomes. Beyond relatedness, attraction plays a role in psychological distress, empathy, envy, and shame. This effect is especially prominent when avatars are slim or muscular, with these body types evoking stronger emotional responses.

Attraction to avatars, especially in team-based scenarios, enhances relatedness and improves performance, particularly in terms of time efficiency. However, attraction to slim avatars or male avatars may also distract participants, leading to lower exertion during tasks. This suggests that while attraction can enhance connection and performance in some scenarios, it may also hinder performance due to distraction or a sense of support provided by the attractive avatar.
Participants attracted to slim avatars or male avatars experience heightened envy and shame, especially in combination with high levels of cooperation, competition, or sport motivation. This complex emotional response suggests that attraction can be both beneficial and detrimental, depending on the specific dynamics of the exercise scenario.

\textit{F2d: Attraction towards an avatar can positively influence relatedness and performance but can also heighten envy and shame in certain contexts.}

\subsubsection{Influence of Identifying with the Avatar}
Previous research has demonstrated that identification with avatars that share similar physical traits, such as hair color, skin color, or facial features, significantly enhances motivation and performance in VR exercise scenarios\cite{fittonDancingAvatarsMinimal2023,clarkeFakeForwardUsingDeepfake2023}. In contrast to these findings, our study did not observe similar results when focusing on body type alone. The influence of body type on identification appears to be less pronounced compared to more personal and recognizable features like hair or facial characteristics.

One possible explanation for this discrepancy could be the limited representation of certain body types within our sample. For example, we had a small number of participants identifying with obese avatars, which may have reduced the statistical power to detect significant effects. Additionally, body type may not have the same inherent identification level as features like skin or hair color, which are more central to an individual’s self-perception. Another factor could be the graphical resolution of the avatars used in our study. According to the uncanny valley \citet{latoschikEffectAvatarRealism2017}, subtle imperfections in avatar realism may have disrupted the identification process, making it harder for participants to connect with their virtual counterparts.

Moreover, one participant specifically noted the absence of diverse skin color options, suggesting that broader customization options could have led to stronger identification and engagement with the avatars. While we intentionally chose not to include different skin colors for simplicity, this feedback highlights the importance of incorporating more diverse and personalized avatar characteristics in future studies.

In future work, offering a broader range of identifying features, including both body types and facial or skin characteristics, could lead to more significant results. By integrating such customizations, we may better capture the full potential of avatar identification to enhance motivation, emotional connection, and overall performance in virtual environments.

\textit{F2e: Identification with the avatar primarily influences empathy and envy, with stronger effects in team scenarios.}

\subsubsection{Influence of Sport Motivation Score}
Sport motivation plays a critical role in performance, particularly influencing the number of sit-ups completed by participants. 
We can assume participants with higher sport motivation are actively exercising in their leisure time, and so they were better trained for this task. 

Additionally, sport motivation influences emotional responses, such as empathy and envy.

\textit{F2f: High sport motivation significantly improves performance and reduces negative emotional responses.}

\subsection{Prediction Tree and Participant-Dependent Predictions}

Since participant-dependent variables are hard to interpret because of their complexity, we built a system to get specific user-centered predictions based on our study data. 
On a website input form (GitHub Link \footnote{Code is in Supplementary Materials for now because of anonymity}(\autoref{fig:Prediction}a), users can enter participant-dependent information (Gender, BodyType that they identify with, BodyType that they are attracted to, Sport Motivation (SMS), Competition and Cooperation values (CCPS) and what they want to predict. 
The best prediction with this prediction model is always the highest value. For example, a prediction for intimidation gives a VR exercise setting for high intimidation, so what setup is better not to use?
As a result, the system provides a result (\autoref{fig:Prediction}b) which shows the inserted values and the best prediction for them, including how high the value for this variable will be predicted as a prediction value. Additionally, a prediction tree is created (\autoref{fig:Prediction}c) that shows all combinations and how high the predicted value would be.
Despite the frequent high prediction values, we caution against over-interpreting the data. The predictions were significantly influenced by the small and rather specific sample of computer science participants (N = 48) and even more so by the over-representation of certain participant identifications (e.g., Slim Muscular) compared to others (e.g., Female Obese). Therefore, predictions for higher-represented samples are likely to be more precise.

\begin{figure}[!h]
    \centering
    \includegraphics[width=\textwidth]{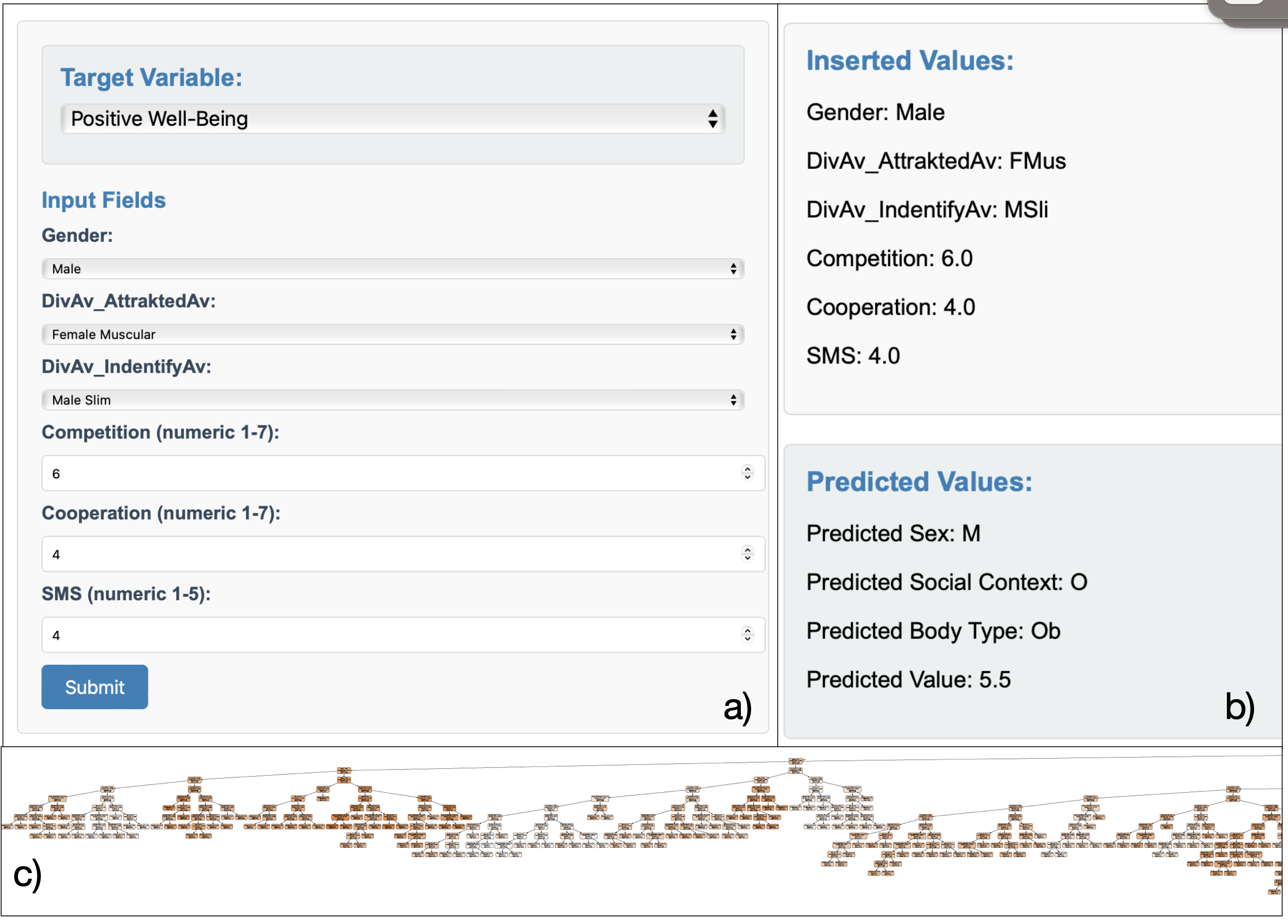}
    \caption{a) The website input form is used to put in the target variable that should be predicted (e.g., Intimidation) and information about a user (here, randomly generated data). b) The result page with the information inserted before and the final prediction for Sex, Body Type, and Social Context, as well as the predicted value for the target variable. c) A complete prediction tree will be further created and saved as a pdf.}
    \label{fig:Prediction}
    \Description{a) On top, there is a drop-down menu for the target variable, then below are drop-down menus for Gender, Attratedness Avatar, and Avatar Identification, as well as text input fields for Competition, Cooperation, and Sport Motivation (SMS). The submit button below sends the information to the Python script. b) The result page first shows the inserted values like Gender, Attraced Avatar, Identification Avatar, Competition, Cooperation, and SMS. And the predicted values of Sex, Social Context, Body Type, and Predicted values. In c), there is a miniature picture of the prediction tree since it is very small and, therefore best visible separately.}
\end{figure}

\subsection{Practical Implications}
In the following, we summarize our main recommendations, highlighting the practical implications of our results. Recommendations are shown as \textbf{R1} - \textbf{R5}.

First, we emphasize that it is important to think about one's goals when exercising: Do users struggle with motivation and need positive emotions to keep going, or do they predominantly want to improve performance? Is it important to improve the relation between the exercising people, because they engage in a team sport or is that not the main priority? These questions need to be asked to determine what avatar would be best in an exercise scenario and to understand the effects thereof.
Furthermore, we consider two different scenarios: either we know about users' characteristics, like how motivated they are to do sports or the body type they identify with or are attracted to, or we do not know anything and need to build a default scenario. 

In the default scenario (participant-independent), we can focus on creating a competitive or cooperative exercise scenario or focus on the body shape of the avatars.

A teammate condition as a way of collaboration could motivate users through a social factor and growing relatedness that motivates them to keep on exercising and, for example, losing weight in the longer run \cite{staianoAdolescentExergamePlay2013a}. It could even contribute to users' meeting outside of VR and exercising together, grounded in social motivation through relatedness \cite{allenSocialMotivationYouth2003}.

Considering team building interests, for example, in sport teams or work teams, exercising together could enhance relatedness.
Regarding body shape, working out with an avatar of slim and not muscular body shape (we do not want to lower responsibility) could support focusing on each other and building stronger connections. What effects regarding attractiveness occur and how long they will hold on after exercising must be considered carefully in future work.

\textbf{R1: To increase relatedness in a social VR default exercise scenario, a teammate with a slim body type would be a balanced default choice.}

In some cases, VR exercise users are mainly interested in increasing their performance. Here, slim or muscular body shapes are strongly recommended. If users struggle with high heart rates, a teammate scenario could help to be more relaxing while still pushing to improve performance.

\textbf{R2: Working out with slim or muscular body-shaped avatars would likely improve performance in a social VR default exercise scenario.}

One of the most important factors for users staying motivated to work out would be positive emotions \cite{popaEMOTIONSROLEMOTIVATION2013,jicolPredictiveModelUnderstanding2023}. 
Here, our results indicate obese avatars as being a good choice to lower intimidation, jealousy, shame, or embarrassment.  In some cases, slim avatars could be a good choice as well, as they can increase enjoyment. 
Additionally, cis-male avatars will lower psychological distress and could be a good choice when building up positive emotions for exercise.

\textbf{R3: In a social VR default exercise scenario with a most comfortable emotional experience, we suggest an obese body type and/or male avatar to work out with.}

If more information about the participant is available (participant-dependent), we could build a more user-tailored scenario, adding information about, for example, attractiveness preferences.

\textbf{R4: In a social VR scenario, participants benefit in performance and lower psychological distress if an avatar is chosen to which they feel attracted. Relatedness benefits if the attracted avatar is in the teammate scenario.}

\textbf{R5: A sport motivation mindset, in general, helps to improve performance very strongly.}

To build better user-tailored scenarios, such a website as we introduced using a prediction tree can be used. However, the prediction would much improve if that data set grew and additional exercises would be included to make it more generalizable.

\subsection{Ethical Implications}
In this paragraph, we critically reflect on implications of our results on society and how the information we gained in our paper can be used/misused. 


We did not find very strong stereotypes for sex and gender. There are specific interaction effects that could indicate underlying stereotypical attitudes about traditional male and female roles, but they can not be proven.
For example, male avatars show a slightly positive effect in emotional states but overall we would cautiously argue that visual sex representation is less important than being attracted towards the avatar. 

However, for body type we did find stereotypical implications. Muscular and slim body types create stronger relatedness and performance compared to obese body types. 
We could assume that attractiveness or identification with obese avatars could play a role. Unfortunately, our sample only had two participants identifying with obese avatars and none stating they are attracted towards obese body types.


Considering individuals possess the capacity to manipulate the visual representation of other bodies in a virtual fitness scenario, several consequences can arise. 
Such manipulation introduces the potential for distortions in one's perception of another person.
Over time, individuals may struggle to discern the correctness of their perceptions, leading to different emotions and relatedness and possibly even behavior changes towards the visually manipulated person. 
Consequently, the dynamics of their relationships may deviate from what they would naturally be without manipulation.
The affected individual may even remain unaware of the distortions depending on the implementation. 
However, properly used, it may result in individuals improving their performance or finding themselves in closer relatedness to others. 
Less controversial in this scenario would be that the manipulated person is the manipulator itself and not others. 


A slightly different scenario emerges where individuals have the ability to manipulate their own bodies. 
The creation of avatars is a ubiquitous and indispensable feature within the digital world. 
Permitting individuals to design and fashion their own avatars as they wish across various digital platforms, ranging from video games to messaging applications.
While this can be used to empower individuals to protect themselves from unwanted sexualization through their body shape, it simultaneously can be used to intentionally intimidate others to gain a physical advantage (e.g., in a competitive scenario) or someone to feel more related to yourself. 
That manipulation through physical body shape in happening video games was already investigated \cite{rodriguesPersonalizationImprovesGamification2021}, but now we know that in VR fitness scenarios this is possible as well. 
That customizing one's own avatar can even trigger social anxiety \cite{dechantHowAvatarCustomization2021} should probably also be taken into account when considering freedom for social VR fitness applications. 
The overarching ethical question that arises relates to prioritizing the disadvantages versus the advantages that these scenarios bring. 
Deciphering whether the negative consequences of such capabilities should take precedence over the potential benefits or vice versa is a complex and multifaceted ethical dilemma that necessitates careful deliberation. 
While manipulation through visual representation happens all the time in real everyday life, at least active manipulation involves a lot of effort or cost (e.g., doing a lot of sports, buying and wearing certain clothes). 

If we think of NPCs/avatars who assist or train us in a virtual training, the ethical conflicts mentioned above diminish. 
However, we should wonder whether manipulated agents of our liking would reinforce stereotypes or not. 
These suspected long-term developments should be looked at closely in future work.

\section{Limitations and Future Work}

Since the study sample consists of university students between 20 and 30 the results might only be applicable beyond this population. 
Regarding the deeper level analysis (RQ2), our sample might be too small and unbalanced for participants' characteristics like body type identification. Future work would be interesting, exploring a bigger group of participants with specific characteristics to see if significance will emerge. 

We chose a White skin color of the avatars which matched the skin color of our participant's sample, thus increasing identification. However, one participant criticized that the White skin color is not neutral. We agree that limiting the skin color limits the generalizability of our results and more studies have to be conducted to proof its validity beyond these restrictions.
In this study, we were aiming for visual body shape influences, which is why we tried to remove as many additional factors as possible (e.g., avatars had no hair). 
Simultaneously, we had to keep a balance between presence, realism and no uncanny valley effect, which is why we decided on a common skin tone in our university's country instead of a more neutral color like yellow. 


It was only in the scope to investigate three body types, which is why we decided on three more distinctive variations in body shape (slim, muscular, and obese). 

Further, we decided to use only two clearly different visual gender appearances. However, we note that there are also many different manifestations and body shapes. 

In future work, many different combinations could be investigated, we were limited in due to complexity and time.
However, we want to list a few suggestions. While we could only investigate three levels of body shapes, a more granular differentiation or a combination with other visual cues like hair and skin color would be interesting.
Here, not only identification with the avatar could be considered but also attractiveness towards the avatar.

Our competition and collaborative social context were created indirectly by putting one's mindset into the social context context. 
Different social contexts due to a changed task or environment would have possibly brought in a new influencing factor.
However, putting participants into a mindset could have worked differently well, with different participants skewing the results. 
Nevertheless, we are very confident that it worked as expected overall since we have significant results between scenarios that logically make sense. 

Compared to previous studies in VR exercise (e.g., \cite{shawCompetitionCooperationVirtual2016}, we did not find a performance increase for competitive scenarios compared to collaborative.
It is unclear where this difference comes from, and we encourage further work towards exercise motivation to explore more cooperative exercise scenarios.
It could be investigated if the type of exercise (endurance vs. strength training) or the setup with a virtual trainer instead of a mate could have an influence.

 We actively decided against an exergame and for an actual sports exercise to see if the effects would occur for general exercise, even if performance in a game would have been easier to measure. 
 However, choosing a sports exercise across 12 scenarios without major performance losses is not particularly easy. 
 In the end, we chose sit-ups because they are easy to count, have little risk of injury, and, unlike push-ups, most people can do a few without losing too much performance over time. 
 Of course, we have tried to compensate for the fact that the performance decreases as much as possible by counterbalancing. 

Eventually, the variance overall is quite small to find many significant effects in the number of sit-ups, but the tendencies we found can and should be further investigated.

\section{Conclusion}

With this work, we took a first step into investigating body shapes of a teammate or opponent in virtual reality workouts and how they could affect relatedness, emotional experience, and performance, and thus keep workout motivation higher.
While our implications for design and society we provide a next step in the HCI community to further understand the complex interplay of social context and body shapes in VR and help designers to implement ethical and social VR exercise scenarios.
We found that the teammate condition is overall a good choice for increasing relatedness, positive emotions, and better performance in a virtual workout. While we found that working out with a slim or muscular avatar increases performance, we propose obese avatars to reduce emotions like intimidation, jealousy, and embarrassment. 
Overall, more research is needed to follow up on these interesting new findings. For example, future studies should investigate more attractedness and not only identification with a social avatar since it seems to strongly contribute to the influence some effects. 
However, especially in designing social VR exercise applications, it should not be forgotten that the ethical implications of our findings need to be carefully considered.

\begin{acks}

\end{acks}

\bibliographystyle{ACM-Reference-Format}
\bibliography{Paper_SocialVR}

\newpage
\input{appendixnew}

\end{document}

%% file: appendixnew.tex
\section{Appendix}

\subsection{Additional Plots}
\label{AdditionalPlots}

\begin{figure}[!h]
  \centering
  \includegraphics[width=0.8\textwidth]{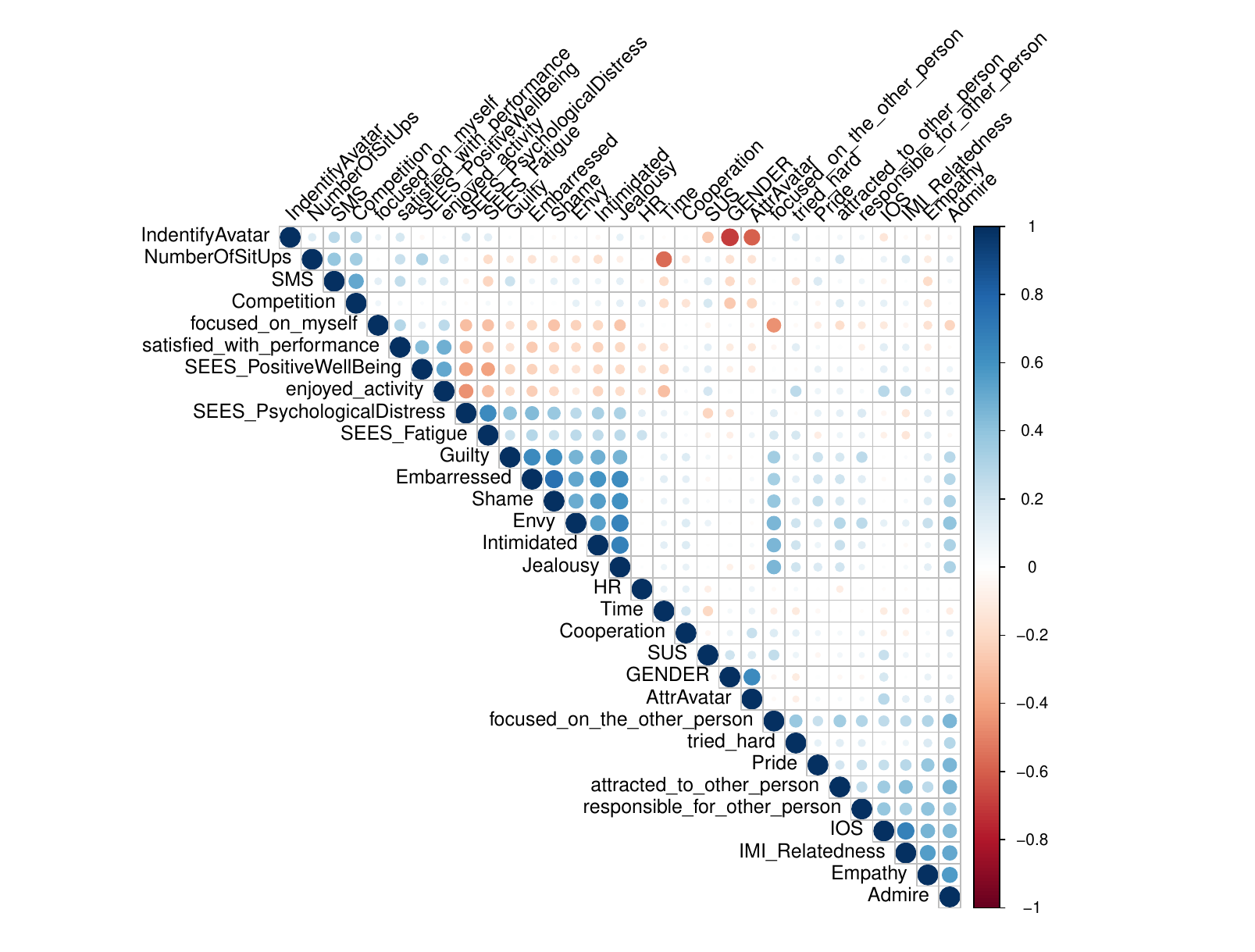}
  \caption{Correlation between all measured dependent and overall participant related demographics.}
  \Description{}
\end{figure}

\newpage

\begin{figure}[!h]
  \centering
  \includegraphics[width=0.8\textwidth]{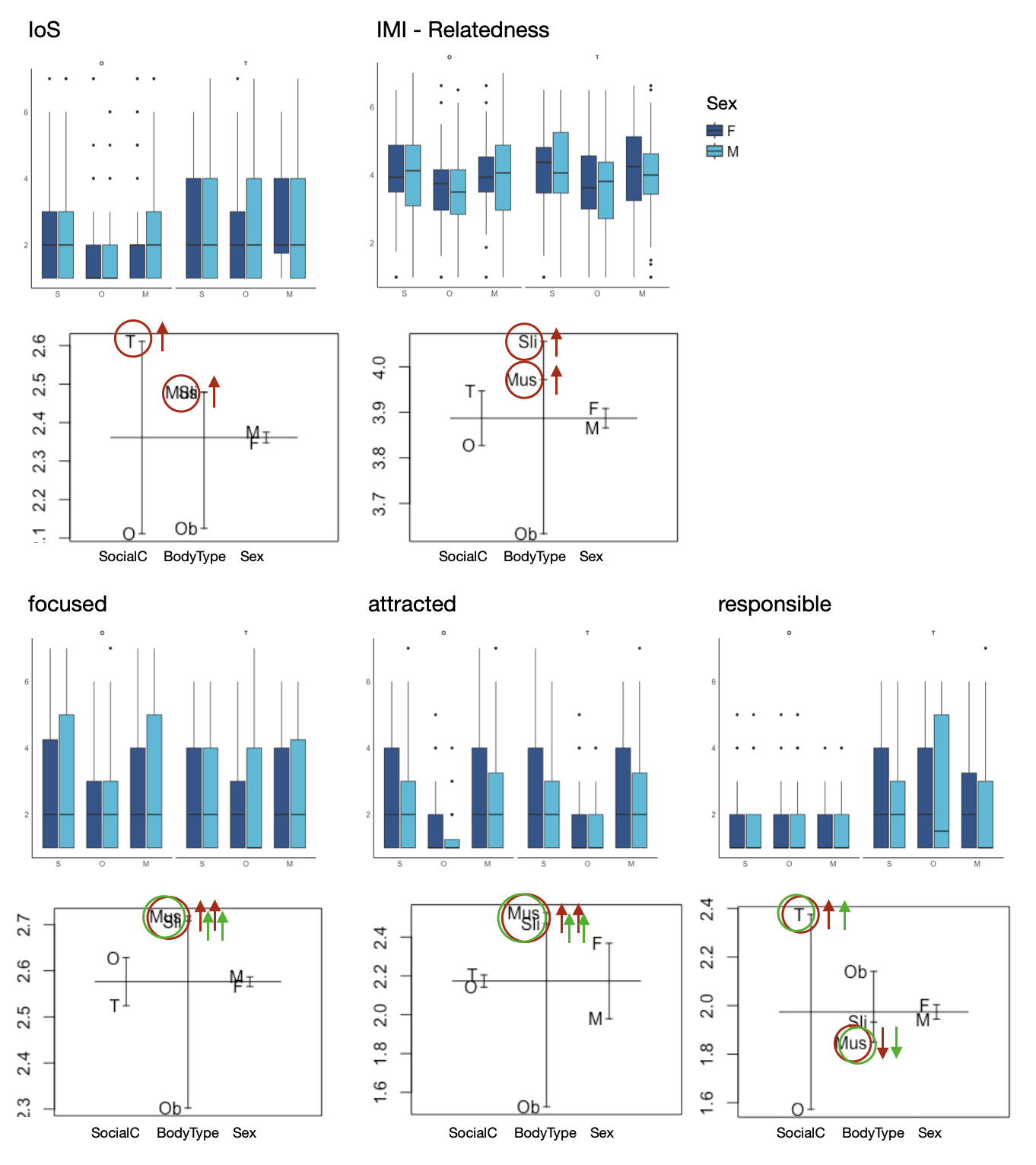}
  \caption{Mean scored and box plot for relatedness results.}
  \Description{}
\end{figure}

\begin{figure}[!h]
  \centering
  \includegraphics[width=0.8\textwidth]{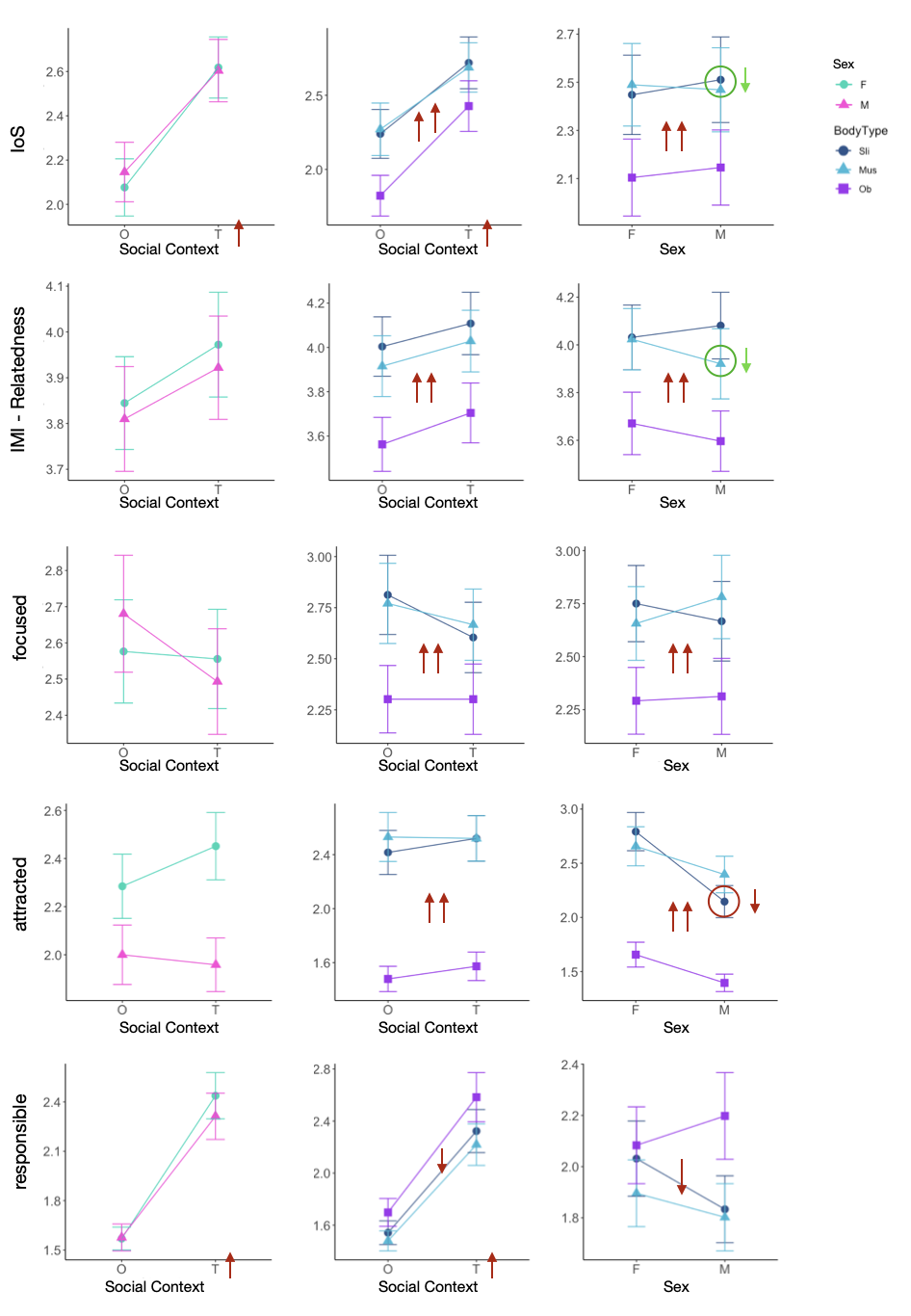}
  \caption{Mean scored and box plot for relatedness results.}
  \Description{}
\end{figure}

\begin{figure}[!h]
  \centering
  \includegraphics[width=0.8\textwidth]{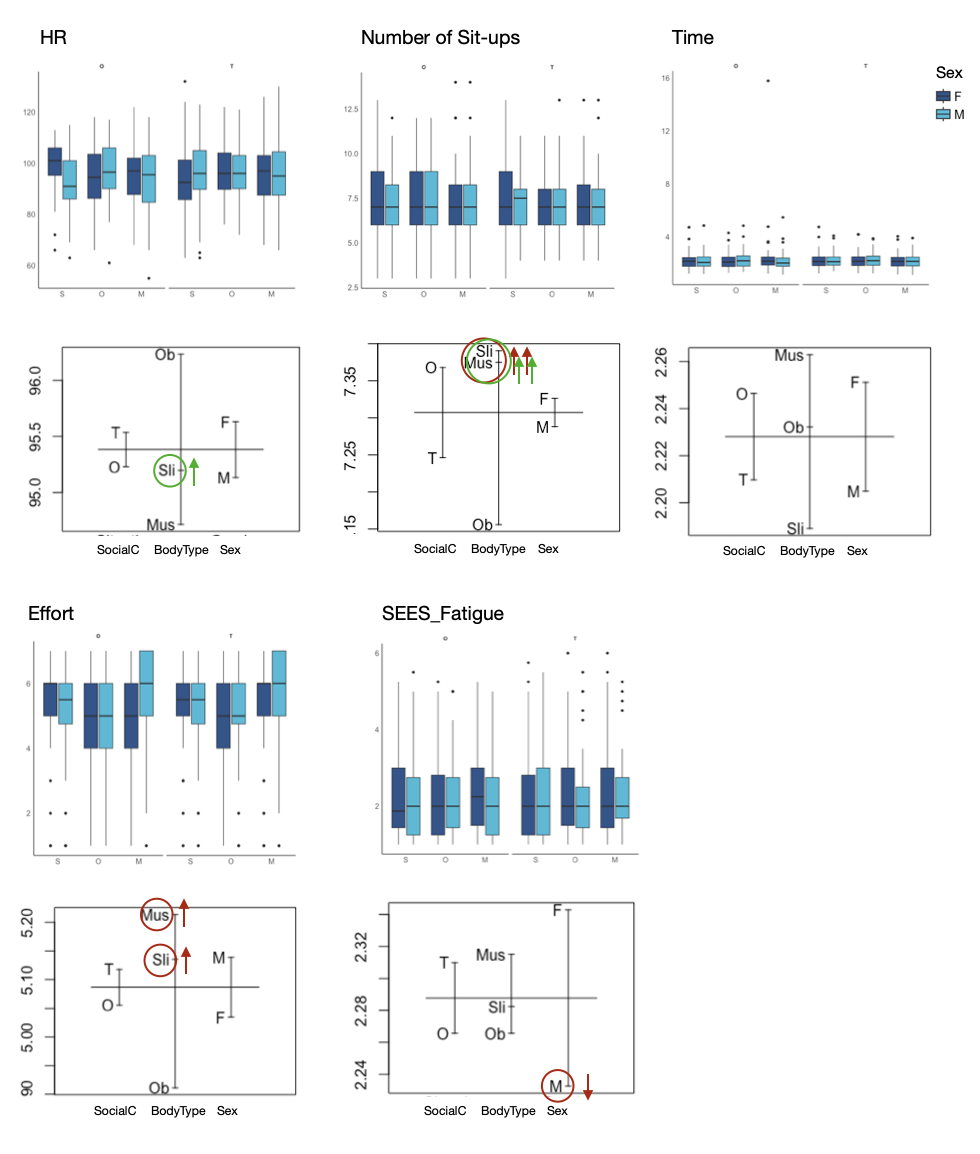}
  \caption{Mean scored and box plot for relatedness results.}
  \Description{}
\end{figure}

\begin{figure}[!h]
  \centering
  \includegraphics[width=0.8\textwidth]{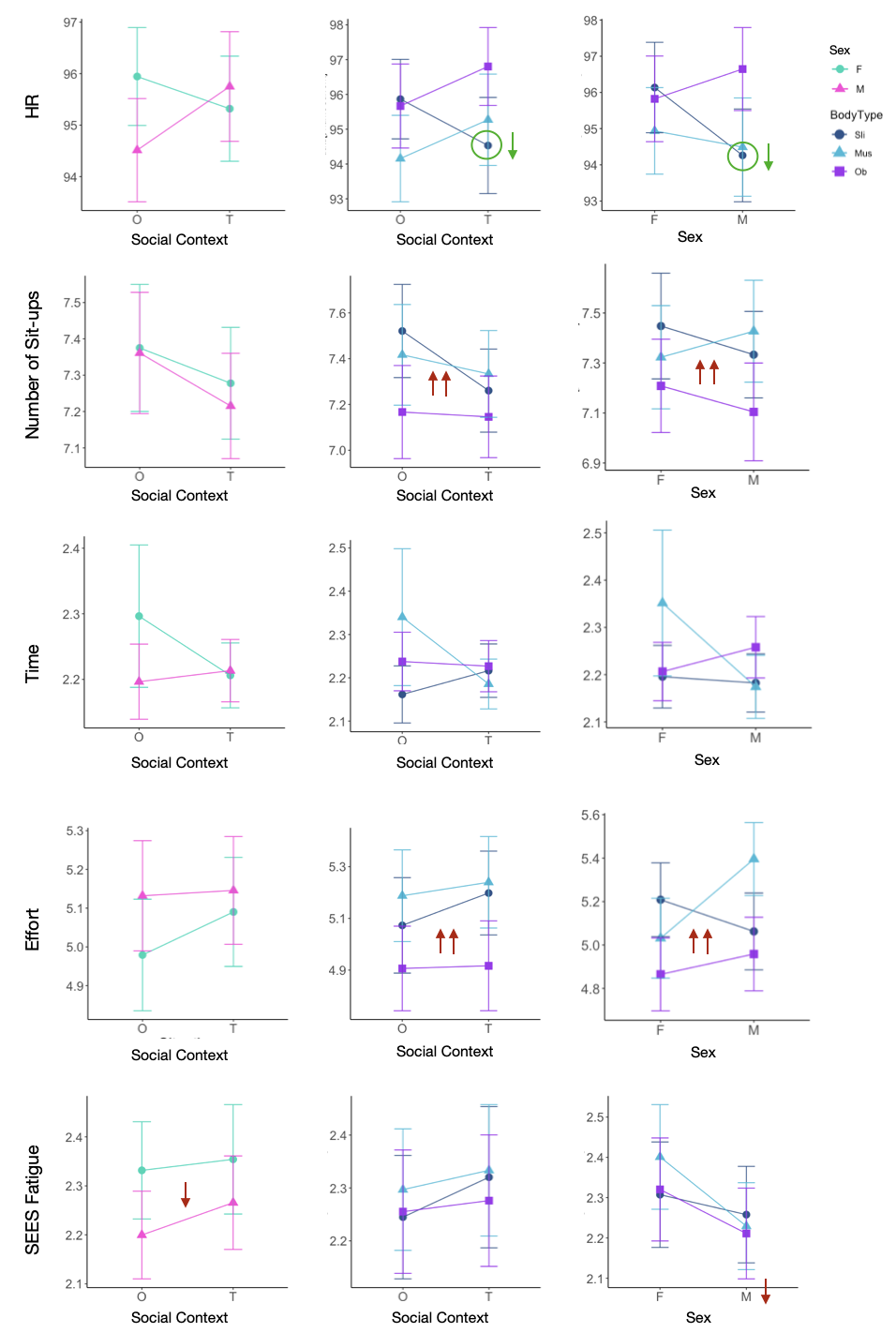}
  \caption{Mean scored and box plot for relatedness results.}
  \Description{}
\end{figure}

\begin{figure}[!h]
  \centering
  \includegraphics[width=0.8\textwidth]{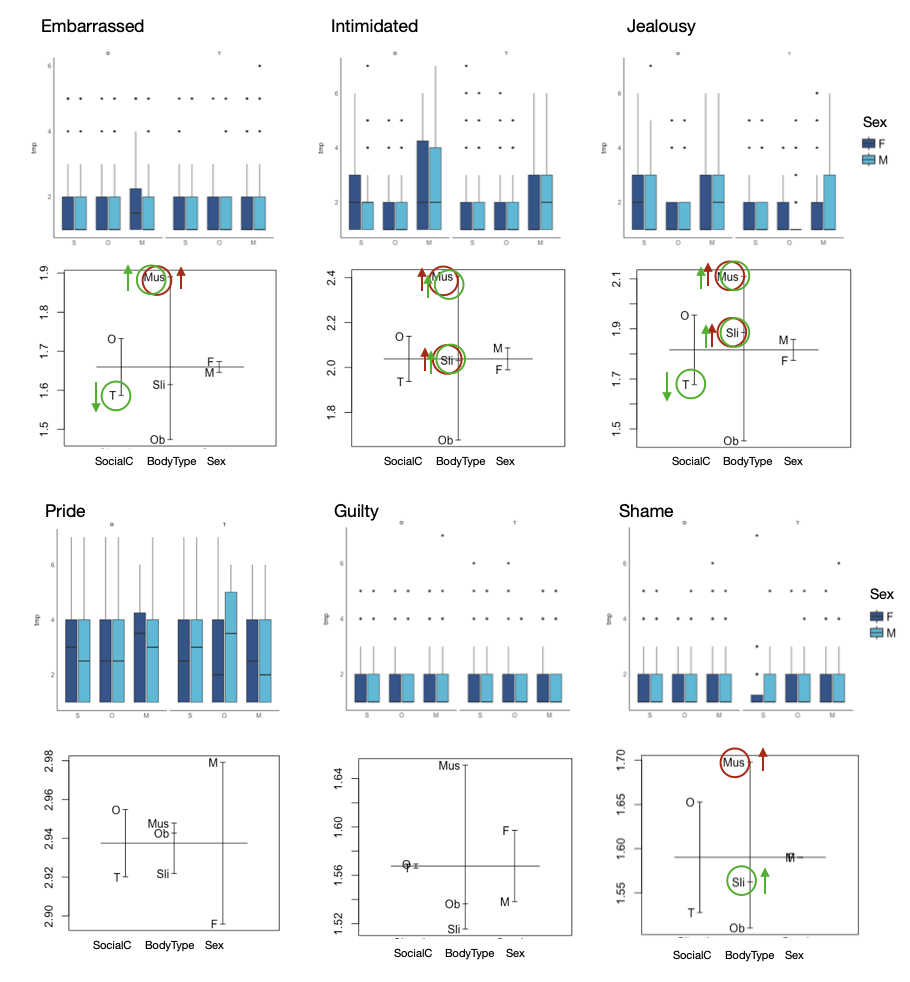}
  \caption{Mean scored and box plot for relatedness results.}
  \Description{}
\end{figure}

\begin{figure}[!h]
  \centering
  \includegraphics[width=0.8\textwidth]{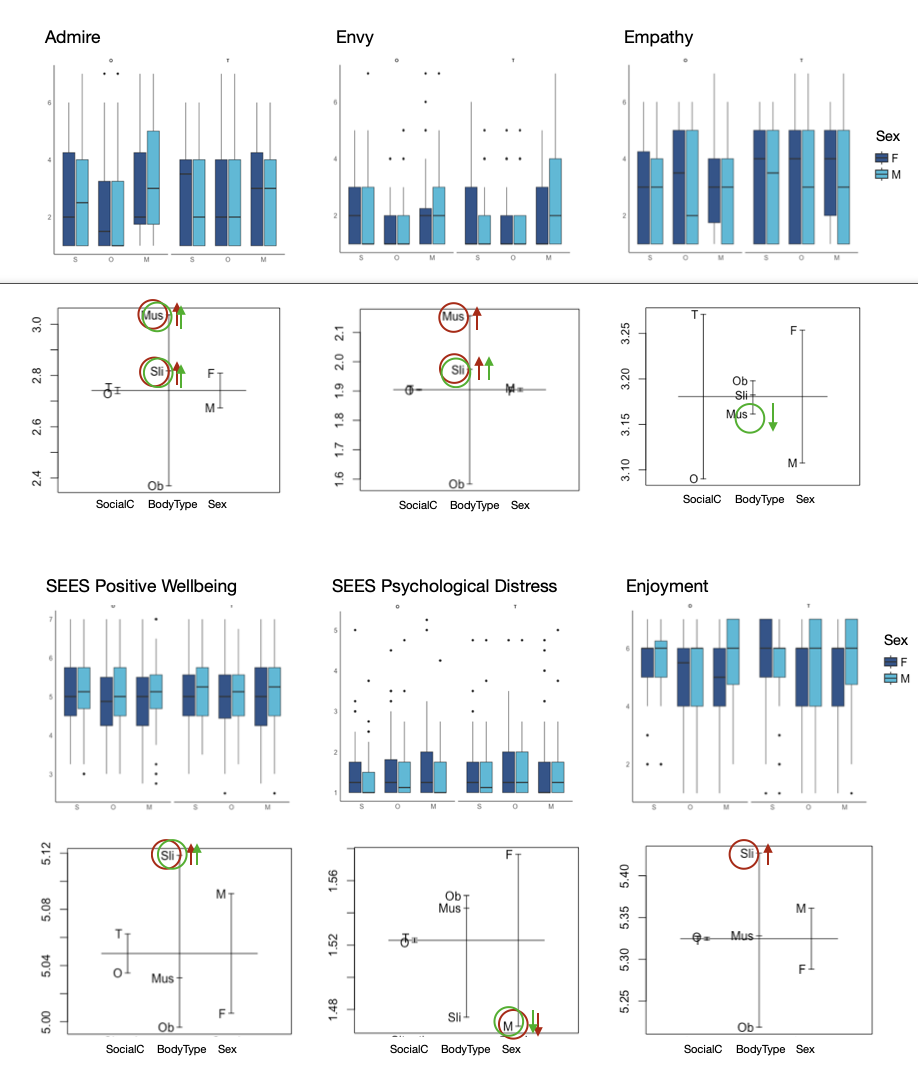}
  \caption{Mean scored and box plot for relatedness results.}
  \Description{}
\end{figure}

\begin{figure}[!h]
  \centering
  \includegraphics[width=0.8\textwidth]{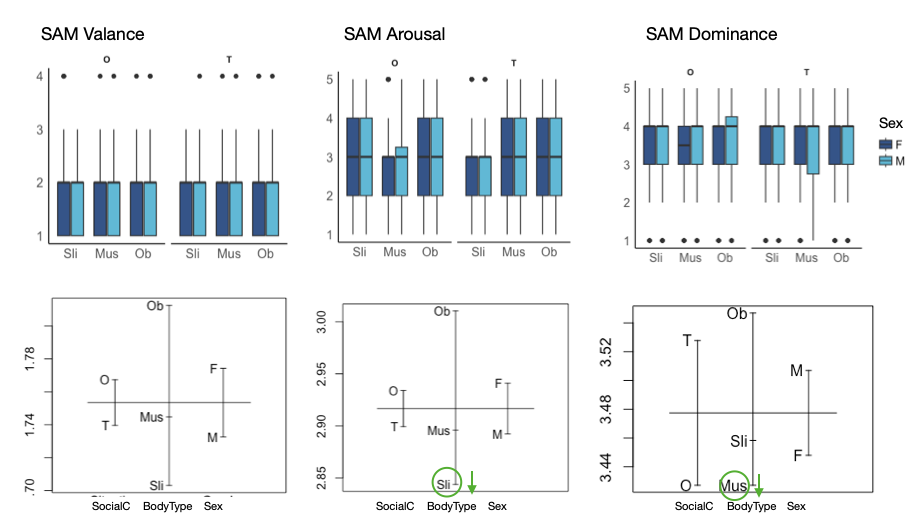}
  \caption{Mean scored and box plot for relatedness results.}
  \Description{}
\end{figure}

\begin{figure}[!h]
  \centering
  \includegraphics[width=0.8\textwidth]{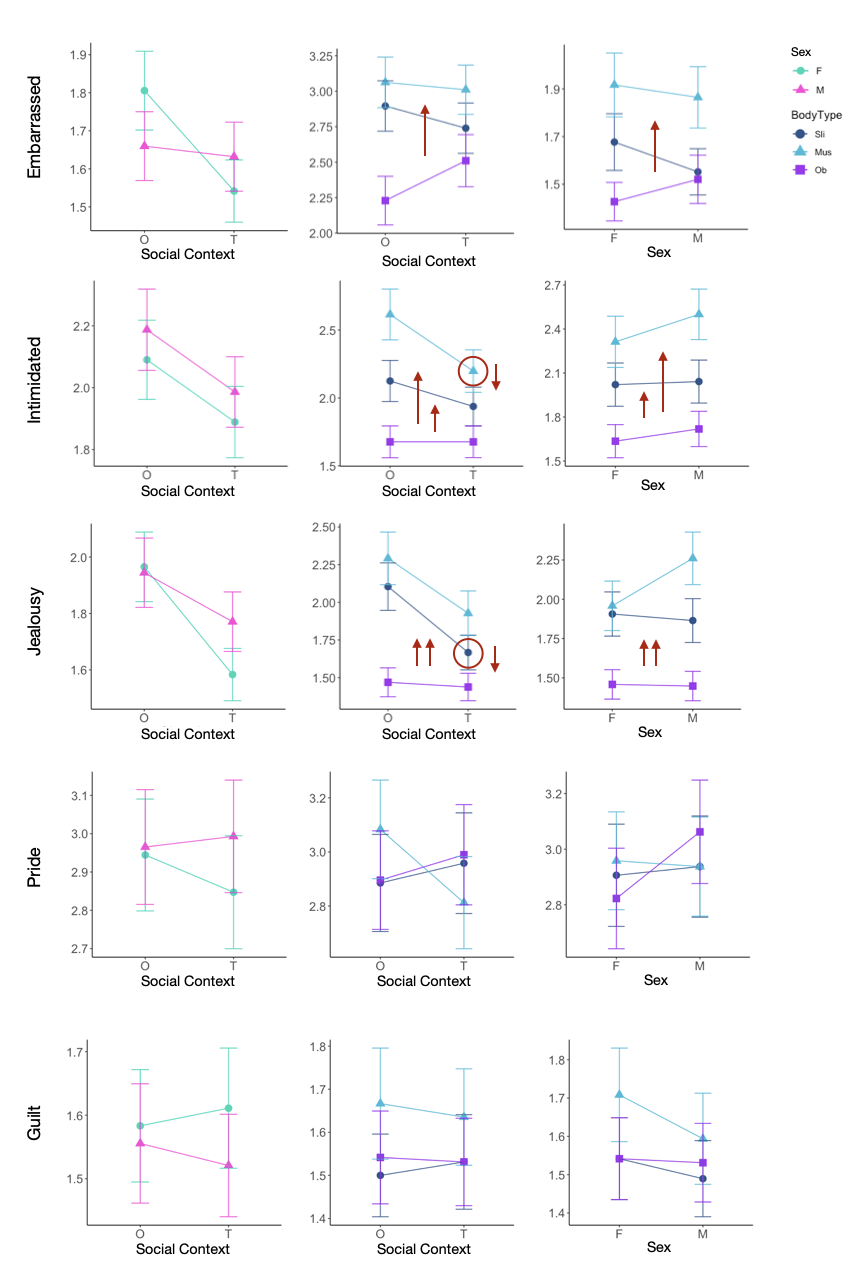}
  \caption{Mean scored and box plot for relatedness results.}
  \Description{}
\end{figure}

\begin{figure}[!h]
  \centering
  \includegraphics[width=0.8\textwidth]{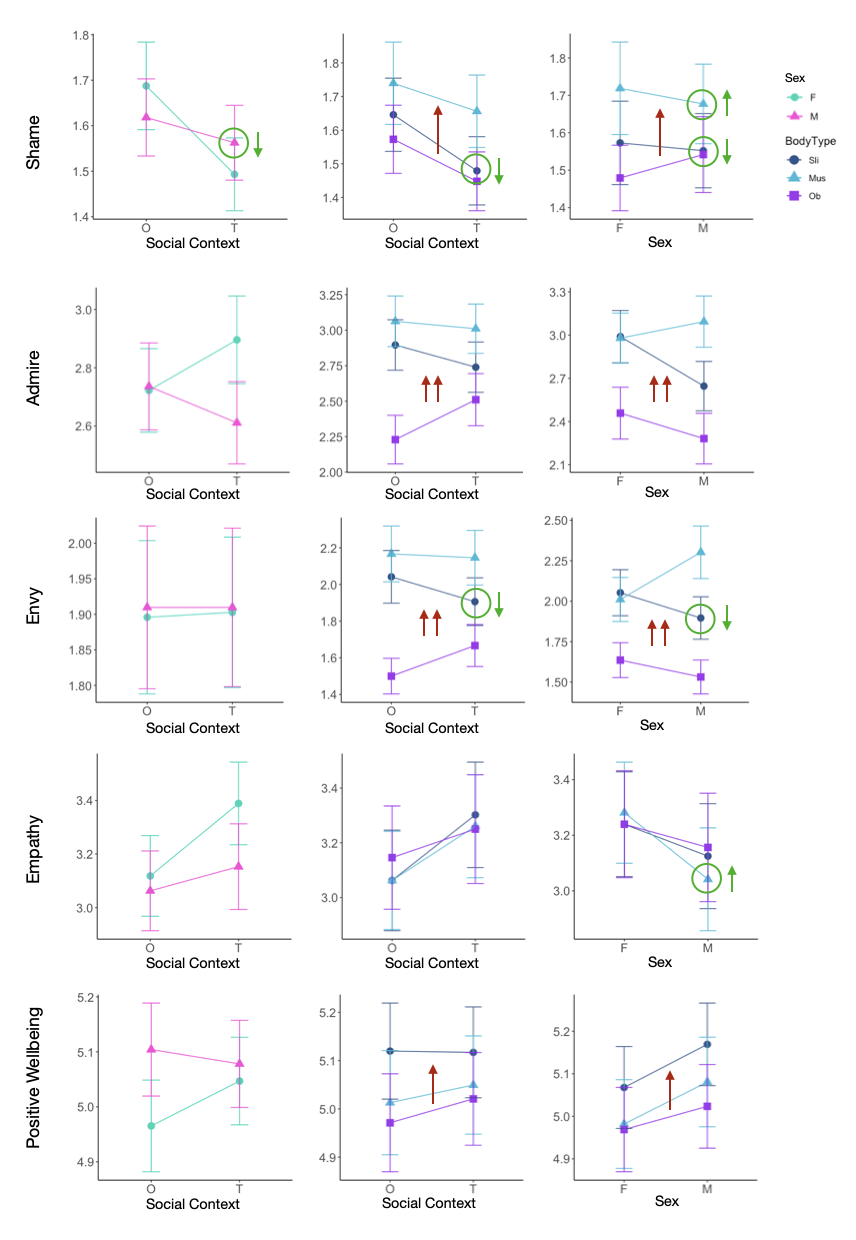}
  \caption{Mean scored and box plot for relatedness results.}
  \Description{}
\end{figure}

\begin{figure}[!h]
  \centering
  \includegraphics[width=0.8\textwidth]{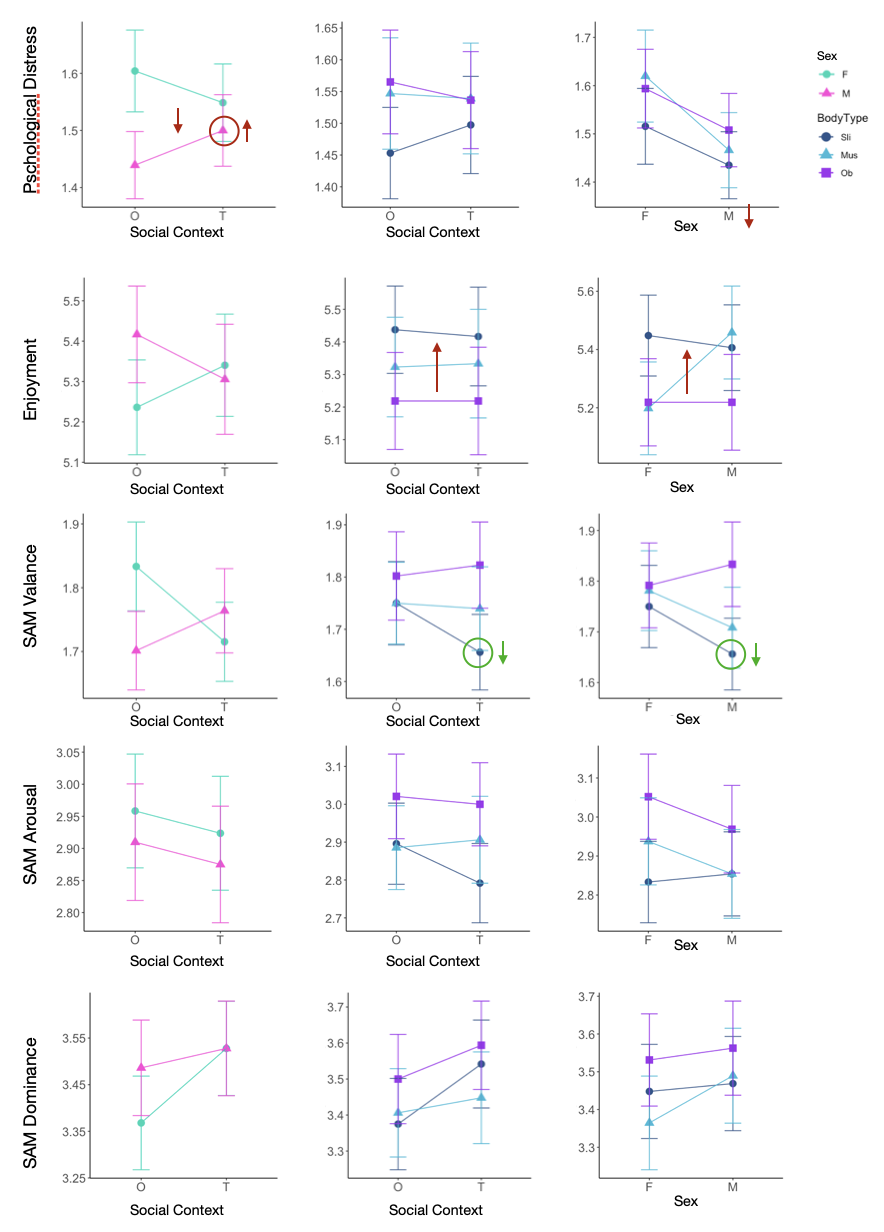}
  \caption{Mean scored and box plot for relatedness results.}
  \Description{}
\end{figure}

\subsection{Model Comparison of Participant-Independent Variables}
\label{ModelComparisionPI}

We compared the following models for participant-independent variables: 

$null.mod <- lmer(IOS ~ 1 + (1|Condition), data = df_final)
lmm_model11 <- lmer(IOS ~ SocialContext + (1 + SocialContext | UserID), data = df_final,REML = FALSE)
lmm_model12 <- lmer(IOS ~ BodyType + (1 + BodyType | UserID), data = df_final,REML = FALSE)
lmm_model13 <- lmer(IOS ~ Sex + (1 + Sex | UserID), data = df_final,REML = FALSE)$

$lmm_model21 <- lmer(IOS ~ SocialContext + BodyType + (1 + SocialContext | UserID) + (1 + BodyType | UserID), data = df_final,REML = FALSE)
lmm_model22 <- lmer(IOS ~ BodyType + Sex + (1 + BodyType | UserID) + (1 + Sex | UserID), data = df_final,REML = FALSE)
lmm_model23 <- lmer(IOS ~ SocialContext + Sex + (1 + SocialContext | UserID) + (1 + Sex | UserID), data = df_final,REML = FALSE)$

$lmm_model31 <- lmer(IOS ~ SocialContext + BodyType + Sex + (1 + SocialContext | UserID) + (1 + BodyType | UserID) + (1 + Sex | UserID), data = df_final,REML = FALSE)$

$lmm_model41 <- lmer(IOS ~ SocialContext * BodyType + (1 + SocialContext | UserID) + (1 + BodyType | UserID), data = df_final,REML = FALSE)
lmm_model42 <- lmer(IOS ~ BodyType * Sex + (1 + BodyType | UserID) + (1 + Sex | UserID), data = df_final,REML = FALSE)
lmm_model43 <- lmer(IOS ~ SocialContext * Sex + (1 + SocialContext | UserID) + (1 + Sex | UserID), data = df_final,REML = FALSE)$

$lmm_model5 <- lmer(IOS ~ SocialContext * BodyType * Sex + (1 + SocialContext | UserID) + (1 + BodyType | UserID) + (1 + Sex | UserID), data = df_final,REML = FALSE) $

\begin{table}[]
\caption{The table includes the final best fitting models (AIC smallest) that were used for each specific variable.}
\label{tab:models3L}
\begin{tabular}{l|l|p{11cm}|l}
Variable & Nr. & Model: Participant-Independent Variables & AIC \\
\hline
IOS & 21 & IOS $\sim$SocialContext + BodyType + (1 + SocialContext | UserID) + (1 + BodyType | UserID) & 1642.848 \\
IMI & 31 & IMI\_Relatedness $\sim$SocialContext + BodyType + Sex + (1 + SocialContext | UserID) + (1 + BodyType | UserID) + (1 + Sex | UserID) & 1331.335 \\
responsible & 21 & responsible\_for\_other\_person $\sim$SocialContext + BodyType + (1 + SocialContext | UserID) + (1 + BodyType | UserID) & 1693.202 \\
attracted & 42 & attracted\_to\_other\_person $\sim$BodyType * Sex + (1 + BodyType | UserID) + (1 + Sex | UserID) & 1812.669 \\
focused & 31 & focused\_on\_the\_other\_person $\sim$SocialContext + BodyType + Sex + (1 + SocialContext | UserID) + (1 + BodyType | UserID) + (1 + Sex | UserID) & 1882.031 \\
NoS & 22 & NumberOfSitUps $\sim$BodyType + Sex + (1 + BodyType | UserID) + (1 + Sex | UserID) & 1626.169 \\
Time & 12 & Time $\sim$BodyType + (1 + BodyType | UserID) & 947.4505 \\
HR & 21 & HR $\sim$SocialContext + BodyType + (1 + SocialContext | UserID) + (1 + BodyType | UserID) & 4396.757 \\
Effort & 21 & tried\_hard $\sim$SocialContext + BodyType + (1 + SocialContext | UserID) + (1 + BodyType | UserID) & 1710.788 \\
Fatigue & 23 & SEES\_Fatigue $\sim$SocialContext + Sex + (1 + SocialContext | UserID) + (1 + Sex | UserID) & 1005.378 \\
Intimidated & 41 & Intimidated $\sim$SocialContext * BodyType + (1 + SocialContext | UserID) + (1 + BodyType | UserID), & 1708.143 \\
Jealousy & 41 & Jealousy $\sim$SocialContext * BodyType + (1 + SocialContext | UserID) + (1 + BodyType | UserID) & 1666.6 \\
Empathy & 21 & Empathy $\sim$SocialContext + BodyType + (1 + SocialContext | UserID) + (1 + BodyType | UserID) & 1938.209 \\
Envy & 21 & Envy $\sim$SocialContext + BodyType + (1 + SocialContext | UserID) + (1 + BodyType | UserID) & 1718.698 \\
Embarressed & 21 & Embarressed $\sim$SocialContext + BodyType + (1 + SocialContext | UserID) + (1 + BodyType | UserID) & 1461.135 \\
Guilty & 21 & Guilty $\sim$SocialContext + BodyType + (1 + SocialContext | UserID) + (1 + BodyType | UserID) & 1444.684 \\
Pride & 11 & Pride $\sim$SocialContext + (1 + SocialContext | UserID) & 1754.571 \\
Shame & 21 & Shame $\sim$SocialContext + BodyType + (1 + SocialContext | UserID) + (1 + BodyType | UserID) & 1358.386 \\
Admire & 12 & Admire $\sim$BodyType + (1 + BodyType | UserID) & 1941.673 \\
PosWellBeing & 31 & SEES\_PositiveWellBeing $\sim$SocialContext + BodyType + Sex + (1 + SocialContext | UserID) + (1 + BodyType | UserID) + (1 + Sex | UserID) & 1020.873 \\
PsyDistress & 43 & SEES\_PsychologicalDistress $\sim$SocialContext * Sex + (1 + SocialContext | UserID & 531.6807 \\
Enjoyment & 31 & enjoyed\_activity $\sim$SocialContext + BodyType + Sex + (1 + SocialContext | UserID) + (1 + BodyType | UserID) + (1 + Sex | UserID) & 1621.901 \\
Arousal & 11 & (SAM\_Arousal $\sim$SocialContext + (1 + SocialContext | UserID) & 1110.802 \\
Dominance & 11 & SAM\_Dominance $\sim$SocialContext + (1 + SocialContext | UserID) & 1211.111 \\
Valance & 11 & SAM\_Valance $\sim$SocialContext + (1 + SocialContext | UserID) & 946.0022
\end{tabular}
\end{table}

\subsection{Model Comparison of Participant-Dependent Variables}
\label{ModelComparisionPD}

And the following models for participant-dependent variables: 
$
lmm_modelA <- lmer(IOS ~ SocialContext + BodyType + Sex + (1 + SocialContext | UserID) + (1 + BodyType | UserID) + (1 + Sex | UserID), data = df_final_G,REML = FALSE)
lmm_modelA1 <- lmer(IOS ~ SocialContext + BodyType + Sex + Gender + (1 + SocialContext | UserID) + (1 + BodyType | UserID) + (1 + Sex | UserID), data = df_final_G,REML = FALSE)
lmm_modelA2 <- lmer(IOS ~ SocialContext + BodyType + Sex + Gender + DivAv_AttraktedAv + (1 + SocialContext | UserID) + (1 + BodyType | UserID) + (1 + Sex | UserID), data = df_final_G,REML = FALSE)
lmm_modelA3 <- lmer(IOS ~ SocialContext + BodyType + Sex + Gender + DivAv_IndentifyAv + (1 + SocialContext | UserID) + (1 + BodyType | UserID) + (1 + Sex | UserID), data = df_final_G,REML = FALSE)
lmm_modelA4 <- lmer(IOS ~ SocialContext + BodyType + Sex + Cooperation + Competition + (1 + SocialContext | UserID) + (1 + BodyType | UserID) + (1 + Sex | UserID), data = df_final_G,REML = FALSE)
lmm_modelA5 <- lmer(IOS ~ SocialContext + BodyType + Sex + SMS + (1 + SocialContext | UserID) + (1 + BodyType | UserID) + (1 + Sex | UserID), data = df_final_G,REML = FALSE)
lmm_modelA6 <- lmer(IOS ~ SocialContext + BodyType + Sex + Gender + DivAv_AttraktedAv + DivAv_IndentifyAv + Cooperation + Competition + SMS + (1 + SocialContext | UserID) + (1 + BodyType | UserID) + (1 + Sex | UserID), data = df_final_G,REML = FALSE)$

$lmm_modelB <- lmer(IOS ~ SocialContext * BodyType * Sex + (1 + SocialContext | UserID) + (1 + BodyType | UserID) + (1 + Sex | UserID), data = df_final_G,REML = FALSE)
lmm_modelB1 <- lmer(IOS ~ SocialContext * BodyType * Sex * Gender + (1 + SocialContext | UserID) + (1 + BodyType | UserID) + (1 + Sex | UserID), data = df_final_G,REML = FALSE)
lmm_modelB2 <- lmer(IOS ~ SocialContext * BodyType * Sex * Gender * DivAv_AttraktedAv + (1 + SocialContext | UserID) + (1 + BodyType | UserID) + (1 + Sex | UserID), data = df_final_G,REML = FALSE)
lmm_modelB3 <- lmer(IOS ~ SocialContext * BodyType * Sex * Gender * DivAv_IndentifyAv + (1 + SocialContext | UserID) + (1 + BodyType | UserID) + (1 + Sex | UserID), data = df_final_G,REML = FALSE)
lmm_modelB4 <- lmer(IOS ~ SocialContext * BodyType * Sex * Cooperation + Competition + (1 + SocialContext | UserID) + (1 + BodyType | UserID) + (1 + Sex | UserID), data = df_final_G,REML = FALSE)
lmm_modelB5 <- lmer(IOS ~ SocialContext * BodyType * Sex * SMS + (1 + SocialContext | UserID) + (1 + BodyType | UserID) + (1 + Sex | UserID), data = df_final_G,REML = FALSE)$

$lmm_modelC <- lmer(IOS ~ SocialContext * BodyType * Sex * Gender * DivAv_AttraktedAv * DivAv_IndentifyAv + (1 + SocialContext | UserID) + (1 + BodyType | UserID) + (1 + Sex | UserID), data = df_final_G,REML = FALSE)
lmm_modelC1 <- lmer(IOS ~ SocialContext * BodyType * Sex * Gender * Cooperation * Competition + (1 + SocialContext | UserID) + (1 + BodyType | UserID) + (1 + Sex | UserID), data = df_final_G,REML = FALSE)
lmm_modelD <- lmer(IOS ~ SocialContext * BodyType * Sex * Gender * DivAv_AttraktedAv * DivAv_IndentifyAv * Cooperation * Competition * SMS + (1 + SocialContext | UserID) + (1 + BodyType | UserID) + (1 + Sex | UserID), data = df_final_G,REML = FALSE)
$

\begin{table}[]
\caption{}
\label{tab:resultsAll9}
\begin{tabular}{l|l|p{11cm}|l}
Variable & Nr. & Model: Participant-Dependent Variables & AIC \\
\hline
IOS & D & IOS $\sim$SocialContext * BodyType * Sex * Gender * DivAv\_AttraktedAv * DivAv\_IndentifyAv * Cooperation * Competition * SMS + (1 + SocialContext | UserID) + (1 + BodyType | UserID) + (1 + Sex | UserID) & 1,535.712 \\
IMI & D & IMI\_Relatedness $\sim$SocialContext * BodyType * Sex * Gender * DivAv\_AttraktedAv * DivAv\_IndentifyAv * Cooperation * Competition * SMS + (1 + SocialContext | UserID) + (1 + BodyType | UserID) + (1 + Sex | UserID) & 1,263.467 \\
responsible & A & responsible\_for\_other\_person $\sim$SocialContext + BodyType + Sex + (1 + SocialContext | UserID) + (1 + BodyType | UserID) + (1 + Sex | UserID) & 1,669.343 \\
attracted & A2 & attracted\_to\_other\_person $\sim$SocialContext + BodyType + Sex + Gender + DivAv\_AttraktedAv + (1 + SocialContext | UserID) + (1 + BodyType | UserID) + (1 + Sex | UserID) & 1,722.855 \\
focused & A & focused\_on\_the\_other\_person $\sim$SocialContext + BodyType + Sex + (1 + SocialContext | UserID) + (1 + BodyType | UserID) + (1 + Sex | UserID) & 1,843.045 \\
NoS & A5 & NumberOfSitUps $\sim$SocialContext + BodyType + Sex + SMS + (1 + SocialContext | UserID) + (1 + BodyType | UserID) + (1 + Sex | UserID) & 1,476.985 \\
Time & D & Time $\sim$SocialContext * BodyType * Sex * Gender * DivAv\_AttraktedAv * DivAv\_IndentifyAv * Cooperation * Competition * SMS + (1 + SocialContext | UserID) + (1 + BodyType | UserID) + (1 + Sex | UserID) & 304.168 \\
HR & B & HR $\sim$SocialContext * BodyType * Sex + (1 + SocialContext | UserID) + (1 + BodyType | UserID) + (1 + Sex | UserID) & 4,319.033 \\
Effort & B & tried\_hard $\sim$SocialContext * BodyType * Sex + (1 + SocialContext | UserID) + (1 + BodyType | UserID) + (1 + Sex | UserID) & 1586.518 \\
Fatigue & A & SEES\_Fatigue $\sim$SocialContext + BodyType + Sex + (1 + SocialContext | UserID) + (1 + BodyType | UserID) + (1 + Sex | UserID) & 955.7268 \\
Intimidated & A & Intimidated $\sim$SocialContext + BodyType + Sex + (1 + SocialContext | UserID) + (1 + BodyType | UserID) + (1 + Sex | UserID) & 1689.120 \\
Jealousy & A & Jealousy $\sim$SocialContext + BodyType + Sex + (1 + SocialContext | UserID) + (1 + BodyType | UserID) + (1 + Sex | UserID) & 1648.132 \\
Empathy & D & Empathy $\sim$SocialContext * BodyType * Sex * Gender * DivAv\_AttraktedAv * DivAv\_IndentifyAv * Cooperation * Competition * SMS + (1 + SocialContext | UserID) + (1 + BodyType | UserID) + (1 + Sex | UserID) & 1874.277 \\
Envy & D & Envy $\sim$SocialContext * BodyType * Sex * Gender * DivAv\_AttraktedAv * DivAv\_IndentifyAv * Cooperation * Competition * SMS + (1 + SocialContext | UserID) + (1 + BodyType | UserID) + (1 + Sex | UserID) & 1647.912 \\
Embarressed & A & Embarressed $\sim$SocialContext + BodyType + Sex + (1 + SocialContext | UserID) + (1 + BodyType | UserID) + (1 + Sex | UserID) & 1444.610 \\
Guilty & A5 & Guilty $\sim$SocialContext + BodyType + Sex + SMS + (1 + SocialContext | UserID) + (1 + BodyType | UserID) + (1 + Sex | UserID) & 1423.765 \\
Pride & A & Pride $\sim$SocialContext + BodyType + Sex + (1 + SocialContext | UserID) + (1 + BodyType | UserID) + (1 + Sex | UserID) & 1712.418 \\
Shame & D & Shame $\sim$SocialContext * BodyType * Sex * Gender * DivAv\_AttraktedAv * DivAv\_IndentifyAv * Cooperation * Competition * SMS + (1 + SocialContext | UserID) + (1 + BodyType | UserID) + (1 + Sex | UserID) & 1310.719 \\
Admire & B & Admire $\sim$SocialContext * BodyType * Sex + (1 + SocialContext | UserID) + (1 + BodyType | UserID) + (1 + Sex | UserID) & 1901.389 \\
PosWellBeing & A & SEES\_PositiveWellBeing $\sim$SocialContext + BodyType + Sex + (1 + SocialContext | UserID) + (1 + BodyType | UserID) + (1 + Sex | UserID) & 960.1621 \\
PsyDistress & A2 & SEES\_PsychologicalDistress $\sim$SocialContext + BodyType + Sex + Gender + DivAv\_AttraktedAv + (1 + SocialContext | UserID) + (1 + BodyType | UserID) + (1 + Sex | UserID) & 517.7136 \\
Enjoyment & A & enjoyed\_activity $\sim$SocialContext + BodyType + Sex + (1 + SocialContext | UserID) + (1 + BodyType | UserID) + (1 + Sex | UserID) & 1581.556 \\
Arousal & A1 & SAM\_Arousal $\sim$SocialContext + BodyType + Sex + Gender + (1 + SocialContext | UserID) + (1 + BodyType | UserID) + (1 + Sex | UserID) & 1081.399 \\
Dominance & A & SAM\_Dominance $\sim$SocialContext + BodyType + Sex + (1 + SocialContext | UserID) + (1 + BodyType | UserID) + (1 + Sex | UserID) & 1211.195 \\
Valance & B & SAM\_Valance $\sim$SocialContext * BodyType * Sex + (1 + SocialContext | UserID) + (1 + BodyType | UserID) + (1 + Sex | UserID) & 932.6200
\end{tabular}
\end{table}

\subsection{Overview Main and Interaction Effects of Participant-Dependent Data}
\label{sec:DiscEffectPartDependent}

\begin{table}[]
\caption{The table shows an overview of all significant main effects (above the line) and interaction between two effects (below the line) for the linear mixed model with social context (SC), body type (BT) and sex of the other avatar and participants gender, attractiveness towards avatar (AttrAv), identification with avatar (IdenAv), competitiveness (Comp), cooperation (Coop) and sport motivation score (SMS). Bold effects, match with results from models without including participant dependent variables.}
\label{tab:DiscEffectPartDependent}
\resizebox{\textwidth}{!}{%
\begin{tabular}{l|l|l|l|l|l|l|l|l|p{1.8cm}|p{1.8cm}|p{3.3cm}}
SC & BT  & Sex & Gender & Attr. & Iden. & Comp & Coop & SMS & Relatedness                                                            & Performance                                                              & Emotion                                                                                    \\
\hline
T &     &     &        &        &        &      &      &     & \textcolor{darkgreen}{ \textbf{responsible ↑}}                          &                                                                          & \textcolor{darkred}{\textbf{Jealousy ↑;Embarrassed ↑; Envy ↑;}} \textcolor{darkgreen}{ \textbf{Empathy ↑}}                                       \\
         & Mus &     &        &        &        &      &      &     & {\color[HTML]{000000} \textbf{responsible ↓ ; attracted ↑; focused ↑}} & {\color[HTML]{000000} \textbf{NumSitups ↑, Fatigue↑}}                    & \textbf{Pride↑;Envy↓;PosWellbeing ↑; PsyDistress↓; Arousal↓}                               \\
         & Sli &     &        &        &        &      &      &     & \textcolor{darkgreen}{ \textbf{attracted ↑; focused ↑}}                 & {\color[HTML]{000000} \textbf{NumSitups ↑, HR↑,Fatigue ↑}}               & \textbf{Intimidated↓; Jealousy↓; Embarrassed↓;Shame↓PosWellbeing ↑; Arousal ↑;Dominance ↑} \\
         &     & M   &        &        &        &      &      &     &                                                                        & \textcolor{darkred}{ \textbf{}}                                         & \textbf{Empathy↓; Envy↓ ; Shame↓; PsyDistress ↑}                                           \\
         &     &     & M      &        &        &      &      &     &                                                                        & \textcolor{darkgreen}{ \textbf{}}                                         & \textcolor{darkred}{ \textbf{Shame ↑; PsyDistress↓;Arousal ↑}}                            \\
         &     &     &        & T      &        &      &      &     &                                                                        & \textcolor{darkgreen}{ \textbf{Time ↓}}                                   & \textcolor{darkred}{ \textbf{Shame ↑}}                                                    \\
         &     &     &        &        & T      &      &      &     &                                                                        &                                                                          & \textcolor{darkgreen}{ \textbf{Envy↓}}                                                      \\
         &     &     &        &        &        & T    &      &     &                                                                        &                                                                          & \textcolor{darkred}{ \textbf{Shame ↑}}                                                    \\
         &     &     &        &        &        &      & T    &     &                                                                        &                                                                          & \textcolor{darkred}{ \textbf{Shame ↑}}                                                    \\
         &     &     &        &        &        &      &      & T   &                                                                        &                                                                          & \textcolor{darkred}{ \textbf{Shame ↑}}                                                    \\
\hline
T & Mus &     &        &        &        &      &      &     & \textcolor{darkgreen}{ \textbf{IOS ↑}}                                  &                                                                          & \textbf{}                                                                                  \\
T & Sli &     &        &        &        &      &      &     & \textbf{}                                                              & \textcolor{darkgreen}{ \textbf{HR↓}}                                      & \textbf{Empathy ↑; Envy ↑}                                                                 \\
T &     & M   &        &        &        &      &      &     & \textbf{}                                                              & \textbf{}                                                                & \textcolor{darkgreen}{ \textbf{Empathy ↑; Valence ↑}}                                       \\
T &     &     & M      &        &        &      &      &     & \textcolor{darkgreen}{ \textbf{IMI ↑}}                                  & \textbf{}                                                                & \textbf{}                                                                                  \\
T &     &     &        & T      &        &      &      &     & \textcolor{darkgreen}{ \textbf{IMI ↑}}                                  & \textcolor{darkred}{ \textbf{Time ↑}}                                   & \textcolor{darkred}{ \textbf{Empathy↓}}                                                   \\
T &     &     &        &        & T      &      &      &     &                                                                        & \textbf{}                                                                & \textcolor{darkred}{ \textbf{Empathy↓}}                                                   \\
T &     &     &        &        &        & T    &      &     &                                                                        &                                                                          & \textcolor{darkred}{ \textbf{Empathy↓}}                                                   \\
T &     &     &        &        &        &      & T    &     &                                                                        &                                                                          & \textbf{Empathy↓; Envy↓}                                                                   \\
T &     &     &        &        &        &      &      & T   &                                                                        &                                                                          & \textcolor{darkred}{ \textbf{Empathy↓}}                                                   \\
         & Mus & M   &        &        &        &      &      &     & \textcolor{darkred}{ \textbf{IMI↓}}                                   &                                                                          & \textcolor{darkred}{ \textbf{Envy ↑;Shame ↑}}                                             \\
         & Mus &     & M      &        &        &      &      &     & \textcolor{darkgreen}{ \textbf{IOS ↑}}                                  &                                                                          & \textcolor{darkred}{ \textbf{Shame ↑}}                                                    \\
         & Mus &     &        &        & T      &      &      &     &                                                                        &                                                                          & \textcolor{darkgreen}{ \textbf{Envy↓}}                                                      \\
         & Mus &     &        &        &        & T    &      &     &                                                                        & \textbf{}                                                                & \textcolor{darkgreen}{ \textbf{Shame↓}}                                                     \\
         & Mus &     &        &        &        &      & T    &     & \textbf{}                                                              &                                                                          & \textcolor{darkred}{ \textbf{Envy ↑}}                                                     \\
         & Mus &     &        &        &        &      &      & T   & \textbf{}                                                              &                                                                          & \textcolor{darkgreen}{ \textbf{Shame↓}}                                                     \\
         & Sli & M   &        &        &        &      &      &     &                                                                        & \textcolor{darkgreen}{ \begin{tabular}[c]{@{}l@{}}HR\\ ↓\end{tabular}}    & \textcolor{darkgreen}{ \textbf{Shame↓}}                                                     \\
         & Sli &     & M      &        &        &      &      &     & \textcolor{darkred}{ \textbf{IOS ↓}}                                  &                                                                          & \textcolor{darkred}{ \textbf{Empathy↓}}                                                   \\
         & Sli &     &        & T      &        &      &      &     &                                                                        & \textcolor{darkred}{ \begin{tabular}[c]{@{}l@{}}Time \\ ↑\end{tabular}} & \textbf{Empathy↑; Envy↑; Shame ↑}                                                          \\
         & Sli &     &        &        & T      &      &      &     &                                                                        &                                                                          & \textcolor{darkgreen}{ \textbf{Envy↓}}                                                      \\
         & Sli &     &        &        &        & T    &      &     & \textbf{}                                                              &                                                                          & \textcolor{darkred}{ \textbf{Shame ↑}}                                                    \\
         & Sli &     &        &        &        &      & T    &     & \textbf{}                                                              &                                                                          & \textcolor{darkred}{ \textbf{Shame↑}}                                                     \\
         & Sli &     &        &        &        &      &      & T   &                                                                        &                                                                          & \textcolor{darkred}{ \textbf{Shame↑}}                                                     \\
         &     & M   &        & T      &        &      &      &     &                                                                        & \textcolor{darkred}{ \begin{tabular}[c]{@{}l@{}}Time \\ ↑\end{tabular}} & \textbf{Empathy↑; Envy ↑; Shame↑}                                                          \\
         &     & M   &        &        & T      &      &      &     & \textcolor{darkgreen}{ \textbf{IOS↑; IMI ↑}}                            & \textcolor{darkgreen}{ Time ↓}                                            & \textcolor{darkred}{ \textbf{Envy ↑;Shame ↑}}                                             \\
         &     & M   &        &        &        & T    &      &     &                                                                        &                                                                          & \textbf{Empathy ↑; Envy↑;Shame↑}                                                           \\
         &     & M   &        &        &        &      & T    &     &                                                                        &                                                                          & \textbf{Empathy ↑; Envy ↑;Shame ↑}                                                         \\
         &     & M   &        &        &        &      &      & T   & \textbf{}                                                              &                                                                          & \textbf{Empathy ↑ ;Envy↑ ;Shame ↑}                                                         \\
         \hline
T & Mus & M   &        &        &        &      &      &     &                                                                        &                                                                          & {\color[HTML]{000000} \textbf{Valance ↑}}                                                  \\
T & Sli & M   &        &        &        &      &      &     &                                                                        & \textcolor{darkgreen}{ \textbf{HR↓}}                                      & \textbf{Valance↓}                                                                         
\end{tabular}}
\end{table}

%% file: Paper_SocialVR.bib
@article{allenSocialMotivationYouth2003,
  title = {Social {{Motivation}} in {{Youth Sport}}},
  author = {Allen, Justine B.},
  year = {2003},
  month = dec,
  journal = {Journal of Sport and Exercise Psychology},
  volume = {25},
  number = {4},
  pages = {551--567},
  publisher = {{Human Kinetics, Inc.}},
  issn = {1543-2904, 0895-2779},
  doi = {10.1123/jsep.25.4.551},
  urldate = {2022-02-25},
  chapter = {Journal of Sport and Exercise Psychology},
  langid = {american}
}

@article{banakouVirtuallyBeingEinstein2018,
  title = {Virtually {{Being Einstein Results}} in an {{Improvement}} in {{Cognitive Task Performance}} and a {{Decrease}} in {{Age Bias}}},
  author = {Banakou, Domna and Kishore, Sameer and Slater, Mel},
  year = {2018},
  journal = {Frontiers in Psychology},
  volume = {9},
  issn = {1664-1078},
  urldate = {2023-06-22}
}

@article{bergstromFirstPersonPerspectiveVirtual2016a,
  title = {First-{{Person Perspective Virtual Body Posture Influences Stress}}: {{A Virtual Reality Body Ownership Study}}},
  shorttitle = {First-{{Person Perspective Virtual Body Posture Influences Stress}}},
  author = {Bergstr{\"o}m, Ilias and Kilteni, Konstantina and Slater, Mel},
  year = {2016},
  month = feb,
  journal = {PLOS ONE},
  volume = {11},
  number = {2},
  pages = {e0148060},
  publisher = {{Public Library of Science}},
  issn = {1932-6203},
  doi = {10.1371/journal.pone.0148060},
  urldate = {2023-07-26},
  langid = {english}
}

@article{bornMotivatingPlayersPerform2021,
  title = {Motivating {{Players}} to {{Perform}} an {{Optional Strenuous Activity}} in a {{Virtual Reality Exergame Using Virtual Performance Augmentation}}},
  author = {Born, Felix and Rygula, Adrian and Masuch, Maic},
  year = {2021},
  month = oct,
  journal = {Proceedings of the ACM on Human-Computer Interaction},
  volume = {5},
  number = {CHI PLAY},
  pages = {225:1--225:21},
  doi = {10.1145/3474652},
  urldate = {2022-05-03}
}

@article{burnettDevelopmentAdolescenceNeural2009,
  title = {Development during {{Adolescence}} of the {{Neural Processing}} of {{Social Emotion}}},
  author = {Burnett, Stephanie and Bird, Geoffrey and Moll, Jorge and Frith, Chris and Blakemore, Sarah-Jayne},
  year = {2009},
  month = sep,
  journal = {Journal of Cognitive Neuroscience},
  volume = {21},
  number = {9},
  pages = {1736--1750},
  issn = {0898-929X},
  doi = {10.1162/jocn.2009.21121},
  urldate = {2023-08-29}
}

@inproceedings{clarkeFakeForwardUsingDeepfake2023,
  title = {{{FakeForward}}: {{Using Deepfake Technology}} for {{Feedforward Learning}}},
  shorttitle = {{{FakeForward}}},
  booktitle = {Proceedings of the 2023 {{CHI Conference}} on {{Human Factors}} in {{Computing Systems}}},
  author = {Clarke, Christopher and Xu, Jingnan and Zhu, Ye and Dharamshi, Karan and McGill, Harry and Black, Stephen and Lutteroth, Christof},
  year = {2023},
  month = apr,
  pages = {1--17},
  publisher = {{ACM}},
  address = {{Hamburg Germany}},
  doi = {10.1145/3544548.3581100},
  urldate = {2023-06-22},
  isbn = {978-1-4503-9421-5},
  langid = {english}
}

@article{dechantHowAvatarCustomization2021,
  title = {How {{Avatar Customization Affects Fear}} in a {{Game-based Digital Exposure Task}} for {{Social Anxiety}}},
  author = {Dechant, Martin Johannes and Birk, Max V. and Shiban, Youssef and Schnell, Knut and Mandryk, Regan L.},
  year = {2021},
  month = oct,
  journal = {Proceedings of the ACM on Human-Computer Interaction},
  volume = {5},
  number = {CHI PLAY},
  pages = {1--27},
  issn = {2573-0142},
  doi = {10.1145/3474675},
  urldate = {2023-06-22},
  langid = {english}
}

@inproceedings{deighanSocialVirtualReality2023,
  title = {Social {{Virtual Reality}} as a {{Mental Health Tool}}: {{How People Use VRChat}} to {{Support Social Connectedness}} and {{Wellbeing}}},
  shorttitle = {Social {{Virtual Reality}} as a {{Mental Health Tool}}},
  booktitle = {Proceedings of the 2023 {{CHI Conference}} on {{Human Factors}} in {{Computing Systems}}},
  author = {Deighan, Mairi Therese and Ayobi, Amid and O'Kane, Aisling Ann},
  year = {2023},
  month = apr,
  pages = {1--13},
  publisher = {{ACM}},
  address = {{Hamburg Germany}},
  doi = {10.1145/3544548.3581103},
  urldate = {2023-06-22},
  isbn = {978-1-4503-9421-5},
  langid = {english}
}

@article{dimenichiPowerCompetitionEffects2015,
  title = {The Power of Competition: {{Effects}} of Social Motivation on Attention, Sustained Physical Effort, and Learning},
  shorttitle = {The Power of Competition},
  author = {DiMenichi, Brynne C. and Tricomi, Elizabeth},
  year = {2015},
  journal = {Frontiers in Psychology},
  volume = {6},
  issn = {1664-1078},
  urldate = {2023-06-22}
}

@inproceedings{dollingerAreEmbodiedAvatars2023,
  title = {Are {{Embodied Avatars Harmful}} to Our {{Self-Experience}}? {{The Impact}} of {{Virtual Embodiment}} on {{Body Awareness}}},
  shorttitle = {Are {{Embodied Avatars Harmful}} to Our {{Self-Experience}}?},
  booktitle = {Proceedings of the 2023 {{CHI Conference}} on {{Human Factors}} in {{Computing Systems}}},
  author = {D{\"o}llinger, Nina and Wolf, Erik and Botsch, Mario and Latoschik, Marc Erich and Wienrich, Carolin},
  year = {2023},
  month = apr,
  pages = {1--14},
  publisher = {{ACM}},
  address = {{Hamburg Germany}},
  doi = {10.1145/3544548.3580918},
  urldate = {2023-06-22},
  isbn = {978-1-4503-9421-5},
  langid = {english}
}

@article{edwardsImpactActivePassive2018,
  title = {Impact of Active and Passive Social Facilitation on Self-Paced Endurance and Sprint Exercise: Encouragement Augments Performance and Motivation to Exercise},
  shorttitle = {Impact of Active and Passive Social Facilitation on Self-Paced Endurance and Sprint Exercise},
  author = {Edwards, Andrew Mark and {Dutton-Challis}, Lia and Cottrell, David and Guy, Joshua H. and Hettinga, Florentina Johanna},
  year = {2018},
  month = jul,
  journal = {BMJ Open Sport \& Exercise Medicine},
  volume = {4},
  number = {1},
  pages = {e000368},
  publisher = {{BMJ Specialist Journals}},
  issn = {2055-7647},
  doi = {10.1136/bmjsem-2018-000368},
  urldate = {2023-06-22},
  chapter = {Original article},
  copyright = {\textcopyright{} Author(s) (or their employer(s)) 2018. Re-use permitted under CC BY-NC. No commercial re-use. See rights and permissions. Published by BMJ.. This is an open access article distributed in accordance with the Creative Commons Attribution Non Commercial (CC BY-NC 4.0) license, which permits others to distribute, remix, adapt, build upon this work non-commercially, and license their derivative works on different terms, provided the original work is properly cited, appropriate credit is given, any changes made indicated, and the use is non-commercial. See: http://creativecommons.org/licenses/by-nc/4.0/},
  langid = {english}
}

@inproceedings{fittonDancingAvatarsMinimal2023,
  title = {Dancing with the {{Avatars}}: {{Minimal Avatar Customisation Enhances Learning}} in a {{Psychomotor Task}}},
  shorttitle = {Dancing with the {{Avatars}}},
  booktitle = {Proceedings of the 2023 {{CHI Conference}} on {{Human Factors}} in {{Computing Systems}}},
  author = {Fitton, Isabel and Clarke, Christopher and Dalton, Jeremy and Proulx, Michael J and Lutteroth, Christof},
  year = {2023},
  month = apr,
  pages = {1--16},
  publisher = {{ACM}},
  address = {{Hamburg Germany}},
  doi = {10.1145/3544548.3580944},
  urldate = {2023-06-22},
  isbn = {978-1-4503-9421-5},
  langid = {english}
}

@misc{FitXRBoxenHIITa,
  title = {{FitXR \textendash{} Boxen, HIIT und Tanz-Workouts f\"ur Oculus Quest 2}},
  journal = {Oculus},
  urldate = {2023-08-29},
  howpublished = {https://www.oculus.com/experiences/quest/2327205800645550/},
  langid = {ngerman},
  note = {(Accessed 29.08.2023)}
}

@article{frankenWhyPeopleCompetition1995,
  title = {Why Do People like Competition? {{The}} Motivation for Winning, Putting Forth Effort, Improving One's Performance, Performing Well, Being Instrumental, and Expressing Forceful/Aggressive Behavior},
  shorttitle = {Why Do People like Competition?},
  author = {Franken, Robert E. and Brown, Douglas J.},
  year = {1995},
  month = aug,
  journal = {Personality and Individual Differences},
  volume = {19},
  number = {2},
  pages = {175--184},
  issn = {0191-8869},
  doi = {10.1016/0191-8869(95)00035-5},
  urldate = {2023-06-22},
  langid = {english}
}

@article{gillGenderDifferencesCompetitive1988,
  title = {Gender Differences in Competitive Orientation and Sport Participation},
  author = {Gill, Diane L.},
  year = {1988},
  journal = {International Journal of Sport Psychology},
  volume = {19},
  number = {2},
  pages = {145--159},
  publisher = {{Edizioni Luigi Pozzi}},
  address = {{Italy}},
  issn = {0047-0767}
}

@inproceedings{hamalainenMartialArtsArtificial2005,
  title = {Martial Arts in Artificial Reality},
  booktitle = {Proceedings of the {{SIGCHI Conference}} on {{Human Factors}} in {{Computing Systems}}},
  author = {H{\"a}m{\"a}l{\"a}inen, Perttu and Ilmonen, Tommi and H{\"o}ysniemi, Johanna and Lindholm, Mikko and Nyk{\"a}nen, Ari},
  year = {2005},
  month = apr,
  series = {{{CHI}} '05},
  pages = {781--790},
  publisher = {{Association for Computing Machinery}},
  address = {{New York, NY, USA}},
  doi = {10.1145/1054972.1055081},
  urldate = {2023-06-22},
  isbn = {978-1-58113-998-3}
}

@article{hamiltonTooLittleExercise2008,
  title = {Too Little Exercise and Too Much Sitting: {{Inactivity}} Physiology and the Need for New Recommendations on Sedentary Behavior},
  shorttitle = {Too Little Exercise and Too Much Sitting},
  author = {Hamilton, Marc T. and Healy, Genevieve N. and Dunstan, David W. and Zderic, Theodore W. and Owen, Neville},
  year = {2008},
  month = jul,
  journal = {Current Cardiovascular Risk Reports},
  volume = {2},
  number = {4},
  pages = {292--298},
  issn = {1932-9520, 1932-9563},
  doi = {10.1007/s12170-008-0054-8},
  urldate = {2022-02-14},
  langid = {english}
}

@article{hareliWhatSocialSocial2008,
  title = {What's {{Social About Social Emotions}}?},
  author = {Hareli, Shlomo and Parkinson, Brian},
  year = {2008},
  journal = {Journal for the Theory of Social Behaviour},
  volume = {38},
  number = {2},
  pages = {131--156},
  issn = {1468-5914},
  doi = {10.1111/j.1468-5914.2008.00363.x},
  urldate = {2023-08-29},
  copyright = {\textcopyright{} 2008 The Authors. Journal compilation \textcopyright{} 2008 The Executive Management Committee/Blackwell Publishing Ltd},
  langid = {english}
}

@article{hoegBuddyBikingUser2023,
  title = {Buddy Biking: A User Study on Social Collaboration in a Virtual Reality Exergame for Rehabilitation},
  shorttitle = {Buddy Biking},
  author = {H{\o}eg, Emil Rosenlund and {Bruun-Pedersen}, Jon Ram and Cheary, Shannon and Andersen, Lars Koreska and Paisa, Razvan and Serafin, Stefania and Lange, Belinda},
  year = {2023},
  month = mar,
  journal = {Virtual Reality},
  volume = {27},
  number = {1},
  pages = {245--262},
  issn = {1434-9957},
  doi = {10.1007/s10055-021-00544-z},
  urldate = {2023-06-22},
  langid = {english}
}

@inproceedings{hudsonAvatarTypesMatter2016,
  title = {Avatar {{Types Matter}}: {{Review}} of {{Avatar Literature}} for {{Performance Purposes}}},
  shorttitle = {Avatar {{Types Matter}}},
  booktitle = {Virtual, {{Augmented}} and {{Mixed Reality}}},
  author = {Hudson, Irwin and Hurter, Jonathan},
  editor = {Lackey, Stephanie and Shumaker, Randall},
  year = {2016},
  series = {Lecture {{Notes}} in {{Computer Science}}},
  pages = {14--21},
  publisher = {{Springer International Publishing}},
  address = {{Cham}},
  doi = {10.1007/978-3-319-39907-2_2},
  isbn = {978-3-319-39907-2},
  langid = {english}
}

@inproceedings{kocurEffectsAvatarEnvironment2023,
  title = {The {{Effects}} of {{Avatar}} and {{Environment}} on {{Thermal Perception}} and {{Skin Temperature}} in {{Virtual Reality}}},
  booktitle = {Proceedings of the 2023 {{CHI Conference}} on {{Human Factors}} in {{Computing Systems}}},
  author = {Kocur, Martin and Jackermeier, Lukas and Schwind, Valentin and Henze, Niels},
  year = {2023},
  month = apr,
  pages = {1--15},
  publisher = {{ACM}},
  address = {{Hamburg Germany}},
  doi = {10.1145/3544548.3580668},
  urldate = {2023-06-22},
  isbn = {978-1-4503-9421-5},
  langid = {english}
}

@inproceedings{kocurFlexingMusclesVirtual2020b,
  title = {Flexing {{Muscles}} in {{Virtual Reality}}: {{Effects}} of {{Avatars}}' {{Muscular Appearance}} on {{Physical Performance}}},
  shorttitle = {Flexing {{Muscles}} in {{Virtual Reality}}},
  booktitle = {Proceedings of the {{Annual Symposium}} on {{Computer-Human Interaction}} in {{Play}}},
  author = {Kocur, Martin and Kloss, Melanie and Schwind, Valentin and Wolff, Christian and Henze, Niels},
  year = {2020},
  month = nov,
  pages = {193--205},
  publisher = {{ACM}},
  address = {{Virtual Event Canada}},
  doi = {10.1145/3410404.3414261},
  urldate = {2023-06-22},
  isbn = {978-1-4503-8074-4},
  langid = {english}
}

@inproceedings{kocurPhysiologicalPerceptualResponses2021a,
  title = {Physiological and {{Perceptual Responses}} to {{Athletic Avatars}} While {{Cycling}} in {{Virtual Reality}}},
  booktitle = {Proceedings of the 2021 {{CHI Conference}} on {{Human Factors}} in {{Computing Systems}}},
  author = {Kocur, Martin and Habler, Florian and Schwind, Valentin and Wo{\'z}niak, Pawe{\l} W. and Wolff, Christian and Henze, Niels},
  year = {2021},
  month = may,
  pages = {1--18},
  publisher = {{ACM}},
  address = {{Yokohama Japan}},
  doi = {10.1145/3411764.3445160},
  urldate = {2023-06-22},
  isbn = {978-1-4503-8096-6},
  langid = {english}
}

@article{kowalczykAgerelatedDifferencesMotives2017,
  title = {Age-Related {{Differences}} in {{Motives}} for and {{Barriers}} to {{Exercise Among Women Exercising}} in {{Fitness Centers}}},
  author = {Kowalczyk, Agnieszka and Nowicka, Mirela and {Sas-Nowosielski}, Krzysztof},
  year = {2017},
  month = sep,
  journal = {The New Educational Review},
  volume = {49},
  number = {null},
  pages = {30--39},
  issn = {1732-6729},
  urldate = {2023-06-22},
  langid = {english}
}

@incollection{lewisSelfconsciousEmotionsEmbarrassment2008,
  title = {Self-Conscious Emotions: {{Embarrassment}}, Pride, Shame, and Guilt},
  shorttitle = {Self-Conscious Emotions},
  booktitle = {Handbook of Emotions, 3rd Ed},
  author = {Lewis, Michael},
  year = {2008},
  pages = {742--756},
  publisher = {{The Guilford Press}},
  address = {{New York, NY, US}},
  isbn = {978-1-59385-650-2}
}

@article{louwExerciseMotivationBarriers2012,
  title = {Exercise Motivation and Barriers among Men and Women of Different Age Groups  Psychology},
  author = {Louw, A. J. and Van, Biljon A. and Mugandani, S. C.},
  year = {2012},
  month = dec,
  journal = {African Journal for Physical Health Education, Recreation and Dance},
  volume = {18},
  number = {41},
  pages = {759--768},
  publisher = {{AFAHPER-SD}},
  doi = {10.10520/EJC128345},
  urldate = {2023-06-23}
}

@article{mansonEscalatingPandemicsObesity2004,
  title = {The {{Escalating Pandemics}} of {{Obesity}} and {{Sedentary Lifestyle}}: {{A Call}} to {{Action}} for {{Clinicians}}},
  shorttitle = {The {{Escalating Pandemics}} of {{Obesity}} and {{Sedentary Lifestyle}}},
  author = {Manson, JoAnn E. and Skerrett, Patrick J. and Greenland, Philip and VanItallie, Theodore B.},
  year = {2004},
  month = feb,
  journal = {Archives of Internal Medicine},
  volume = {164},
  number = {3},
  pages = {249--258},
  issn = {0003-9926},
  doi = {10.1001/archinte.164.3.249},
  urldate = {2022-06-08}
}

@article{millerSocialInteractionAugmented2019,
  title = {Social Interaction in Augmented Reality},
  author = {Miller, Mark Roman and Jun, Hanseul and Herrera, Fernanda and Villa, Jacob Yu and Welch, Greg and Bailenson, Jeremy N.},
  year = {2019},
  month = may,
  journal = {PLOS ONE},
  volume = {14},
  number = {5},
  pages = {e0216290},
  publisher = {{Public Library of Science}},
  issn = {1932-6203},
  doi = {10.1371/journal.pone.0216290},
  urldate = {2023-06-22},
  langid = {english}
}

@article{mouattUseVirtualReality2020,
  title = {The {{Use}} of {{Virtual Reality}} to {{Influence Motivation}}, {{Affect}}, {{Enjoyment}}, and {{Engagement During Exercise}}: {{A Scoping Review}}},
  shorttitle = {The {{Use}} of {{Virtual Reality}} to {{Influence Motivation}}, {{Affect}}, {{Enjoyment}}, and {{Engagement During Exercise}}},
  author = {Mouatt, Brendan and Smith, Ashleigh E. and Mellow, Maddison L. and Parfitt, Gaynor and Smith, Ross T. and Stanton, Tasha R.},
  year = {2020},
  journal = {Frontiers in Virtual Reality},
  volume = {1},
  issn = {2673-4192},
  urldate = {2023-06-22}
}

@article{penaIncreasingExergamePhysical2014a,
  title = {Increasing Exergame Physical Activity through Self and Opponent Avatar Appearance},
  author = {Pe{\~n}a, Jorge and Kim, Eunice},
  year = {2014},
  month = dec,
  journal = {Computers in Human Behavior},
  volume = {41},
  pages = {262--267},
  issn = {0747-5632},
  doi = {10.1016/j.chb.2014.09.038},
  urldate = {2023-06-22},
  langid = {english}
}

@article{pengPlayingParallelEffects2013,
  title = {Playing in {{Parallel}}: {{The Effects}} of {{Multiplayer Modes}} in {{Active Video Game}} on {{Motivation}} and {{Physical Exertion}}},
  shorttitle = {Playing in {{Parallel}}},
  author = {Peng, Wei and Crouse, Julia},
  year = {2013},
  month = jun,
  journal = {Cyberpsychology, Behavior, and Social Networking},
  volume = {16},
  number = {6},
  pages = {423--427},
  publisher = {{Mary Ann Liebert, Inc., publishers}},
  issn = {2152-2715},
  doi = {10.1089/cyber.2012.0384},
  urldate = {2023-08-11}
}

@article{pengPlayingParallelEffects2013a,
  title = {Playing in {{Parallel}}: {{The Effects}} of {{Multiplayer Modes}} in {{Active Video Game}} on {{Motivation}} and {{Physical Exertion}}},
  shorttitle = {Playing in {{Parallel}}},
  author = {Peng, Wei and Crouse, Julia},
  year = {2013},
  month = jun,
  journal = {Cyberpsychology, Behavior, and Social Networking},
  volume = {16},
  number = {6},
  pages = {423--427},
  publisher = {{Mary Ann Liebert, Inc., publishers}},
  issn = {2152-2715},
  doi = {10.1089/cyber.2012.0384},
  urldate = {2023-08-29}
}

@article{piryankovaOwningOverweightUnderweight2014,
  title = {Owning an {{Overweight}} or {{Underweight Body}}: {{Distinguishing}} the {{Physical}}, {{Experienced}} and {{Virtual Body}}},
  shorttitle = {Owning an {{Overweight}} or {{Underweight Body}}},
  author = {Piryankova, Ivelina V. and Wong, Hong Yu and Linkenauger, Sally A. and Stinson, Catherine and Longo, Matthew R. and B{\"u}lthoff, Heinrich H. and Mohler, Betty J.},
  year = {2014},
  month = aug,
  journal = {PLOS ONE},
  volume = {9},
  number = {8},
  pages = {e103428},
  publisher = {{Public Library of Science}},
  issn = {1932-6203},
  doi = {10.1371/journal.pone.0103428},
  urldate = {2023-07-26},
  langid = {english}
}

@article{popaEMOTIONSROLEMOTIVATION2013,
  title = {{{THE EMOTIONS}}' {{ROLE IN THE MOTIVATION PROCESS}}},
  author = {Popa, Mirela and Salanț{\u a}, Irina},
  journal = {},
  volume = {},
  number = {},
  pages = {},
  year = {2013},
  month = jan
}

@article{portela-pinoGenderDifferencesMotivation2020,
  title = {Gender {{Differences}} in {{Motivation}} and {{Barriers}} for {{The Practice}} of {{Physical Exercise}} in {{Adolescence}}},
  author = {{Portela-Pino}, Iago and {L{\'o}pez-Castedo}, Antonio and {Mart{\'i}nez-Pati{\~n}o}, Mar{\'i}a Jos{\'e} and {Valverde-Esteve}, Teresa and {Dom{\'i}nguez-Alonso}, Jos{\'e}},
  year = {2020},
  month = jan,
  journal = {International Journal of Environmental Research and Public Health},
  volume = {17},
  number = {1},
  pages = {168},
  publisher = {{Multidisciplinary Digital Publishing Institute}},
  issn = {1660-4601},
  doi = {10.3390/ijerph17010168},
  urldate = {2023-06-22},
  copyright = {http://creativecommons.org/licenses/by/3.0/},
  langid = {english}
}

@article{rheuEnhancingHealthyBehaviors2020,
  title = {Enhancing {{Healthy Behaviors Through Virtual Self}}: {{A Systematic Review}} of {{Health Interventions Using Avatars}}},
  shorttitle = {Enhancing {{Healthy Behaviors Through Virtual Self}}},
  author = {Rheu, Minjin (MJ) and Jang, Youjin and Peng, Wei},
  year = {2020},
  month = apr,
  journal = {Games for Health Journal},
  volume = {9},
  number = {2},
  pages = {85--94},
  publisher = {{Mary Ann Liebert, Inc., publishers}},
  issn = {2161-783X},
  doi = {10.1089/g4h.2018.0134},
  urldate = {2023-06-22}
}

@article{rodriguesPersonalizationImprovesGamification2021,
  title = {Personalization {{Improves Gamification}}: {{Evidence}} from a {{Mixed-methods Study}}},
  shorttitle = {Personalization {{Improves Gamification}}},
  author = {Rodrigues, Luiz and Palomino, Paula T. and Toda, Armando M. and Klock, Ana C. T. and Oliveira, Wilk and {Avila-Santos}, Anderson P. and Gasparini, Isabela and Isotani, Seiji},
  year = {2021},
  month = oct,
  journal = {Proceedings of the ACM on Human-Computer Interaction},
  volume = {5},
  number = {CHI PLAY},
  pages = {287:1--287:25},
  doi = {10.1145/3474714},
  urldate = {2022-05-03}
}

@inproceedings{sadekSuperheroPoseEnhancing2022,
  title = {The {{Superhero Pose}}: {{Enhancing Physical Performance}} in {{Exergames}} by {{Embodying Celebrity Avatars}} in {{Virtual Reality}}},
  shorttitle = {The {{Superhero Pose}}},
  booktitle = {Nordic {{Human-Computer Interaction Conference}}},
  author = {Sadek, Nouran and Elagroudy, Passant and Khalil, Ali and Abdennadher, Slim},
  year = {2022},
  month = oct,
  pages = {1--11},
  publisher = {{ACM}},
  address = {{Aarhus Denmark}},
  doi = {10.1145/3546155.3546707},
  urldate = {2022-10-31},
  isbn = {978-1-4503-9699-8},
  langid = {english}
}

@article{shahSocialVRbasedCollaborative2022,
  title = {A Social {{VR-based}} Collaborative Exergame for Rehabilitation: Codesign, Development and User Study},
  shorttitle = {A Social {{VR-based}} Collaborative Exergame for Rehabilitation},
  author = {Shah, Syed Hammad Hussain and Karlsen, Anniken Susanne T. and Solberg, Mads and Hameed, Ibrahim A.},
  year = {2022},
  month = nov,
  journal = {Virtual Reality},
  volume = {},
  pages = {},
  number = {},
  issn = {1434-9957},
  doi = {10.1007/s10055-022-00721-8},
  urldate = {2023-06-22},
  langid = {english}
}

@article{shawCompetitionCooperationVirtual2016,
  title = {Competition and Cooperation with Virtual Players in an Exergame},
  author = {Shaw, Lindsay A. and Buckley, Jude and Corballis, Paul M. and Lutteroth, Christof and Wuensche, Burkhard C.},
  year = {2016},
  month = oct,
  journal = {PeerJ Computer Science},
  volume = {2},
  pages = {e92},
  number = {},
  issn = {2376-5992},
  doi = {10.7717/peerj-cs.92},
  urldate = {2023-06-22},
  langid = {english}
}

@article{staianoAdolescentExergamePlay2013a,
  title = {Adolescent Exergame Play for Weight Loss and Psychosocial Improvement: {{A}} Controlled Physical Activity Intervention},
  shorttitle = {Adolescent Exergame Play for Weight Loss and Psychosocial Improvement},
  author = {Staiano, Amanda E. and Abraham, Anisha A. and Calvert, Sandra L.},
  year = {2013},
  journal = {Obesity},
  volume = {21},
  number = {3},
  pages = {598--601},
  issn = {1930-739X},
  doi = {10.1002/oby.20282},
  urldate = {2023-08-29},
  copyright = {Copyright \textcopyright{} 2012 The Obesity Society},
  langid = {english}
}

@inproceedings{taoEmbodyingPhysicsAwareAvatars2023,
  title = {Embodying {{Physics-Aware Avatars}} in {{Virtual Reality}}},
  booktitle = {Proceedings of the 2023 {{CHI Conference}} on {{Human Factors}} in {{Computing Systems}}},
  author = {Tao, Yujie and Wang, Cheng Yao and Wilson, Andrew D and Ofek, Eyal and {Gonzalez-Franco}, Mar},
  year = {2023},
  month = apr,
  pages = {1--15},
  publisher = {{ACM}},
  address = {{Hamburg Germany}},
  doi = {10.1145/3544548.3580979},
  urldate = {2023-06-22},
  isbn = {978-1-4503-9421-5},
  langid = {english}
}

@article{taylorRelationPhysicalActivity1985,
  title = {The Relation of Physical Activity and Exercise to Mental Health.},
  author = {Taylor, C B and Sallis, J F and Needle, R},
  year = {1985},
  journal = {Public Health Reports},
  volume = {100},
  number = {2},
  pages = {195--202},
  issn = {0033-3549},
  urldate = {2022-02-15},
  pmcid = {PMC1424736},
  pmid = {3920718}
}

@article{vallerandIntegrativeAnalysisIntrinsic1999,
  title = {An Integrative Analysis of Intrinsic and Extrinsic Motivation in Sport},
  author = {Vallerand, Robert J. and Losier, Ga{\'e}tan F.},
  year = {1999},
  month = mar,
  journal = {Journal of Applied Sport Psychology},
  volume = {11},
  number = {1},
  pages = {142--169},
  publisher = {{Routledge}},
  issn = {1041-3200},
  doi = {10.1080/10413209908402956},
  urldate = {2023-06-22}
}

@article{xiaoExerciseCardiovascularProtection2021,
  title = {Exercise and Cardiovascular Protection: {{Update}} and Future},
  shorttitle = {Exercise and Cardiovascular Protection},
  author = {Xiao, Junjie and Rosenzweig, Anthony},
  year = {2021},
  month = dec,
  journal = {Journal of Sport and Health Science},
  volume = {10},
  number = {6},
  pages = {607--608},
  issn = {2095-2546},
  doi = {10.1016/j.jshs.2021.11.001},
  urldate = {2023-06-22},
  pmcid = {PMC8724613},
  pmid = {34793994}
}

@article{woosnam2010inclusion,
  title={The inclusion of other in the self (IOS) scale},
  author={Woosnam, Kyle M and others},
  journal={Annals of Tourism Research},
  volume={37},
  number={3},
  pages={857--860},
  year={2010},
  publisher={Elsevier Ltd}
}

@inproceedings{ostrow2018testing,
  title={Testing the validity and reliability of intrinsic motivation inventory subscales within assistments},
  author={Ostrow, Korinn S and Heffernan, Neil T},
  booktitle={Artificial Intelligence in Education: 19th International Conference, AIED 2018, London, UK, June 27--30, 2018, Proceedings, Part I 19},
  pages={381--394},
  year={2018},
  organization={Springer}
}

@article{bradley1994measuring,
  title={Measuring emotion: the self-assessment manikin and the semantic differential},
  author={Bradley, Margaret M and Lang, Peter J},
  journal={Journal of behavior therapy and experimental psychiatry},
  volume={25},
  number={1},
  pages={49--59},
  year={1994},
  publisher={Elsevier}
}

@article{meauley1994subjective,
  title={The subjective exercise experiences scale (SEES): Development and preliminary validation},
  author={MeAuley, Edward and Courneya, Kerry S},
  journal={Journal of Sport and Exercise Psychology},
  volume={16},
  number={2},
  pages={163--177},
  year={1994},
  publisher={Human Kinetics, Inc.}
}

@article{mallett2007sport,
  title={Sport motivation scale-6 (SMS-6): A revised six-factor sport motivation scale},
  author={Mallett, Clifford and Kawabata, Masato and Newcombe, Peter and Otero-Forero, Andres and Jackson, Susan},
  journal={Psychology of sport and exercise},
  volume={8},
  number={5},
  pages={600--614},
  year={2007},
  publisher={Elsevier}
}

@article{lu2013cooperativeness,
  title={Cooperativeness and competitiveness as two distinct constructs: Validating the Cooperative and Competitive Personality Scale in a social dilemma context},
  author={Lu, Su and Au, Wing-Tung and Jiang, Feng and Xie, Xiaofei and Yam, Paton},
  journal={International Journal of Psychology},
  volume={48},
  number={6},
  pages={1135--1147},
  year={2013},
  publisher={Taylor \& Francis}
}

@article{slater2000virtual,
  title={A virtual presence counter},
  author={Slater, Mel and Steed, Anthony},
  journal={Presence},
  volume={9},
  number={5},
  pages={413--434},
  year={2000},
  publisher={MIT Press}
}

@misc{who_sex,   
    title = {Sexual health},   
    url = {https://www.who.int/health-topics/sexual-health#tab=tab_2},   
    author = {World Health Organization},   
    year = {2023},   
    note = {Accessed on September 14, 2023} 
}

@article{RODGERS2022284,
title = {Social media and body image: Modulating effects of social identities and user characteristics},
journal = {Body Image},
volume = {41},
pages = {284-291},
year = {2022},
issn = {1740-1445},
doi = {https://doi.org/10.1016/j.bodyim.2022.02.009},
url = {https://www.sciencedirect.com/science/article/pii/S1740144522000407},
author = {Rachel F. Rodgers and Ann Rousseau}
}

@article{battertonLikertScaleWhat2017,
  title = {The {{Likert Scale What It Is}} and {{How To Use It}}},
  author = {Batterton, Katherine A. and Hale, Kimberly N.},
  year = {2017},
  journal = {Phalanx},
  volume = {50},
  number = {2},
  eprint = {26296382},
  eprinttype = {jstor},
  pages = {32--39},
  publisher = {{Military Operations Research Society}},
  issn = {0195-1920},
  urldate = {2023-11-20}
}

@article{brondiEvaluatingEffectsCompetition2015,
  title = {Evaluating the Effects of Competition vs Collaboration on User Engagement in an Immersive Game Using Natural Interaction},
  author = {Brondi, R. and Avveduto, G. and Alem, L. and Faita, C. and Carrozzino, M. and Tecchia, F. and Pisan, Y. and Bergamasco, M.},
  year = {2015},
  month = nov,
  journal = {Proceedings of the 21st ACM Symposium on Virtual Reality Software and Technology},
  pages = {191--191},
  publisher = {{ACM}},
  address = {{Beijing China}},
  doi = {10.1145/2821592.2821643},
  urldate = {2023-11-20},
  isbn = {9781450339902},
  langid = {english}
}

@article{edwardsImpactActivePassive2018b,
  title = {Impact of Active and Passive Social Facilitation on Self-Paced Endurance and Sprint Exercise: Encouragement Augments Performance and Motivation to Exercise},
  shorttitle = {Impact of Active and Passive Social Facilitation on Self-Paced Endurance and Sprint Exercise},
  author = {Edwards, Andrew Mark and {Dutton-Challis}, Lia and Cottrell, David and Guy, Joshua H and Hettinga, Florentina Johanna},
  year = {2018},
  month = jul,
  journal = {BMJ Open Sport \& Exercise Medicine},
  volume = {4},
  number = {1},
  pages = {e000368},
  issn = {2055-7647},
  doi = {10.1136/bmjsem-2018-000368},
  urldate = {2023-11-20},
  langid = {english}
}

@article{jicolPredictiveModelUnderstanding2023,
  title = {A Predictive Model for Understanding the Role of Emotion for the Formation of Presence in Virtual Reality},
  author = {Jicol, Crescent and Cheng, Hoi Ying and Petrini, Karin and O'Neill, Eamonn},
  year = {2023},
  month = mar,
  journal = {PLOS ONE},
  volume = {18},
  number = {3},
  pages = {e0280390},
  publisher = {{Public Library of Science}},
  issn = {1932-6203},
  doi = {10.1371/journal.pone.0280390},
  urldate = {2023-11-20},
  langid = {english}
}

@inproceedings{kocurEffectsSelfExternal2020,
  title = {The {{Effects}} of {{Self-}} and {{External Perception}} of {{Avatars}} on {{Cognitive Task Performance}} in {{Virtual Reality}}},
  booktitle = {Proceedings of the 26th {{ACM Symposium}} on {{Virtual Reality Software}} and {{Technology}}},
  author = {Kocur, Martin and Schauhuber, Philipp and Schwind, Valentin and Wolff, Christian and Henze, Niels},
  year = {2020},
  month = nov,
  series = {{{VRST}} '20},
  pages = {1--11},
  publisher = {{Association for Computing Machinery}},
  address = {{New York, NY, USA}},
  doi = {10.1145/3385956.3418969},
  urldate = {2023-11-20},
  isbn = {978-1-4503-7619-8}
}

@article{maisterChangingBodiesChanges2015,
  title = {Changing Bodies Changes Minds: Owning Another Body Affects Social Cognition},
  shorttitle = {Changing Bodies Changes Minds},
  author = {Maister, Lara and Slater, Mel and {Sanchez-Vives}, Maria V. and Tsakiris, Manos},
  year = {2015},
  month = jan,
  journal = {Trends in Cognitive Sciences},
  volume = {19},
  number = {1},
  pages = {6--12},
  publisher = {{Elsevier}},
  issn = {1364-6613, 1879-307X},
  doi = {10.1016/j.tics.2014.11.001},
  urldate = {2023-11-20},
  langid = {english},
  pmid = {25524273}
}

@article{mcneillSelfmodelledSkilledpeerModelled2020,
  title = {Self-Modelled versus Skilled-Peer Modelled {{AO}}+{{MI}} Effects on Skilled Sensorimotor Performance},
  author = {McNeill, Eoghan and Toth, Adam J. and Ramsbottom, Niall and Campbell, Mark J.},
  year = {2020},
  month = jul,
  journal = {Psychology of Sport and Exercise},
  volume = {49},
  pages = {101683},
  issn = {1469-0292},
  doi = {10.1016/j.psychsport.2020.101683},
  urldate = {2023-11-20}
}

@article{mcneillSelfmodelledSkilledpeerModelled2021,
  title = {Self-Modelled versus Skilled-Peer Modelled {{AO}}+{{MI}} Effects on Skilled Sensorimotor Performance: {{A}} Stage 2 Registered Report},
  shorttitle = {Self-Modelled versus Skilled-Peer Modelled {{AO}}+{{MI}} Effects on Skilled Sensorimotor Performance},
  author = {McNeill, Eoghan and Toth, Adam J. and Ramsbottom, Niall and Campbell, Mark J.},
  year = {2021},
  month = may,
  journal = {Psychology of Sport and Exercise},
  volume = {54},
  pages = {101910},
  issn = {1469-0292},
  doi = {10.1016/j.psychsport.2021.101910},
  urldate = {2023-11-20}
}

@inproceedings{michaelRaceYourselvesLongitudinal2020,
  title = {Race {{Yourselves}}: {{A Longitudinal Exploration}} of {{Self-Competition Between Past}}, {{Present}}, and {{Future Performances}} in a {{VR Exergame}}},
  shorttitle = {Race {{Yourselves}}},
  booktitle = {Proceedings of the 2020 {{CHI Conference}} on {{Human Factors}} in {{Computing Systems}}},
  author = {Michael, Alexander and Lutteroth, Christof},
  year = {2020},
  month = apr,
  series = {{{CHI}} '20},
  pages = {1--17},
  publisher = {{Association for Computing Machinery}},
  address = {{New York, NY, USA}},
  doi = {10.1145/3313831.3376256},
  urldate = {2023-11-20},
  isbn = {978-1-4503-6708-0}
}

@article{murrayEffectsPresenceOthers2016,
  title = {The Effects of the Presence of Others during a Rowing Exercise in a Virtual Reality Environment},
  author = {Murray, Edward G. and Neumann, David L. and Moffitt, Robyn L. and Thomas, Patrick R.},
  year = {2016},
  month = jan,
  journal = {Psychology of Sport and Exercise},
  volume = {22},
  pages = {328--336},
  issn = {1469-0292},
  doi = {10.1016/j.psychsport.2015.09.007},
  urldate = {2023-11-20}
}

@inproceedings{nunesMotivatingPeoplePerform2014,
  title = {Motivating People to Perform Better in Exergames: Competition in Virtual Environments},
  shorttitle = {Motivating People to Perform Better in Exergames},
  booktitle = {Proceedings of the 29th {{Annual ACM Symposium}} on {{Applied Computing}}},
  author = {Nunes, Mateus and Nedel, Luciana and Roesler, Valter},
  year = {2014},
  month = mar,
  series = {{{SAC}} '14},
  pages = {970--975},
  publisher = {{Association for Computing Machinery}},
  address = {{New York, NY, USA}},
  doi = {10.1145/2554850.2555009},
  urldate = {2023-11-20},
  isbn = {978-1-4503-2469-4}
}

@article{panVirtualCharacterPersonality2015,
  title = {Virtual {{Character Personality Influences Participant Attitudes}} and {{Behavior}} \textendash{} {{An Interview}} with a {{Virtual Human Character}} about {{Her Social Anxiety}}},
  author = {Pan, Xueni and Gillies, Marco and Slater, Mel},
  year = {2015},
  journal = {Frontiers in Robotics and AI},
  pages = {},
  number = {},
  volume = {2},
  issn = {2296-9144},
  urldate = {2023-11-20}
}

@article{partonEffectsCompetitivenessChallenge2019,
  title = {The Effects of Competitiveness and Challenge Level on Virtual Reality Rowing Performance},
  author = {Parton, Brett J. and Neumann, David L.},
  year = {2019},
  month = mar,
  journal = {Psychology of Sport and Exercise},
  volume = {41},
  pages = {191--199},
  issn = {1469-0292},
  doi = {10.1016/j.psychsport.2018.06.010},
  urldate = {2023-11-20}
}

@article{pataneExploringEffectCooperation2020,
  title = {Exploring the {{Effect}} of {{Cooperation}} in {{Reducing Implicit Racial Bias}} and {{Its Relationship With Dispositional Empathy}} and {{Political Attitudes}}},
  author = {Patan{\'e}, Ivan and Lelgouarch, Anne and Banakou, Domna and Verdelet, Gregoire and Desoche, Clement and Koun, Eric and Salemme, Romeo and Slater, Mel and Farn{\`e}, Alessandro},
  year = {2020},
  journal = {Frontiers in Psychology},
  volume = {11},
  number = {},
  pages = {},
  issn = {1664-1078},
  urldate = {2023-11-20}
}

@article{penaAmWhatSee2016,
  title = {I {{Am What I See}}: {{How Avatar}} and {{Opponent Agent Body Size Affects Physical Activity Among Men Playing Exergames}}},
  shorttitle = {I {{Am What I See}}},
  author = {Pe{\~n}a, Jorge and Khan, Subuhi and Alexopoulos, Cassandra},
  year = {2016},
  month = may,
  journal = {Journal of Computer-Mediated Communication},
  volume = {21},
  number = {3},
  pages = {195--209},
  issn = {1083-6101},
  doi = {10.1111/jcc4.12151},
  urldate = {2023-11-20}
}

@article{planteDoesVirtualReality2003,
  title = {Does Virtual Reality Enhance the Psychological Benefits of Exercise?},
  author = {Plante, Thomas G. and Frazier, S. and Tittle, A. and Babula, M. and Ferlic, E. and Riggs, E.},
  year = {2003},
  journal = {Journal of Human Movement Studies},
  volume = {45},
  number = {6},
  pages = {485--507},
  publisher = {{Teviot Scientific Publications}},
  address = {{United Kingdom}},
  issn = {0306-7297}
}

@inproceedings{praetoriusUserAvatarRelationshipsVarious2021,
  title = {User-{{Avatar Relationships}} in {{Various Contexts}}: {{Does Context Influence}} a {{Users}}' {{Perception}} and {{Choice}} of an {{Avatar}}?},
  shorttitle = {User-{{Avatar Relationships}} in {{Various Contexts}}},
  booktitle = {Proceedings of {{Mensch}} Und {{Computer}} 2021},
  author = {Praetorius, Anna Samira and Krautmacher, Lara and Tullius, Gabriela and Curio, Crist{\'o}bal},
  year = {2021},
  month = sep,
  series = {{{MuC}} '21},
  pages = {275--280},
  publisher = {{Association for Computing Machinery}},
  address = {{New York, NY, USA}},
  doi = {10.1145/3473856.3474007},
  urldate = {2023-11-20},
  isbn = {978-1-4503-8645-6}
}

@article{rahillEffectsAvatarPlayersimilarity2021,
  title = {Effects of {{Avatar}} Player-Similarity and Player-Construction on Gaming Performance},
  author = {Rahill, Katherine M. and Sebrechts, Marc M.},
  year = {2021},
  month = aug,
  journal = {Computers in Human Behavior Reports},
  volume = {4},
  pages = {100131},
  issn = {2451-9588},
  doi = {10.1016/j.chbr.2021.100131},
  urldate = {2023-11-20}
}

@inproceedings{satoCollaborativeDigitalSports2014a,
  title = {Collaborative {{Digital Sports Systems That Encourage Exercise}}},
  booktitle = {Human-{{Computer Interaction}}. {{Applications}} and {{Services}}},
  author = {Sato, Ayaka and Yokokubo, Anna and Siio, Itiro and Rekimoto, Jun},
  editor = {Kurosu, Masaaki},
  year = {2014},
  series = {Lecture {{Notes}} in {{Computer Science}}},
  pages = {332--340},
  publisher = {{Springer International Publishing}},
  address = {{Cham}},
  doi = {10.1007/978-3-319-07227-2_32},
  isbn = {978-3-319-07227-2},
  langid = {english}
}

@article{sekhavatCollaborationBattleMinds2020,
  title = {Collaboration or Battle between Minds? {{An}} Attention Training Game through Collaborative and Competitive Reinforcement},
  shorttitle = {Collaboration or Battle between Minds?},
  author = {Sekhavat, Yoones A.},
  year = {2020},
  month = may,
  journal = {Entertainment Computing},
  volume = {34},
  pages = {100360},
  issn = {1875-9521},
  doi = {10.1016/j.entcom.2020.100360},
  urldate = {2023-11-20}
}

@article{tammylinExercisingEmbodiedYoung2021,
  title = {Exercising {{With Embodied Young Avatars}}: {{How Young}} vs. {{Older Avatars}} in {{Virtual Reality Affect Perceived Exertion}} and {{Physical Activity Among Male}} and {{Female Elderly Individuals}}},
  shorttitle = {Exercising {{With Embodied Young Avatars}}},
  author = {Tammy Lin, Jih-Hsuan and Wu, Dai-Yun},
  year = {2021},
  journal = {Frontiers in Psychology},
  volume = {12},
  number = {},
  pages = {},
  issn = {1664-1078},
  urldate = {2023-11-20}
}

@article{mcwhorterObeseChildMotivation2003,
  title = {The Obese Child: {{Motivation}} as a Tool for Exercise},
  shorttitle = {The Obese Child},
  author = {McWhorter, J. Wesley and Wallmann, Harvey W. and Alpert, Patricia T.},
  year = {2003},
  month = jan,
  journal = {Journal of Pediatric Health Care},
  volume = {17},
  number = {1},
  pages = {11--17},
  issn = {0891-5245},
  doi = {10.1067/mph.2003.25},
  urldate = {2023-11-30}
}

@misc{RelativeEffectsPositive,
  title = {On the Relative Effects of Positive and Negative Verbal Feedback on Males' and Females' Intrinsic Motivation.},
  urldate = {2023-11-30},
  howpublished = {https://psycnet.apa.org/record/1989-11791-001?casa\_token=oPuHr2zHIFYAAAAA:2-E2ZkfTfguxH70WIukQhxQ4\_8CRutX3\_H-4VN4NZW1x5n44h2eMUtdzXoOKDbB2t3T5g0Ev-LGTqA5suSE2gtdr},
  note = {(Accessed 30.11.2023)}
}

@article{savikkoPsychosocialGroupRehabilitation2010,
  title = {Psychosocial Group Rehabilitation for Lonely Older People: Favourable Processes and Mediating Factors of the Intervention Leading to Alleviated Loneliness},
  shorttitle = {Psychosocial Group Rehabilitation for Lonely Older People},
  author = {Savikko, Niina and Routasalo, Pirkko and Tilvis, Reijo and Pitk{\"a}l{\"a}, Kaisu},
  year = {2010},
  journal = {International Journal of Older People Nursing},
  volume = {5},
  number = {1},
  pages = {16--24},
  issn = {1748-3743},
  doi = {10.1111/j.1748-3743.2009.00191.x},
  urldate = {2023-11-30},
  copyright = {{\textcopyright} 2009 Blackwell Publishing Ltd},
  langid = {english}
}

@article{weinsteinYouHaveHear2018,
  title = {You `Have' to Hear This: {{Using}} Tone of Voice to Motivate Others},
  shorttitle = {You `Have' to Hear This},
  author = {Weinstein, Netta and Zougkou, Konstantina and Paulmann, Silke},
  year = {2018},
  journal = {Journal of Experimental Psychology: Human Perception and Performance},
  volume = {44},
  number = {6},
  pages = {898-913},
  publisher = {{American Psychological Association}},
  address = {{US}},
  issn = {1939-1277},
  doi = {10.1037/xhp0000502}
}

@inproceedings{latoschikEffectAvatarRealism2017,
  title = {The Effect of Avatar Realism in Immersive Social Virtual Realities},
  booktitle = {Proceedings of the 23rd {{ACM Symposium}} on {{Virtual Reality Software}} and {{Technology}}},
  author = {Latoschik, Marc Erich and Roth, Daniel and Gall, Dominik and Achenbach, Jascha and Waltemate, Thomas and Botsch, Mario},
  year = {2017},
  month = nov,
  series = {{{VRST}} '17},
  pages = {1--10},
  publisher = {{Association for Computing Machinery}},
  address = {{New York, NY, USA}},
  doi = {10.1145/3139131.3139156},
  urldate = {2020-10-19},
  isbn = {978-1-4503-5548-3}
}
